\documentclass{CSML}

\def\dOi{11(4:17)2015}
\lmcsheading%
{\dOi}
{1--44}
{}
{}
{Apr.~19, 2015}
{Dec.~29, 2015}
{}

\ACMCCS{[{\bf Theory of computation}]: Logic---Finite Model Theory}
\subjclass{F.4.1 [\textit{Mathematical Logic and Formal Languages}]:
  Mathematical Logic---Computational Logic, Model Theory}

\usepackage{hyperref}
\usepackage{hyphenat}
\usepackage[shortcuts]{extdash}
\usepackage{url} 
\usepackage{stmaryrd,mathrsfs}
\usepackage{amsmath,amssymb,latexsym}
\usepackage{xspace,ifthen} 
\usepackage{calc}
\usepackage{framed}
\usepackage{graphicx}
\usepackage[noadjust]{cite}
\usepackage{tikz}
\usepackage{subfigure}
\usepackage{mathtools}
\usepackage{mathdots}
\usepackage{amsmath,amssymb,amsfonts,mathrsfs,latexsym,stmaryrd,bbold}
\usepackage{amsthm}

\theoremstyle{plain}

\newtheorem{lemma}[thm]{Lemma}

\newcounter{claimcounter}

\newcommand{\newclaims}{\setcounter{claimcounter}{0}}
\newcommand{\uproof}[1]{\noindent\textit{Proof of #1.\ }}

\theoremstyle{definition}

\theoremstyle{remark}

\usepackage{todonotes}

\newcommand{\isom}{\cong}

\newcommand{\bit}{\ensuremath{\mathop{\text{bit}}}}

\newcommand{\sph}{\textit{sph}}
\newcommand{\isdef}{\mathrel{\mathop:}=}

\newcommand{\Tower}{\text{Tower}}
\newcommand{\bigO}{\ensuremath{O}}
\newcommand{\feq}{\textit{eq}}
\renewcommand{\phi}{\varphi}
\newcommand{\lin}[1]{\bigO(#1)}
\newcommand{\poly}[1]{\operatorname{poly}(#1)}

\newcommand*{\ov}[1]{\ensuremath{\overline{#1}}}

\newcommand*{\ar}{\ensuremath{\operatorname{ar}}} 
\newcommand*{\qr}{\ensuremath{\operatorname{qr}}} 
\newcommand*{\size}[1]{\ensuremath{|\!|#1|\!|}}
\newcommand{\Root}{\ensuremath{\textit{root}}}
\newcommand{\bin}{\operatorname{bin}}

\newcommand{\free}{\text{free}}

\newcommand*{\A}{\ensuremath{\mathcal{A}}}
\newcommand*{\B}{\ensuremath{\mathcal{B}}}
\newcommand*{\C}{\ensuremath{\mathcal{C}}}
\newcommand*{\D}{\ensuremath{\mathcal{D}}}
\newcommand*{\F}{\ensuremath{\mathcal{F}}}

\newcommand*{\T}{\ensuremath{\mathcal{T}}}
\newcommand*{\N}{\ensuremath{\mathcal{N}}}

\newcommand*{\class}[1]{\ensuremath{\textrm{\upshape{#1}}}}
\newcommand{\Class}{\ensuremath{\mathfrak{C}}}

\newcommand*{\FO}{\class{FO}}

\renewcommand*{\geq}{\geqslant}
\renewcommand*{\leq}{\leqslant}

\newcommand*{\FOMOD}[1]{\class{FO+MOD$_{#1}$}}

\newcommand*{\ZZ}{\ensuremath{\mathbb{Z}}}
\newcommand*{\NN}{\ensuremath{\mathbb{N}}}
\newcommand*{\RR}{\ensuremath{\mathbb{R}}}
\newcommand*{\NNpos}{\ensuremath{\NN_{\mbox{\tiny $\scriptscriptstyle
        \geq 1$}}}}
\newcommand*{\RRpos}{\ensuremath{\RR_{\mbox{\tiny $\scriptscriptstyle
        \geq 0$}}}}
\newcommand*{\set}[1]{\ensuremath{\{ #1 \}}}
\newcommand*{\setc}[2]{\set{#1 \,:\, #2}}

\newcommand*{\und}{\ensuremath{\wedge}}
\newcommand*{\Und}{\ensuremath{\bigwedge}}
\newcommand*{\oder}{\ensuremath{\vee}}
\newcommand*{\Oder}{\ensuremath{\bigvee}}

\newcommand*{\nicht}{\ensuremath{\neg}}
\newcommand*{\impl}{\ensuremath{\rightarrow}}
\newcommand*{\gdw}{\ensuremath{\leftrightarrow}}

\newcommand{\dist}{\ensuremath{\textit{dist}}}

\newcommand{\dcup}{\ensuremath{\mathop{\dot{\cup}}}}

\newcommand{\fmin}{\textit{min}}
\newcommand{\fmax}{\textit{max}}

\newcommand{\modex}[1]{\ensuremath{\exists^{0\,\textup{mod}\,#1}}}

\newcommand{\existm}[2]{\ensuremath{\exists^{#1\,\textup{mod}\,#2}}}

\newcommand{\AVARS}{X}
\newcommand{\AVAR}{X}

\begin{document}

\title[Preservation and decomposition theorems for bounded degree structures]{Preservation and decomposition theorems \\ for bounded degree structures\rsuper*}

\author[F.~Harwath]{Frederik Harwath\rsuper a}
\address{{\lsuper a}Institut f\"ur Informatik\\ Goethe-Universit\"at Frankfurt am Main}
\email{harwath@cs.uni-frankfurt.de}

\author[L.~Heimberg]{Lucas Heimberg\rsuper b}
\address{{\lsuper{b,c}}Institut f\"ur Informatik\\ Humboldt-Universit\"at zu Berlin}
\email{\{lucas.heimberg, schweika\}@informatik.hu-berlin.de}

\author[N.~Schweikardt]{Nicole Schweikardt\rsuper c}
\address{\vspace{-18 pt}}
%\email{schweika@informatik.hu-berlin.de}

\keywords{computational logic,
first-order logic,
modulo counting quantifiers,
structures of bounded degree, Hanf locality, 
elementary algorithms,
preservation theorems, existential preservation, homomorphism preservation,
\L{}o\'{s}-Tarski,
Feferman-Vaught}

\titlecomment{{\lsuper*}The present paper is the full version of the conference contribution \cite{HHS_CSLLICS14}}

\begin{abstract}
  We provide elementary algorithms for two preservation theorems for first-order sentences ($\FO$) on the class $\Class_d$ of all finite structures of degree at most $d$: For each $\FO$-sentence that is preserved under extensions (homomorphisms) on $\Class_d$, a $\Class_d$-equivalent existential (existential-positive) $\FO$-sentence can be constructed in 5\=/fold (4\=/fold) exponential time. This is complemented by lower bounds showing that a $3$-fold exponential blow-up of the computed existential (existential-positive) sentence is unavoidable.
  Both algorithms can be extended (while maintaining the upper and lower bounds on their time complexity) to input first-order sentences with modulo $m$ counting quantifiers~($\FOMOD{m}$). 

Furthermore, we show that for an input $\FO$\=/formula, a $\mathfrak{C}_d$\=/equivalent 
Feferman\=/Vaught decomposition can be computed in 3\=/fold exponential time. We also
provide a matching lower bound.
\end{abstract}

\maketitle

\section{Introduction}\label{Introduction}

Classical \emph{preservation theorems} studied in model theory relate
syntactic restrictions of formulas with structural properties of the
classes of structures defined. 
For example, the \emph{\L{}o\'{s}-Tarski theorem} states that a first-order
sentence is preserved under extensions on the class of all structures
if, and only if, it is equivalent, on this class, to an
existential first-order sentence.
The \emph{homomorphism preservation theorem} states that a first-order
sentence is preserved under homomorphisms on the class of all
structures if, and only if, it is equivalent, on this class, to an
existential-positive first-order sentence.

In the last decade, variants of both theorems have been obtained, where
the class of all structures is replaced by restricted classes
that meet certain requirements.
For example,
\cite{Rossman08-homomorphism,AtseriasDK06-homomorphism-preservation,Dawar10-homomorphism-quasi-wide}
obtained that the homomorphism preservation theorem holds for the
class of all finite structures, as well as for the classes of all
finite structures of degree at most $d$ or of treewidth at most $k$, and, in
general, for \emph{quasi-wide} classes of structures that are closed
under taking substructures and disjoint unions (this includes classes
of bounded expansion and classes that locally
exclude minors).
While the \L{}o\'{s}-Tarski theorem is known to fail on the class of
all finite structures,
in \cite{ADG-Preservation-2008} 
it was shown to hold for various classes of structures, including
the class of all finite structures of degree at most $d$,
the class of all finite structures of  treewidth at most $k$,
and all \emph{wide} classes of structures that are 
closed under taking substructures and disjoint unions.

For most of these results, it is known that the equivalent
existential or existential-positive sentence may be non-elementarily
larger than the corresponding first-order sentence \cite{DGKS07}.
A notable exception affects the \L{}o\'{s}-Tarski theorem for the class
of acyclic finite structures of degree at most $d$, for which
\cite{DGKS07} obtained a 5-fold exponential upper bound on the size of the
equivalent existential first-order sentence.

The present paper's first main result
(Theorem~\ref{thm:ep-upper-bound}) generalises the latter in three ways:
(1)~We show that the 5-fold exponential upper bound for the
\L{}o\'{s}-Tarski theorem holds for \emph{every} 
class $\mathfrak{C}$ of structures of degree at most $d$ that is closed under taking
induced substructures and disjoint unions (this includes, e.g.,
the class of all finite structures of degree at most $d$).
(2)~We provide an \emph{algorithmic} version of the theorem, showing
that for a given first-order sentence, the existential sentence can be
constructed in 5-fold exponential time.
(3)~Our algorithm also works for input sentences of the extension
$\FOMOD{m}$ of first-order logic with modulo $m$ counting quantifiers.
The main ingredient of our proof is a new, technically challenging
upper bound on the size of \emph{minimal} models of sentences that are
preserved under extensions on $\mathfrak{C}$
(Theorem~\ref{thm:ep-bound-minimal-model}).

Our second main result (Theorem~\ref{thm:hp-upper-bound}) provides an
algorithmic version of the homomorphism preservation theorem over any 
class $\mathfrak{C}$ of structures of bounded degree that
is closed under taking induced substructures and disjoint unions, and
that is decidable
in 1-fold exponential time (e.g., the
class of all finite structures of degree at most $d$). Specifically, 
we show that for a given $\FOMOD{m}$-sentence that is preserved under 
homomorphisms on $\mathfrak{C}$, an equivalent existential-positive first-order
sentence can be constructed in 4-fold exponential time.
The proof, again, relies on a new upper bound on the size of minimal
models (Theorem~\ref{thm:hp-bound-minimal-model}).

Two counterexamples (Theorem~\ref{thm:counterexample-disjoint-union} and Theorem~\ref{thm:counterexample-substructures}) show that the closure properties of the classes of structures considered, required by our preservation theorems, are indeed necessary.

We complement our preservation theorems by 
lower bounds (Theorem~\ref{thm:hp-lower-bound} and
Theorem~\ref{thm:ep-lower-bound}), providing a sequence of 
first-order sentences that are preserved under extensions
(homomorphisms) for which the smallest equivalent 
existential (respectively, existential-positive) sentences are 3-fold exponentially larger.
Both lower bound proofs use particular encodings of numbers by binary trees
introduced in \cite{HKS13-LICS}.

Our third main result deals with \emph{Feferman-Vaught decompositions}
of first-order formulas. 
The classical Feferman-Vaught theorem states that for certain forms of
compositions of structures, the theory of a structure composed from
simpler structures is determined by the theories of the simpler structures.
This applies, for example, to 
disjoint sums and direct products (also known as cartesian products or as tensor products) of structures (cf., e.g., \cite{Hodges}).
Feferman-Vaught-like theorems find application in results about the \emph{decidability} of theories, as well as in results about \emph{model checking} and \emph{satisfiability checking} \cite{Makowsky2004159,GoeJL12}. Another use of Feferman-Vaught decompositions is within the proof of \emph{Gaifman's theorem} \cite{Gaifman}, which is an important tool for inexpressibility results as well as for so-called algorithmic meta-theorems~\cite{Kreutzer-AMT-Survey}.

Algorithmic versions of decomposition theorems \`{a} la Feferman-Vaught 
are typically of the following form (cf., \cite{Makowsky2004159,GoeJL12}):
A given first-order sentence $\phi$ that shall be evaluated in the disjoint sum or the
direct product $\A$ of $s$ structures $\A_1,\ldots,\A_s$, can be transformed into 
a finite set $\Delta$ of formulas and a propositional formula $\beta$
whose propositions are tests of the form ``the $i$-th structure $\A_i$
 satisfies the $j$-th formula in $\Delta$'', such that $\A$ is a model of $\phi$ iff $\beta$
is true. It is known that the Feferman-Vaught decomposition $(\Delta, \beta)$ may be non-elementarily larger than $\phi$ \cite{DGKS07}.
Our third main result (Theorem~\ref{thm:fv-upper-bound} and
Corollary~\ref{cor:fv-upper-bound-tensorprod}) shows that for any class
$\mathfrak{C}$ of structures of degree at most~$d$, such 
Feferman-Vaught decompositions for disjoint sums and direct
products can be computed in 3\hbox{-}fold exponential time. This is
complemented by a matching lower bound
(Theorem~\ref{thm:fv-lower-bound-BF}).
Our lower bound proof, again, relies on encodings of numbers by binary
trees, now along with a method of \cite{GoeJL12}. Our algorithm
produces a set $\Delta$ of so-called \emph{Hanf-formulas} and relies
on a result of \cite{BolligKuske-Hanf-2012} that transforms the given
sentence $\phi$ into Hanf normal form.

%\medskip
The rest of this paper is structured as follows: 
Section~\ref{Preliminaries} fixes the basic notations.
Section~\ref{section:PreservationTheorems} presents our algorithms concerning preservation theorems.
Section~\ref{fv} presents our results concerning Feferman-Vaught decompositions.
Section~\ref{section:lower-bounds} contains lower bounds that complement
our results of Section~\ref{section:PreservationTheorems} and Section~\ref{fv}.
Section~\ref{section:conclusion} gives a short conclusion and provides directions for future work.

\section{Preliminaries}\label{Preliminaries}

We write $\ZZ$ for the set of integers and $\NN$ for the set of non-negative integers.
For all $m,n\in\NN$ with $m \leq n$, we denote the set $\setc{i\in\NN}{m\leq i\leq n}$ by $[m,n]$. 
The set of non-negative real numbers is denoted by $\RRpos$.
For $r>0$, by $\log(r)$ (respectively, $\log_b(r)$) we denote the logarithm of $r$ with
respect to base~$2$ (respectively, base $b$, for $b\geq 2$).

If $f$ is a function from~$\NN$ to $\RRpos$, then
$\poly{f(n)}$ denotes the class of all functions~$g\colon\NN\to\RRpos$ for
which there is a number~$c>0$ such that 
$g(n)\leq (f(n))^c$ is true for all sufficiently large $n\in\NN$.

We say that a function $f$ from $\NN$ to $\RRpos$ is \emph{at most $k$-fold exponential}, for some $k\geq 1$, 
if there exists a number $c>0$ such that for all sufficiently large $n\in\NN$ we have 
$f(n)\leq \textit{T}(k,n^c)$, where  $T(k,m)$ is a tower of 2s of height $k$ with an $m$ on top
(i.e., $\textit{T}(1,m)= 2^m$ and 
$\textit{T}(k{+}1,m)= 2^{\textit{T}(k,m)}$ for all $k\geq 1$ and $m\geq 0$). 

For a function $f \colon A \to B$ and a subset $A' \subseteq A$, we denote by $f(A')$ the set $\setc{ f(a) }{a \in A'}$. 
For $n\in\NN$, we write $\ov{x}$ to denote the tuple
$(x_1,\ldots,x_n)$. 
For a tuple $\ov{x} \in A^n$ we
write $f(\ov{x})$ to denote the tuple $(f(x_1),\ldots,f(x_n))$.
Sometimes we treat tuples as if they were sets;
e.g., \mbox{$a\in \ov{x}$} means $a\in \set{x_1,\ldots,x_n}$. 

\smallskip

\subsection{Structures and formulas}

A \emph{signature} $\sigma$ is a finite set of relation and constant
symbols. Associated with every relation symbol $R$ is a positive
integer $\ar(R)$ called the \emph{arity} of $R$. 
The \emph{size} $\size{\sigma}$ of $\sigma$ is the number of its
constant symbols plus the sum of the arities of its relation symbols.
A signature $\sigma$ is called \emph{relational} if it does not contain any constant symbol.

A \emph{$\sigma$-structure} $\A$ consists of a non-empty set $A$ called the
\emph{universe} of $\A$, a relation $R^\A\subseteq A^{\ar(R)}$ for
each relation symbol $R\in\sigma$, and an element $c^\A\in A$ for each
constant symbol $c\in\sigma$.
The size $\size{\A}$ of $\A$ is the size of a reasonable
representation of $\A$ as a binary string (cf., e.g.,
\cite{EbbinghausFlum, Libkin-FMT}); 
in particular,
$\size{\A}\in \bigO(|A|^{\size{\sigma}})$.

For a relational signature $\sigma$ and $\sigma$-structures $\A$ and
$\B$, we say that $\B$ is a \emph{substructure} of $\A$ if $B \subseteq A$ and
$R^{\B} \subseteq R^{\A}$  for each  $R \in \sigma$. The structure $\B$ is an
\emph{induced substructure} of $\A$ if $\B$ is a substructure of $\A$
and $R^{\B} = R^{\A} \cap B^{\ar(R)}$ for each $R\in\sigma$. 
We then say that $\B$ is the substructure of $\A$ \emph{induced} by the set $B$. 

For every non-empty set $B$ such that $A \cap B \not=\emptyset$, we write
$\A[B]$ to denote the substructure of $\A$ induced by $A \cap B$. 
Furthermore, if $A \setminus B \not=\emptyset$ then $\A - B$ is the induced substructure $\A[A \setminus B]$ of $\A$ obtained by deleting all elements from $B$. 

We use the standard notation concerning first-order logic and
extensions thereof, cf.\ 
\cite{EbbinghausFlum, Libkin-FMT}. 
By $\qr(\varphi)$ we denote the \emph{quantifier rank} of $\varphi$,
i.e., the maximum nesting depth of quantifiers occurring in
a formula $\varphi$.
By $\free(\varphi)$ we denote the set of all free variables of
$\varphi$. A \emph{sentence} is a formula $\varphi$ with
$\free(\varphi)=\emptyset$. A $\sigma$-structure $\A$ is called a
\emph{model} of a sentence $\varphi$ if $\varphi$ is satisfied in~$\A$.

We write $\varphi(\ov{x})$, for $\ov{x}=(x_1,\ldots,x_n)$ with $n\geq 0$, to
indicate that $\free(\varphi)\subseteq\set{x_1,\ldots,x_n}$. 
If $\A$ is a $\sigma$-structure and $\ov{a} = (a_1,\ldots,a_n)\in A^n$, we write
$\A\models\varphi[\ov{a}]$ to indicate that the formula
$\varphi(\ov{x})$ is satisfied in $\A$ when interpreting the free
occurrences of the variables $x_1,\ldots,x_n$ with the elements
$a_1,\ldots,a_n$.
We write $\phi(\A)$ to denote the set of all tuples $\ov{a} \in A^n$
such that $\A \models \phi[\ov{a}]$. 
For a class $\Class$ of structures,
two formulas $\varphi(\ov{x})$ and $\psi({\ov{x}})$ of signature
$\sigma$ are called
\emph{equivalent on $\Class$} (for short: \emph{$\Class$-equivalent}) if for
all $\sigma$\hbox{-}structures \mbox{$\A\in\Class$} 
we have $\varphi(\A)=\psi(\A)$.

By $\FO(\sigma)$ we denote the class of all first-order formulas
of signature $\sigma$. The extension of $\FO(\sigma)$ by \emph{modulo
  counting quantifiers} is defined as follows: Let $m$ be an integer
such that $m\geq 2$. We write $\modex{m}$ to denote the \emph{modulo $m$ counting quantifier}.
A formula of the form $\modex{m} y\; \psi(\ov{x},y)$
is satisfied by a $\sigma$-structure $\A$ and an interpretation
$\ov{a}$ of the variables $\ov{x}$ if, and only if, the number
of elements $b\in A$ such that $\A\models\psi[\ov{a},b]$ is a multiple
of $m$.
For a fixed number $m$ we write $\FOMOD{m}(\sigma)$ to denote the
extension of $\FO(\sigma)$ with modulo $m$ counting quantifiers.
The quantifier rank $\qr(\varphi)$ of an $\FOMOD{m}$-formula
$\varphi$ is defined as
the maximum nesting depth of \emph{all} quantifiers
(i.e., first-order quantifiers and modulo counting quantifiers).

The size  $\size{\varphi}$ of an $\FOMOD{m}(\sigma)$-formula $\varphi$ is
its length when viewed as a word over the alphabet
$\sigma \cup \set{=} \cup \set{ \modex{m}, \exists, \forall, \nicht, \und, \oder,\impl,\gdw,(,) } \cup \set{,} \cup \textit{Var}$, \allowbreak
where $\textit{Var}$ is a countable set of variable symbols.

\smallskip

\subsection{Gaifman graph}\label{section:preliminaries-gaifman-graph}

For a $\sigma$-structure
$\A$, we write $G_\A$ to denote the Gaifman graph of $\A$, i.e., the
undirected, loop-free graph with vertex set $A$ and an edge between
two distinct vertices $a,b\in A$ iff there exists an $R\in\sigma$ and
a tuple $(a_1,\ldots,a_{\ar(R)})\in R^\A$ such that
$a,b\in\set{a_1,\ldots,a_{\ar(R)}}$.

Given a $\sigma$-structure $\A$ and two elements $a,b\in A$ that are connected in the Gaifman graph $G_{\A}$, the \emph{distance} $\dist^{\A}(a, b)$ between $a$ and $b$ is the minimal length (i.e., the number of edges) of a path from $a$ to $b$ in $G_{\A}$. For $a,b \in A$ that are not connected in $G_{\A}$ we let $\dist^{\A}(a, b) \isdef \infty$.

For $r\geq 0$ and $a\in A$, the \emph{$r$-neighbourhood} of $a$ in $\A$ is the set
\begin{gather*}
  N_r^\A(a) \ \coloneqq \quad \setc{b\in A}{\dist^\A(a,b)\leq r}.
\end{gather*}
The $r$-neighbourhood $N_r^{\A}(W)$ of a set $W\subseteq A$ is the union of the $r$-neighbourhoods $N_r^{\A}(a)$ for all $a\in W$.
For a tuple $\ov{a}=(a_1,\ldots,a_n)$, we write $N_r^\A(\ov{a})$ instead of~$N_r^\A(\set{a_1,\ldots,a_n})$.

\smallskip

\subsection{Bounded structures}\label{bounded-structures}

The \emph{degree} of a $\sigma$-structure $\A$ is the degree of its
Gaifman graph $G_\A$.
Let $\nu\colon\NN\to\NN$ be a function. A $\sigma$-structure $\A$ is
\emph{$\nu$-bounded} if\ $|N_r^\A(a)|\le\nu(r)$\ for all $r\geq 0$ and
all $a\in A$.
Clearly, if $\A$ is $\nu$-bounded, then it has degree at most $\nu(1){-}1$.  On the other hand, if $\A$ has degree at most $d$,
then $\A$ is $\nu_d$-bounded for $\nu_d\colon\NN\to\NN$ with\ $\nu_d(r)=
1+d\cdot\sum_{i=0}^{r-1}(d-1)^i$. Thus, $\A$ has degree at most $d$ iff
it is $\nu_d$-bounded. Note that $\nu_d$ is at most $1$-fold exponential. 

We will restrict attention to at most $1$-fold exponential functions
$\nu\colon\NN\to\NN$ that are \emph{strictly increasing}. This is
reasonable, since then $(r{+}1)$-neighbourhoods may  
contain more elements than $r$-neighbourhoods, and
it excludes pathological cases where $\nu$-boundedness of
a structure implies that the structure is a disjoint 
union of finite structures whose size is bounded by a constant
depending on $\nu$.

Let $\sigma$ be a finite relational signature, let $\A$ be a $\sigma$-structure, let $n\geq 1$, and let
$\ov{a}\in A^n$. Let $c_1,\ldots,c_n$ be distinct  constant
symbols.

For $r\geq 0$, the \emph{$r$-sphere around 
  $\ov{a}$} is defined as the $\sigma\cup\set{c_1,\ldots,c_n}$-structure
$  \N_r^\A(\ov{a}) \isdef \big(\, \A[N^\A_r(\ov{a})]\,,\,\ov{a}\,\big),$
where 
the constant symbols $c_1,\ldots,c_n$
are interpreted by the elements $a_1,\ldots,a_n$.

An \emph{$r$-sphere with $n$ centres} is a
$\sigma\cup\set{c_1,\ldots,c_n}$-structure $\tau=(\B,\ov{b})$ with
$\ov{b}\in B^n$ and universe $B=N_r^\B(\ov{b})$.  We say that
\emph{$\tau$ is realised by $\ov{a}$ in $\A$} iff $\N^\A_r(\ov{a})$ is
isomorphic to~$\tau$. 
By $\tau(\A)$ we denote the set of all $\ov{a}\in A^n$ that realise
$\tau$ in $\A$.

Note that a $\nu$-bounded $r$-sphere $\tau$ with $n$ centres contains
at most $n{\cdot}\nu(r)$ elements.  Thus, there is an $\FO(\sigma)$-formula
$\sph_\tau(\ov{x})$
of size 
\ $(n{\cdot}\nu(r))^{O(\size{\sigma})}$ \
such that for all $\sigma$-structures $\A$
we have $\sph_\tau(\A)=\tau(\A)$.

Unless otherwise indicated, we assume an $r$-sphere to have only one centre.
Up to isomorphism, 
the number of $\sigma$-structures with exactly $n$ elements
is\footnote{We will usually omit brackets and shortly write \ $2^{n^a}$ \ for \ $2^{(n^a)}$.} at most \ $2^{n^a}{\cdot} n^c$, \ where $a$ is the sum of the arities of the relation symbols
in $\sigma$ and $c$ is the number of constant symbols in $\sigma$.
Hence we can bound the number of non-isomorphic $r$-spheres
that can be realised in $\nu$-bounded $\sigma$-structures 
from above by
\begin{gather*}
  2^{\nu(r)^{\size{\sigma}+1}}.
\end{gather*}

\subsection{Disjoint unions}

For a relational signature $\sigma$ and each $s\geq 1$, the \emph{disjoint union $\A_1\dcup\cdots\dcup \A_s$} of
$\sigma$\hbox{-}structures $\A_1,\ldots,\A_s$ is a structure $\A$ that is defined 
(up to isomorphism) as follows: Let $A$
be a set of size $|A_1|+\cdots+|A_s|$ and let, for each $i\in[1,s]$,
$f_i \colon A_i \to A$ be an injective function such that $f_1(A_1), \ldots,
f_s(A_s)$ is a partition of~$A$. Now, $\A$ is the $\sigma$-structure with
universe $A$ where, for each $R\in\sigma$, the relation $R^{\A}$
is the union of the sets $\setc{ f_i(\ov{a}) }{\ov{a}\in R^{\A_i}}$ for all $i \in [1,s]$.
The \emph{mapping of $\A$}
is the function $\pi \colon A \to (A_1 \cup \cdots \cup A_s)$ with $\pi(a) =
b$ where $b$ is chosen such that $ b \in A_i$ for some $i \in [1,s]$ and $f_i(b)=a$.

If the universes of $\A_1,\ldots,\A_s$ are pairwise disjoint, we let 
$A\isdef A_1\cup\cdots\cup A_s$ and we let $f_i$ be the
identity on~$A_i$ for each $i\in [1,s]$, and $\pi$ the identity on
$A$.  

\section{Preservation theorems}\label{section:PreservationTheorems}

Throughout this section, $\sigma$ will always denote a finite relational signature and $\Class$ will always denote a class of $\sigma$-structures.
A $\sigma$-structure $\B$ is an \emph{extension} of a
$\sigma$-structure $\A$ if $\A$ is an \emph{induced} substructure of
$\B$.  
A sentence $\phi$ is \emph{preserved under
  extensions on~$\mathfrak{C}$} if for each model $\A\in\mathfrak{C}$
of $\phi$ and every extension $\B\in\mathfrak{C}$ of~$\A$, \ $\B$ is
also a model of~$\phi$. 
An \emph{existential $\FO(\sigma)$-formula} has the form $\exists
x_1\cdots \exists x_n \, \phi$, where $\phi$ is quantifier-free. 
It is straightforward to see that every existential
$\FO(\sigma)$-sentence is preserved under extensions on arbitrary classes of $\sigma$-structures.

A \emph{homomorphism} of $\sigma$-structures $\A$ and $\B$ is a mapping $h\colon A \to B$
such that for each relation symbol $R\in \sigma$ with $r\isdef \ar(R)$
and all tuples $(a_1, \ldots, a_r)\in A^r$, if $(a_1, \ldots, a_r) \in R^{\A}$ then $(h(a_1), \ldots, h(a_r)) \in R^{\B}$.
A sentence $\phi$ is \emph{preserved under homomorphisms on
$\mathfrak{C}$} if for each model $\A\in \mathfrak{C}$ of $\phi$
and each structure $\B\in \mathfrak{C}$ for which a homomorphism $h$ from $\A$ to $\B$ exists,
$\B$ is also a model of $\phi$.
An \emph{existential-positive} $\FO(\sigma)$-formula is an existential
$\FO(\sigma)$-formula that does not contain any of the symbols 
$\nicht$, $\impl$, $\gdw$. For convenience, we will say that also
\textit{false} is an existential-positive $\FO(\sigma)$-sentence (that is not
satisfied by any $\sigma$-structure).
It is straightforward to see that every existential-positive
$\FO(\sigma)$-sentence is preserved under homomorphisms on arbitrary classes
of $\sigma$-structures.

For the remainder of this section, let $\nu \colon \NN\to\NN$ be a fixed time-constructible
strictly increasing function that is at most 1-fold exponential. 
\newcommand{\Nsph}[1]{\ensuremath{S_\nu(#1, \size{\sigma})}}
Recall that the number of non-isomorphic $r$\hbox{-}spheres (with one centre) that can be realised in $\nu$-bounded $\sigma$-structures is bounded from above by $2^{\nu(r)^{\bigO(\size{\sigma})}}$.
Henceforth, we will 
abbreviate this expression by $\Nsph{r}$, i.e.,
\begin{gather*}
\Nsph{r} \quad \isdef \quad 2^{\nu(r)^{\bigO(\size{\sigma})}}.
\end{gather*}
Thus, $S_{\nu}(\cdot, \cdot)$ is the class of all functions $f \colon \NN\times\NN \to \NN$ for which there exists a $c > 0$ such that $f(r, s) \leq 2^{\nu(r)^{c\cdot s}}$ for all sufficiently large $r, s \geq 1$. 

In this section, we explore the complexity of constructing existential
(respectively, existential-positive) 
$\FO(\sigma)$-sentences for $\FOMOD{m}(\sigma)$-sentences 
that are preserved under extensions
(respectively, homomorphisms) on classes of~$\nu$-bounded $\sigma$\hbox{-}structures
that are closed under disjoint unions and closed under induced
substructures 
(respectively, closed under disjoint unions, closed under induced substructures, 
and decidable in~$1$\hbox{-}fold exponential time).  
It is straightforward to see that
the class $\mathfrak{C}_d$ of all finite $\sigma$-structures of
degree at most $d$, for any fixed $d\geq 0$, meets all these
requirements. Similarly, as we assume that~$\nu$ is
time\hbox{-}constructible and at most $1$\hbox{-}fold exponential, also the class of all $\nu$-bounded structures is
easily seen to be decidable in $1$-fold exponential time, as well as closed
under taking induced substructures and disjoint unions. 

In Subsection~\ref{subsection:closure-properties} we present two examples of classes of structures, 
which show that the closure properties required by our constructions are indeed necessary.
\smallskip

\subsection{Summary of this section's main results}
Table~\ref{table:PresThm:UpperBounds} summarises the time complexity
of our algorithms (depending on the size of an input sentence)
on the class of all $\nu$-bounded structures. 
The summary differentiates between functions $\nu$
with either exponential or polynomial growth.
\begin{table}[h!tbp]
\normalsize
\renewcommand\tabcolsep{4.3pt}
\begin{center}
  \begin{tabular}{l|c|c}
    & \text{extensions} & \text{homomorphisms} \\[1mm] \hline
    exponential $\nu$ & \ 5-exp \ & \ 4-exp \  \\ \hline
    polynomial $\nu$ & 3-exp & 3-exp \\ \hline
  \end{tabular}
\end{center}
\caption{The time complexity of our algorithms. For each fixed $d\geq 3$, the class
  $\mathfrak{C}_d$ of all finite structures of degree $\leq d$ is a
  case of ``exponential $\nu$''. The class $\mathfrak{C}_2$ of all finite structures of degree $\leq 2$ is a case of ``polynomial $\nu$''.}\label{table:PresThm:UpperBounds}
\end{table}

The precise statement of this section's first main result reads as
follows; a proof is given in Subsection~\ref{subsection:ep} below.
\begin{thm}\label{thm:ep-upper-bound}
  Let $\mathfrak{C}_{\nu}$ be a class of $\nu$-bounded $\sigma$-structures that is closed under disjoint unions and induced substructures.
  There is an algorithm that, given an
  $\FOMOD{m}(\sigma)$-sentence $\phi$ of quantifier rank $q\geq 0$ as input, constructs in time
  \begin{gather}\label{expr:ep-upper-bound}
    \size{\phi} \cdot \Nsph{\Nsph{3^q}}^{(\log m)^2}
    \qquad\qquad
    \Bigl(
    =
    \quad
    \size{\phi} \cdot 2^{(\log m)^2 \cdot \nu(2^{\nu(3^q)^{\bigO(\size{\sigma})}})^{\bigO(\size{\sigma})}}\ 
    \Bigr)
  \end{gather}
  an existential $\FO(\sigma)$-sentence $\psi$ such that the following holds:\\
  If $\phi$ is preserved under extensions on $\mathfrak{C}_{\nu}$, then $\phi$ and $\psi$ are equivalent on $\mathfrak{C}_{\nu}$.
\end{thm}

Consequently, if the function $\nu$ is exponential,
$\phi$ is an
$\FOMOD{m}(\sigma)$\=/sentence, and $\sigma$ consists of exactly the relation symbols that occur in $\phi$, then the 
algorithm uses time
$5$-fold exponential in the size of $\phi$. 
In particular, if $\nu=\nu_d$ for a fixed $d\geq 3$, and 
$\mathfrak{C}_{\nu}$ is the class of all finite structures of degree $\leq d$, 
then our algorithm uses time at most
\begin{gather}
  \size{\phi} \cdot 2^{(\log m)^2 \cdot 
     d^{2^{d^{
     2^{\bigO(q+\log \size{\sigma})}}}}}
\quad \leq \quad
  2^{(\log m)^2 \cdot d^{2^{d^{2^{\bigO(\size{\phi})}}}}}
\end{gather}
when given a first-order sentence $\phi$ with modulo $m$ counting quantifiers, quantifier rank $q$, and signature $\sigma$.
Note that the constant suppressed by the
$O$-notation does not depend on the particular signature $\sigma$.
Furthermore, if $\phi$ does not contain any modulo counting quantifier, we can assume $m$ to be $2$, and hence $(\log m)^2 = 1$.

On the other hand, if $\nu$ is polynomial (e.g., $\nu=\nu_2$ and $\mathfrak{C}_\nu$ is the class of all finite structures of
degree $\leq 2$), then expression \eqref{expr:ep-upper-bound} simplifies to
the $3$\=/fold exponential expression
\begin{gather}
  \size{\phi} \cdot 
  2^{(\log m)^2 \cdot 2^{
  2^{\bigO(q\cdot\size{\sigma})}}}
\quad \leq \quad
  2^{(\log m)^2 \cdot 2^{2^{\bigO(\size{\phi}^2)}}}.
\end{gather}
\medskip

This section's second main result reads as follows; a proof is
given in Subsection~\ref{subsection:hp} below.
\begin{thm}\label{thm:hp-upper-bound}
  Let $\mathfrak{C}_{\nu}$ be a class of $\nu$-bounded
   $\sigma$-structures that is closed under disjoint unions and
   induced substructures
   and decidable in time $t(n)$ for some function $t : \NN \to \NN$.
   There is an algorithm which, given an input
   $\FOMOD{m}(\sigma)$-sentence $\phi$ of quantifier rank $q\geq 0$,
   constructs in time 
   \begin{gather} \label{expr:hp-upper-bound}
     2^{\size{\phi} \cdot {\Nsph{2{\cdot}3^q}}}
     \cdot \
     t(\Nsph{2{\cdot}3^q})
     \qquad\qquad
     \Bigl(
     =
     \quad
     2^{\size{\phi} \cdot {2^{\nu(2\cdot 3^q)^{\bigO(\size{\sigma})}}}}
     \cdot\ 
     t(2^{\nu(2\cdot 3^q)^{\bigO(\size{\sigma})}})\ 
     \Bigr)
   \end{gather}
   an existential-positive $\FO(\sigma)$-sentence $\psi$ of size
   $ 2^{\Nsph{2\cdot 3^q}} $
   such that the following holds:\\
  If $\phi$ is preserved under homomorphisms on $\mathfrak{C}_{\nu}$,
  then~$\phi$ and $\psi$ are $\mathfrak{C}_{\nu}$-equivalent. 
\end{thm}

Consequently, if the functions $\nu$ and $t$ are 1-fold exponential, then 
the algorithm uses time 4-fold exponential in the size of the input sentence $\phi$.
In particular, if $\nu=\nu_d$ for a fixed $d\geq 3$ and $\mathfrak{C}_\nu$ is the class of all finite structures of degree
$\leq d$, the algorithm uses time
\begin{gather*}
  2^{\size{\phi}\cdot
   2^{d^{
    2^{O(q+\log\size{\sigma})}
   }}
  }
  \quad \leq \quad
  2^{2^{d^{2^{\bigO(\size{\phi})}}}}.
\end{gather*}
If $\nu$ is polynomial and $t$ is at most 1-fold exponential (this is, in particular, true if $\mathfrak{C}_\nu$ is the class
of all finite structures of degree $\leq 2$), the algorithm's running time is  
\begin{gather*}
  2^{\size{\phi}\cdot 2^{
   2^{O(q\cdot\size{\sigma})}
  }}
  \qquad \leq \qquad
  2^{{2^{2^{\bigO(\size{\phi}^2)}}}},
\end{gather*}
which is $3$\=/fold exponential in the size of the input sentence $\phi$.
\smallskip

\subsection{Introductory notes on the proofs of Theorem~\ref{thm:ep-upper-bound} and Theorem~\ref{thm:hp-upper-bound}}
Consider a class $\Class$ of $\sigma$-structures and an $\FOMOD{m}(\sigma)$-sentence $\phi$. A $\sigma$-structure $\A$ is a \emph{$\Class$-minimal model of $\phi$} if $\A \in \Class$, $\A$ is a model of $\phi$, and there is no proper induced substructure $\B \in \Class$ of $\A$ that is a model of $\phi$. 

Let $\Class_{\nu}$ be a class of $\nu$-bounded $\sigma$-structures that is closed under disjoint unions and closed under induced substructures. 
The main combinatorial parts (Theorem~\ref{thm:ep-bound-minimal-model} and Theorem~\ref{thm:hp-bound-minimal-model}) of the proofs of Theorem~\ref{thm:ep-upper-bound} and Theorem~\ref{thm:hp-upper-bound} provide upper bounds on the size of $\Class_{\nu}$-minimal models for $\FOMOD{m}(\sigma)$-sentences that are preserved under extensions (respectively, homomorphisms) on $\Class_{\nu}$.

The proofs of both upper bounds on the size of the $\Class_{\nu}$-minimal models proceed as follows: Assume that $\phi$ is an $\FOMOD{m}(\sigma)$-sentence that is preserved under extensions (homomorphisms) on $\Class_{\nu}$. We show that for every $\Class_{\nu}$-minimal model whose size exceeds the upper bound, a proper induced subtructure $\A'\in\Class_{\nu}$ of $\A$ can be constructed that is also a model of $\phi$, which is a contradiction to the minimality of~$\A$. 

In both cases, we use the following generalisation of Hanf's theorem (see
e.g. \cite{EbbinghausFlum,Libkin-FMT}) by Nurmonen \cite{Nurmonen2000}
to sentences with modulo counting quantifiers. Two
$\sigma$-structures $\A$ and~$\B$ are \mbox{\emph{$(m,q)$-equivalent}}
($\A \equiv^q_m \B$, for short) if they satisfy the same
$\FOMOD{m}(\sigma)$-sentences of quantifier rank at most~$q$.
\begin{thm}[Theorem 3.4 in
  \cite{Nurmonen2000}]\label{thm:Hanf-FOMOD}
  Let $\A, \B$ be $\sigma$\hbox{-}structures. Let $m\geq 2$ and $q\geq
  0$.
  Suppose that for some $e\geq 0$ each $3^q$-neighbourhood of an
  element in $\A$ or $\B$ has less than $e$ elements and that for each
  $3^q$-sphere~$\tau$ (with one centre), $|\tau(\A)|$ and $|\tau(\B)|
  $ are congruent modulo $m$ and either
  \begin{gather*}
    |\tau(\A)| = |\tau(\B)| 
    \qquad \text{or} \qquad \big( \ \ 
    |\tau(\A)|\geq t 
       \quad \text{and} \quad 
    |\tau(\B)| \geq t \ \ \big) \ ,
  \end{gather*}
  where $t \isdef q {\cdot} e + 1$.  Then $\A \equiv^q_m \B$.\qed
\end{thm}
The proof of Theorem~\ref{thm:ep-bound-minimal-model} employs a novel inductive construction that constructs a sequence of structures from $\Class_{\nu}$ that alternates between proper induced substructures and disjoint extensions of $\A$ and finally stops with two consecutive $(m,q)$-equivalent structures, where $q\geq 0$ is the quantifier rank of $\phi$.

Our proof of Theorem~\ref{thm:hp-bound-minimal-model} is an adaptation of a result by Ajtai and Gurevich (Lemma~7.1 in~\cite{AjtaiGurevich1989}) where we use Nurmonen's theorem (Theorem~\ref{thm:Hanf-FOMOD}) instead of Gaifman's theorem.

Finally, the upper bounds on the size of $\Class_{\nu}$-minimal models are used as an input to algorithms that compute a $\Class_{\nu}$-equivalent existential (existential-positive) $\FO(\sigma)$-sentence for an input $\FOMOD{m}(\sigma)$-sentence $\phi$ that is preserved under extensions (homomorphisms) on $\Class_{\nu}$ (see Lemma~\ref{lem:existential-sentence} and Lemma~\ref{lem:existential-positive-sentence}). 

 The construction for existential sentences generalises Lemma~8.4 in~\cite{DGKS07} to first-order sentences with modulo $m$ counting quantifiers. Here, the handling of the modulo $m$ counting quantifier requires an inductive construction (rather than a straightforward brute-force approach) to ensure the desired time complexity. 

The construction for existential-positive sentences uses an algorithmic version of the Chandra-Merlin Theorem \cite{ChandraMerlin}, which requires an additional assumption on the decidability of $\Class_{\nu}$.

\smallskip
\subsection{Preservation under extensions: Proof of
  Theorem~\ref{thm:ep-upper-bound}}
\label{subsection:ep}
This subsection is devoted to the proof of
Theorem~\ref{thm:ep-upper-bound}.  

A set of elements $B$ in a $\sigma$-structure $\A$ is
\emph{$r$-scattered}, for an $r\geq 0$, if for all distinct elements
$a,b\in B$ we have $N_r^{\A}(a) \cap N_r^{\A}(b) = \emptyset$.  We
will make use of the following easy fact.
\begin{lem}\label{lem:scattered-sets}
  Let $\A$ be a $\nu$-bounded $\sigma$-structure and let
  $m,r\geq 0$.
  If \ $|A|> (m{-}1) \cdot \nu(2r)$, \ then there
  exists an $r$\hbox{-}scattered subset of $A$ of cardinality $m$.
\end{lem}
\proof We show that $|A| \leq (m{-}1) \cdot \nu(2r)$, if there is no
$r$-scattered subset of cardinality $m$ in $\A$.

  Choose a number $n < m$ such that there is an $r$-scattered subset
  $B$ of cardinality $n$ in $\A$, but no $r$-scattered subset of
  cardinality greater than $n$.  Every element $a$ of $A$ has to be contained in
  the $2r$\hbox{-}neighbourhood of $B$ (for otherwise, $B \cup
  \set{a}$ would be an~$r$\hbox{-}scattered subset of $\A$ of
  cardinality~$n{+}1$).  
  Therefore, $|A| = |N_{2r}^{\A}(B)| \leq (m{-}1) \cdot \nu(2r)$. 
\qed

The main combinatorial contribution of this subsection is an upper bound on the size of 
$\Class_{\nu}$-minimal models for $\FOMOD{m}(\sigma)$-sentences, which is provided by
the following theorem: 
\begin{thm}\label{thm:ep-bound-minimal-model}
  Let $\mathfrak{C}_{\nu}$ be a class of $\nu$-bounded
  $\sigma$-structures that is closed under disjoint unions and induced
  substructures. Let $m\geq 1$ and let $\phi$ be an
  $\FOMOD{m}(\sigma)$\hbox{-}sentence of quantifier rank $q \geq 0$
  that is preserved under extensions on~$\mathfrak{C}_{\nu}$.
  There is a number
  \begin{gather*}
    N_\nu(m,q,\size{\sigma})
    \quad \in \quad  
    m\cdot \Nsph{\Nsph{3^q}}
    \qquad\qquad 
    \Bigl( = \quad m \cdot 2^{\nu(2^{\nu(3^q)^{\bigO(\size{\sigma})}})^{\bigO(\size{\sigma})}}\ \Bigr)
  \end{gather*}
  such that every $\mathfrak{C}_{\nu}$-minimal model of $\phi$ has
  size at most $N_\nu(m,q,\size{\sigma})$.
\end{thm}
\proof
  Let $r\isdef 3^q$ and let $s \in \Nsph{r}$ be the number of isomorphism types of
  $r$\hbox{-}spheres (with one centre) that are realised in $\sigma$-structures in
  $\mathfrak{C}_{\nu}$. Let $R \isdef 2sr$ and let $S \in \Nsph{R}$ be the number
  of isomorphism types of $R$-spheres (with one centre) that are realised in $\sigma$-structures in
  $\mathfrak{C}_{\nu}$. Finally, let $t\isdef q\cdot (\nu(r){+}1)+1$
  be the threshold from Theorem~\ref{thm:Hanf-FOMOD} for $\nu$-bounded
  $\sigma$-structures and for quantifier rank $q$.

  Let $\A$ be a $\mathfrak{C}_{\nu}$-minimal model of $\phi$.  Towards
  a contradiction, assume that the universe $A$ of $\A$ has cardinality greater than $(2Stm{-}1)\cdot\nu(2R)$. Then, by
  Lemma~\ref{lem:scattered-sets}, there exists an $R$-scattered subset
  of $A$ of cardinality $2Stm$.  Because there are at most~$S$
  different $R$-spheres realised in $\A$, there is an $R$-scattered
  set $X'$ of~$2tm$ elements in $A$ that realise the same $R$-sphere.

  An $r$-sphere $\tau$ is \emph{frequent in a structure $\A$} if
  \mbox{$|\tau(\A)| \geq t$}. Otherwise, 
  it is \emph{rare in $\A$}.  Note that
  each $r$-sphere realised by an element from the
  $(R{-}r)$\hbox{-}neighbourhood of $X'$ is frequent in~$\A$, because
  it occurs at least $2tm \geq t$ times in $\A$.

  Let $X$ be a subset of $X'$ of cardinality $tm$. Since
  \begin{gather}\label{expr:iso-remove-X}
    \N_R^{\A}(a)\ \cong\ \N_R^{\A}(b) \quad \text{for each $a\in X$ and
      each $b\in X'\setminus X$,}
  \end{gather}
  the following holds for each $r$-sphere $\tau$:
  \begin{center}
    If \ \ $|\tau(\A) \cap N_{R-r}^{\A}(X)| \geq t$, \ \ also \ \ $|\tau(\A - N_{R-r}^{\A}(X))| \geq t$. 
  \end{center}
  I.e., every $r$-sphere $\tau$ that is realised by at
  least $t$ elements of the $(R{-}r)$-neighbourhood of $X$ is still
  frequent in the substructure of $\A$ induced by deleting the
  $(R{-}r)$-neighbourhood of $X$.

  Consider the following sequences $(\C_i)_{i\geq 0}$ and
  $(\D_{2i})_{i > 0}$ of $\sigma$-structures in $\mathfrak{C}_{\nu}$.  Let
  $\C_0 \isdef \A$ and $\C_1 \isdef \A - X$.  For each even
  $i>1$, let $\D_i \isdef \C_{i-1}[N_{2(i-1)r}^{\A}(X)]$. Note
  that for odd $i$, $\D_i$ is neither defined nor required. For each
  $i \geq 2$, let
  \begin{align*}
    \C_i \ \isdef \quad
    \begin{cases}
      \ \A \dcup \D'_i & \text{if $i$ is even,} \\
      \ \bigl(\A - N_{2(i-1)r}^{\A}(X)\bigr) \dcup \D_{i-1} &
      \text{if $i$ is odd,}
    \end{cases}
  \end{align*}
  where $\D'_i$ is a structure isomorphic to $\D_i$ whose universe is
  disjoint to $A$.  For all even $i$, $\C_i$ is a disjoint extension
  of~$\A$, and for all odd $i$, $\C_i$ is a proper induced
  substructure of $\A$.

\begin{figure}
  \centering \tikzstyle{struct} = [draw=black, fill=lightgray]
  \subfigure{
    \begin{tikzpicture}[scale=0.2]
      \path[draw,use as bounding box] (-0.5,-0.5) rectangle
      (19.5,9.5);
      \begin{scope}
        \draw[struct] (0,0) rectangle (9,9);
      \end{scope}
      \node[draw=none] at (18,1) { $\C_0$};
    \end{tikzpicture}
  } \subfigure{
    \begin{tikzpicture}[scale=0.2]
      \path[draw,use as bounding box] (-0.5,-0.5) rectangle
      (19.5,9.5);
      \begin{scope}
        \draw[struct] (0,0) rectangle (9,9); \draw[struct, fill=white]
        (4.5,4.5) circle (0.2cm);
      \end{scope}
      \node[draw=none] at (18,1) { $\C_1$};
    \end{tikzpicture}
  } \subfigure{
    \begin{tikzpicture}[scale=0.2]
      \path[draw,use as bounding box] (-0.5,-0.5) rectangle
      (19.5,9.5);
      \begin{scope}
        \draw[struct] (0,0) rectangle (9,9);
        \begin{scope}[xshift=10cm]
          \draw[struct] (4.5,4.5) circle (1cm); \draw[struct,
          fill=white] (4.5,4.5) circle (0.2cm);
        \end{scope}
      \end{scope}
      \node[draw=none] at (18,1) { $\C_2$};
    \end{tikzpicture}
  } \subfigure{
    \begin{tikzpicture}[scale=0.2]
      \path[draw,use as bounding box] (-0.5,-0.5) rectangle
      (19.5,9.5);
      \begin{scope}
        \draw[struct] (0,0) rectangle (9,9);
        \begin{scope}[xshift=0cm]
          \draw[struct, fill=white] (4.5,4.5) circle (2cm);
          \draw[struct] (4.5,4.5) circle (1cm); \draw[struct,
          fill=white] (4.5,4.5) circle (0.2cm);
        \end{scope}
      \end{scope}
      \node[draw=none] at (18,1) { $\C_3$};
    \end{tikzpicture}
  } \subfigure{
    \begin{tikzpicture}[scale=0.2]
      \path[draw,use as bounding box] (-0.5,-0.5) rectangle
      (19.5,9.5);
      \begin{scope}
        \draw[struct] (0,0) rectangle (9,9);
        \begin{scope}[xshift=10cm]
          \draw[struct] (4.5,4.5) circle (3cm); \draw[struct,
          fill=white] (4.5,4.5) circle (2cm); \draw[struct] (4.5,4.5)
          circle (1cm); \draw[struct, fill=white] (4.5,4.5) circle
          (0.2cm);
        \end{scope}
      \end{scope}
      \node[draw=none] at (18,1) { $\C_4$};
    \end{tikzpicture}
  } \subfigure{
    \begin{tikzpicture}[scale=0.2]
      \path[draw,use as bounding box] (-0.5,-0.5) rectangle
      (19.5,9.5);
      \begin{scope}
        \draw[struct] (0,0) rectangle (9,9);
        \begin{scope}[xshift=0cm]
          \draw[struct, fill=white] (4.5,4.5) circle (4cm);
          \draw[struct] (4.5,4.5) circle (3cm); \draw[struct,
          fill=white] (4.5,4.5) circle (2cm); \draw[struct] (4.5,4.5)
          circle (1cm); \draw[struct, fill=white] (4.5,4.5) circle
          (0.2cm);
        \end{scope}
      \end{scope}
      \node[draw=none] at (18,1) { $\C_5$};
    \end{tikzpicture}
  }
  \caption{The first six elements of the sequence $(\C_i)_{i\geq
      0}$. Note that for all even $i$, $\C_i$ is a disjoint extension
    of $\A$  while for each odd
    $i$, $\C_i$ is a proper induced substructure of $\A$.}
  \label{fig:Ci}
\end{figure}
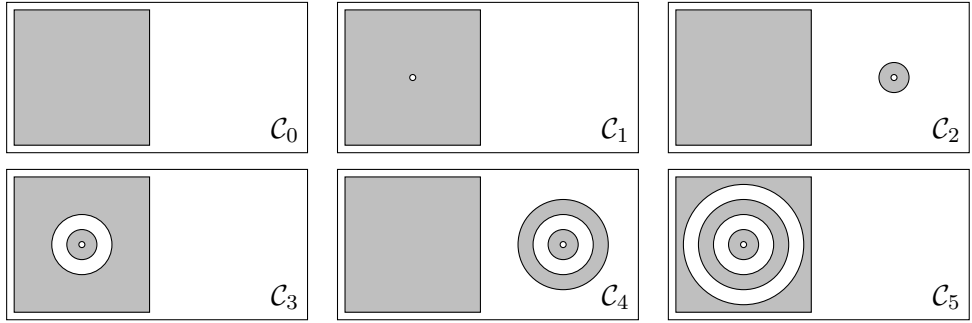

Let $\tau$ be an $r$-sphere. Recall that $|X|=tm$ and that the
$R$\hbox{-}spheres around the elements of $X$ are disjoint and
isomorphic. This implies that, for each $i \in [1,s{-}1]$, there is a
number $k \in \ZZ$ such that $|\tau(\A - N^{\A}_{2(i-1)r}(X))|
= |\tau(\A)| + ktm$ and therefore $|\tau(\A -
N^{\A}_{2(i-1)r}(X))|\equiv |\tau(\A)| \mod m$.  Furthermore, since
$|X|\equiv 0\mod m$, it is straightforward to see that
$|\tau(\D'_i)|=|\tau(\D_i)| \equiv 0 \mod m$ for all even $i \leq s$.

This immediately proves the following Claim~1. 

\newclaims
\begin{clm}
  For all $i < s$ and every $r$-sphere $\tau$, $|\tau(\C_i)|$ and
  $|\tau(\C_{i+1})|$ are congruent modulo $m$.
\end{clm}
A proof of the following Claim~2 is deferred to the end of the proof of Theorem~\ref{thm:ep-bound-minimal-model}.
\begin{clm}\label{claim:ep-upper-bound-1}
  The following holds for all $i < s$ and every
  $r$\hbox{-}sphere~$\tau$: \smallskip
  \begin{enumerate}
  \item[(a)] If $\tau$ is frequent in $\C_i$, it is also frequent in
    $\C_{i+1}$.\smallskip
  \item[(b)] If $\tau$ is rare in $\C_i$ and rare in $\C_{i+1}$, then
    \mbox{$|\tau(\C_i)| = |\tau(\C_{i+1})|$}.
  \end{enumerate}
\end{clm}
While every $r$-sphere that is frequent in $\C_i$ is also frequent
in $\C_{i+1}$, the opposite is not necessarily true: There may be $r$-spheres that
are rare in $\C_{i}$ but that occur frequently in $\C_{i+1}$.
However, since there are only $s$ pairwise non-isomorphic $r$-spheres in
$\sigma$-structures in $\mathfrak{C}_{\nu}$, and $\C_0$ already contains
frequent $r$-spheres, Claim~\ref{claim:ep-upper-bound-1}~(a) implies
that there has to be an $i < s$ such that all frequent $r$-spheres of
$\C_{i+1}$ are frequent already in $\C_i$. Thus, for this particular
$i$ we know that any $r$-sphere is either frequent in $\C_{i+1}$ and
in $\C_i$ or it is rare in $\C_{i+1}$ and in $\C_i$.
Hence, with Claim~\ref{claim:ep-upper-bound-1}~(b) it follows that for
every $r$-sphere~$\tau$, either $|\tau(\C_i)| = |\tau(\C_{i+1})|$ or
$\tau$ is frequent in $\C_i$ and $\C_{i+1}$.

Together with Claim~$1$, we can conclude from
Theorem~\ref{thm:Hanf-FOMOD} that $\C_i \equiv^q_m
\C_{i+1}$. Therefore, by using Claim~$2$, the proof of
Theorem~\ref{thm:ep-bound-minimal-model} can be completed as follows:
In case that $i$ is even, we let $\B \isdef \C_i$ and $\A' \isdef
\C_{i+1}$; and in case that $i$ is odd, we let $\B \isdef \C_{i+1}$
and $\A' \isdef \C_i$. Since $\B$ is a disjoint extension of $\A$ with
an induced substructure of $\A$ (and hence belongs to
$\mathfrak{C}_\nu$, since $\mathfrak{C}_\nu$ is closed under disjoint
unions and induced substructures), $\A\models\phi$, and $\phi$ is
preserved under extensions on $\mathfrak{C}_{\nu}$, we obtain that $\B
\models \phi$. Because $\A' \equiv^q_m \B$ and $\phi$ has quantifier
rank~$q$, we know that $\A' \models \phi$. Since $\A'$ is a proper
induced substructure of $\A$ and $\mathfrak{C}_\nu$ is closed under
induced substructures, we have $\A'\in\mathfrak{C}_{\nu}$. However,
this is a contradiction to the assumption that $\A$ is a
$\mathfrak{C}_{\nu}$-minimal model of $\phi$.

Therefore, the size $|A|$ of the universe of $\A$ is at most $(2Stm{-}1)\cdot \nu(2R)$.\\ We let $N_{\nu}(m,q,\size{\sigma}) \isdef (2Stm{-}1)\cdot\nu(2R)$. To obtain an upper bound on $N_{\nu}(m,q,\size{\sigma})$, recall that $S \in \Nsph{R}$ and $R \in 2r \cdot \Nsph{r}$. Therefore, since $\Nsph{R}$ is an abbreviation for the expression $2^{\nu(R)^{\bigO(\size{\sigma})}}$, we know that 
\begin{gather}
  \label{expr:upper-bound-S}
  S \quad \in \quad
  \Nsph{2r \cdot 2^{\nu(r)^{\bigO(\size{\sigma})}}}
  \quad \subseteq \quad 
  \Nsph{\Nsph{r}}.
\end{gather}
The latter inclusion is correct since $\nu$ is strictly increasing.
Similarly, we have that
\begin{gather}
  \label{expr:upper-bound-nu2R}
  \nu(2R) 
  \quad = \quad
  \nu(2 \cdot 2r\cdot 2^{\nu(r)^{\bigO(\size{\sigma})}})
  \quad \subseteq \quad 
  \nu(\Nsph{r}).
\end{gather}
Using (\ref{expr:upper-bound-S}) and considering that $t \in \bigO(q \cdot \nu(r))$, we conclude that
\begin{gather}
  \label{expr:upper-bound-2Stm}
  2Stm\quad = \quad \bigO(2^{\nu(\Nsph{r})^{\bigO(\size{\sigma})}}\cdot q \cdot \nu(r)\cdot m)
  \quad \subseteq \quad \Nsph{\Nsph{r}} \cdot q \cdot m.  
\end{gather}
Putting (\ref{expr:upper-bound-2Stm}) and (\ref{expr:upper-bound-nu2R}) together, we obtain that 
\begin{gather*}
  \begin{split}
  (2Stm{-}1)\cdot\nu(2R) &
  \quad \subseteq  \quad
  \Nsph{\Nsph{r}}
  \cdot 
  \nu(\Nsph{r})
  \cdot q \cdot m \\
  & \quad \subseteq \quad
  \Nsph{\Nsph{r}} \cdot q \cdot m.
  \end{split}
\end{gather*}
Therefore, recalling that $r \isdef 3^q$, we know that
\begin{gather*}
  N_{\nu}(m,q,\size{\sigma}) 
  \quad \in \quad
  \Nsph{\Nsph{3^q}} \cdot q \cdot m
  \quad \subseteq \quad
  m \cdot \Nsph{\Nsph{3^q}}.
\end{gather*}
All that remains to be done to finish the proof of
Theorem~\ref{thm:ep-bound-minimal-model} is to prove Claim~2.

\medskip
 
  \noindent
  \uproof{Claim~2} Observe that, for all $i,j \leq s$,
  \begin{gather}\label{expr:outer-iso}
    \C_i[A \setminus N_{R-2r}^{\A}(X)] \quad\cong\quad
    \C_j[A \setminus N_{R-2r}^{\A}(X)].
  \end{gather}
  Let $i < s$.  For the proof of Claim~$2$~(b) let $\tau$ be an
  $r$-sphere that is rare in $\C_i$ and $\C_{i+1}$.  Since $X$ is an
  $R$-scattered set  of size $\geq t$, the rareness of $\tau$ implies
  that $\tau(\C_i)$ and $\tau(\C_{i+1})$ are subsets of $A \setminus
  N_{R-r}^{\A}(X)$. Hence, \eqref{expr:outer-iso} implies that
  $|\tau(\C_i)| = |\tau(\C_{i+1})|$. This proves Claim~$2$~(b).
  \\
  For the proof of Claim~$2$~(a), we distinguish between even and odd
  $i$.  \smallskip
  
  \textit{Even $i$:\ } Recall that $\C_0 = \A$ and that for each even $i
  \geq 2$, $\C_i = \A \dcup \D'_i$.  Let $\tau$ be an $r$-sphere that
  is frequent in $\C_i$.
\\
  If $\tau$ is realised in $\C_i$ by an element in $N_{R-r}^{\A}(X)$
  then, since $X$ is $R$-scattered and of cardinality at least $t$, we know that $\tau(\A) \cap
  N_{R-r}^{\A}(X)$ contains at least $t$ elements. Furthermore, we obtain by Observation~\eqref{expr:iso-remove-X}
  that also \mbox{$\tau(\A)\cap N^{\A}_{R-r}(X' \setminus
    X)$} contains at least $t$ elements.  Since $X\subseteq X'$, the set $X'$ is $R$-scattered, and
  $2ir < R$, we obtain that \mbox{$|\tau(\A -
    N^{\A}_{2ir}(X))| \geq t$}. Thus, $\tau$ is also frequent in
  $\C_{i+1}$.
\\
  If $\tau$ is realised in $\C_i$ by an element in $\D'_i$, for $i\geq
  2$, then, since $\D'_i\isom \D_i$, $\tau$ is realised in $\C_{i+1}$
  also by an element in $D_i\subseteq N^\A_{R-r}(X)$. Thus, $\tau$ is
  also frequent in $\C_{i+1}$.
\\
  Otherwise, we know that $|\tau(\A) \setminus N_{R-r}^{\A}(X)|\geq
  t$. Then it follows from Observation~\eqref{expr:outer-iso} that $\tau$ is also
  frequent in $\C_{i+1}$.  \smallskip
  
  \textit{Odd $i$:\ } Recall that $\C_1 = \A - X$ and that,
  for each odd $i \geq 3$, $\C_i = ( \A -
  N_{2(i-1)r}^{\A}(X)) \dcup \D_{i-1}$. Let $\tau$ be an
  $r$-sphere that is frequent in $\C_i$.
\\
  If $\tau$ is realised in $\C_i$ by an element in
  $N^{\A}_{2(i-1)r+r}(X)$ then, since $X$ is $R$-scattered and of cardinality $\geq t$, 
  the set $\tau(\C_i) \cap N^{\A}_{2(i-1)r+r}(X)$ contains at least $t$ elements. Since
  $\D_{i+1}$ is defined as $\C_i[N^{\A}_{2(i-1)r+2r}(X)]$, we obtain 
  $|\tau(\D_{i+1})|\geq t$. Furthermore, since $\D'_{i+1}\isom\D_{i+1}$, we have
  $|\tau(\D'_{i+1})|\geq t$. Consequently, since $\C_{i+1} = \A \dcup
  \D'_{i+1}$, also $|\tau(\C_{i+1})|\geq t$. Thus, $\tau$ is frequent
  in $\C_{i+1}$.
\\
  Otherwise, we know that $\tau(\C_i) \cap (A \setminus
  N^{\A}_{2(i-1)r+r}(X))$ has cardinality $\geq t$. Since the $r$-sphere of each $a \in
  A\setminus N^{\A}_{2(i-1)r+r}(X)$ in $\A -
  N_{2(i-1)r}^{\A}(X)$ is isomorphic to its $r$-sphere in $\A$, $\tau$
  is also frequent in $\C_{i+1}= \A \dcup \D'_{i+1}$. 

  This concludes
  the proof Claim~$2$ and Theorem~\ref{thm:ep-bound-minimal-model}.
\qed
\medskip

For proving Theorem~\ref{thm:ep-upper-bound}, it remains to do the following:
for a given $\FOMOD{m}(\sigma)$-sentence $\psi$ that is preserved under extensions on $\Class_{\nu}$ and for an upper bound on the size of its $\Class_{\nu}$\hbox{-}minimal models (obtained from Theorem~\ref{thm:ep-bound-minimal-model}), construct an existential $\FO(\sigma)$-sentence that is $\Class_{\nu}$-equivalent to~$\psi$. 
This is achieved by the following lemma which is a generalisation of Lemma~8.4
in~\cite{DGKS07} to sentences with modulo counting quantifiers.

\begin{lem}
  \label{lem:existential-sentence}
  Let $\Class$ be a class of $\sigma$-structures that is closed under induced substructures. There is an algorithm which, given a number $N\geq 0$ and an $\FOMOD{m}(\sigma)$-sentence $\psi$ of quantifier rank $q \geq 0$, constructs in time 
  \begin{gather*}
    \size{\psi} \cdot N^{\bigO(q \cdot \log m)}
  \end{gather*}
  an existential $\FO(\sigma)$-sentence $\psi_N$ such that the following holds: \ If $\psi$ is preserved under extensions on $\Class$ and every $\Class$-minimal model of $\psi$ has at most $N$ elements, then $\psi_N$ is $\Class$-equivalent to $\psi$.

  Furthermore, the constant suppressed by the $O$-notation does not depend on the signature $\sigma$.
  If $\psi$ does not contain any  modulo counting quantifier, the construction only takes time $ \bigO(\size{\psi}\cdot N^{q})$. 
\end{lem}

The key ingredient for the proof of Lemma~\ref{lem:existential-sentence} is contained in the following lemma. Here, an \emph{enumeration of a set $A$} is a tuple $(a_1,\ldots,a_M) \in A^M$ of length $M=|A|$ that contains each element of $A$ exactly once (i.e., $A = \set{a_1,\ldots,a_M}$).
\begin{lem}
  \label{lem:existential-formula}
  Let $k \geq 0$ and let $\psi$ be an $\FOMOD{m}(\sigma)$-formula with 
  variables among $x_1,\ldots,x_k$. Let $M \geq 1$ and let $s \colon [1,k] \to [1,M]$. There is a quantifier-free $\FO(\sigma)$-formula $(\psi)_{M,s}$ with 
variables among $y_1,\ldots,y_M$ such that for each $\sigma$-structure $\A$ with exactly $M$ elements and each enumeration $(a_1,\ldots,a_M)$ of $A$, the following equivalence holds:
  \begin{align*}
    &  & (\A, a_{s(1)},\ldots,a_{s(k)}) \quad \models \quad & \psi(x_1,\ldots,x_k) \\
    &\text{iff} & (\A, a_1,\ldots,a_M) \quad \models \quad & (\psi)_{M,s}(y_1,\ldots,y_M).
  \end{align*}   
  Furthermore, if $\psi$ does not contain any modulo counting quantifier, the above equivalence holds more generally for each $\sigma$-structure $\A$ with at most $M$ elements and each tuple $(a_1,\ldots,a_M)\in A^M$ that contains each element of $A$ at least once.

  The formula,  $(\psi)_{M,s}$ can be constructed in time
  \begin{gather*}
    \size{\psi} \cdot M^{\bigO(q \cdot \log m)},
  \end{gather*}
  and the constant suppressed by the $O$-notation does not depend on the signature $\sigma$.
\\
  If $\psi$ does not contain any modulo counting quantifier, the construction of $(\psi)_{M,s}$ only takes time in
  \begin{gather*}
    \bigO(\size{\psi}\cdot M^{q}). 
  \end{gather*}
\end{lem}\medskip

\noindent Before presenting the proof of Lemma~\ref{lem:existential-formula}, we first show how to use Lemma~\ref{lem:existential-formula}
for proving Lemma~\ref{lem:existential-sentence}.
\medskip

\proof[Proof of Lemma~\ref{lem:existential-sentence} using Lemma~\ref{lem:existential-formula}.]\ \\
  Choose $k$ such that the variables occurring in $\psi$ are among $x_1,\ldots,x_k$.
  For applying Lemma~\ref{lem:existential-formula} for any
  $M\geq 1$, we let $s$ be an arbitrary function from $[1,k]$ to $[1,M]$.
  For every $M\in [1,N]$ we apply Lemma~\ref{lem:existential-formula} and let
  \begin{gather*}
    \phi_{M}(y_1, \ldots, y_M) \ \isdef \quad \Und_{1\leq i<j\leq M} \!\!\!\!\neg\, y_i{=}y_j \quad \und \quad (\psi)_{M,s}(y_1,\ldots,y_M).
  \end{gather*}
  Furthermore, we let
  \begin{gather*}
    \psi_N \ \isdef \quad \Oder_{M=1}^N \exists y_1 \cdots \exists y_M\ \phi_{M}(y_1,\ldots,y_M).
  \end{gather*}
   Suppose that $\A' \in \Class$ is a model of $\psi$. Since $\psi$ is preserved under extensions on $\Class$ and every $\Class$-minimal model of $\psi$ has at most $N$ elements, there is an induced substructure $\A \in \Class$ of $\A'$ with a universe of exactly $M \leq N$ pairwise distinct elements $a_1,\ldots,a_M$, such that $\A \models \psi$. Therefore, by Lemma~\ref{lem:existential-formula}, $(\A, a_1,\ldots, a_M)$ is a model of $\phi_{M}(y_1, \ldots, y_M)$. Hence, $\A \models \psi_N$, and since $\psi_N$ is existential, also $\A' \models \psi_N$.  

On the other hand, suppose that $\A' \in \Class$ is a model of $\psi_N$. Then, there is an $M \in [1,N]$ and a substructure $\A\in\Class$ of $\A'$, induced by pairwise distinct elements $a_1,\ldots, a_M$ from $A'$, such that $(\A, a_1,\ldots,a_M)$ is a model of $(\psi)_{M,s}(y_1,\ldots,y_M)$. Hence, $\A \models \psi$ and, since $\A'$ is an extension of~$\A$, also $\A' \models \psi$. 

Thus, $\psi_N$ is $\mathfrak{C}$-equivalent to $\psi$.

  By Lemma~\ref{lem:existential-formula}, $(\psi)_{M,s}$ can be constructed in time $\size{\psi}  \cdot N^{\bigO(q\cdot \log m)}$. Hence, $\psi_N$ can be constructed in time
  \begin{gather*}
    N \cdot (N^2 + \size{\psi} \cdot N^{\bigO(q\cdot\log m)})
    \quad \subseteq \quad \size{\psi} \cdot N^{\bigO(q\cdot \log m)}.
  \end{gather*}

  Furthermore, if $\psi$ does not contain any modulo counting quantifier, it suffices to let
  \begin{gather*}
    \psi_N\ \isdef \quad \exists y_1 \cdots \exists y_N\ (\psi)_{N,s}(y_1,\ldots,y_N).
  \end{gather*}
  Here, the construction of $(\psi)_{N,s}$ and $\psi_N$ takes time only $\bigO(\size{\psi} \cdot N^q)$.
\qed

\medskip

\proof[Proof of Lemma~\ref{lem:existential-formula}.] \ \\
  The proof proceeds by induction on the shape of $\psi$.  
  \smallskip
  \begin{itemize}
  \item For the induction start, let $\psi$ be an atomic formula with variables among $x_1,\ldots,x_k$. In this case, $(\psi)_{M,s}$ is the formula obtained by replacing all occurrences of variables $x_i$ in $\psi$ by $y_{s(i)}$.
  It is easy to check that $(\psi)_{M,s}$ satisfies the condition of Lemma~\ref{lem:existential-formula}.  Assume that $\psi = R(x_{i_1}, \ldots, x_{i_{\ell}})$ for a relation symbol $R\in\sigma$ of arity $\ell\geq 1$ and for $i_1,\ldots,i_{\ell}\in [1,k]$.
    For each $\sigma$-structure $\A$ with at most $M$ elements and each tuple $(a_1,\ldots,a_M) \in A^M$ that contains each element of $A$ at least once, the following equivalence holds: \smallskip
    \begin{itemize}[label={iff\ }]
    \item[] $(\A, a_{s(1)}, \ldots, a_{s(k)}) \ \models\ R(x_{i_1}, \ldots, x_{i_{\ell}})$ \smallskip
    \item[iff\ ] $(a_{s(i_1)}, \ldots, a_{s(i_{\ell})}) \in R^{\A}$ \smallskip
    \item[iff\ ] $(\A, a_1,\ldots,a_N) \ \models\  R(y_{s(i_1)}, \ldots, y_{s(i_{\ell})})$. \smallskip
    \end{itemize}
    For $\psi = (x_i = x_j)$ the argumentation is analogous. \medskip
  \item   If $\psi = \neg\psi'$ then $(\psi)_{M,s} \isdef \neg(\psi')_{M,s}$.
   \\
   Similarly, for each Boolean connective $\star \in \set{\und,\oder,\rightarrow,\leftrightarrow}$, we let 
   \[
      (\psi' \star \psi'')_{M,s} \quad \isdef \quad (\psi')_{M,s} \star (\psi'')_{M,s}. 
   \]
   In each of these cases it is easy to check that the constructed formula satisfies the lemma's condition. \medskip
    \item
      If $\psi = \exists x_i \; \psi'$ then 
      \begin{gather*}
        (\psi)_{M,s} \ \isdef \quad \Oder_{j=1}^M (\psi')_{M,s[i\to j]},
      \end{gather*}
      where $s[i{\to}j]$ is the function $s'$ that agrees with $s$ on all values except $i$, and $s'(i) = j$. Note that $\size{(\psi)_{M,s}} \in \bigO(M \cdot \size{(\psi')_{M,s'}})$.
      For each $\sigma$-structure $\A$ with at most $M$ elements and each tuple $(a_1,\ldots,a_M)\in A^M$ that contains each element of $A$ at least once, the following equivalence holds: \medskip
      \begin{itemize}[label={iff\ }]
      \item[] $(\A, a_{s(1)},\ldots, a_{s(k)}) \ \models\  \exists x_i \; \psi'(x_1, \ldots, x_k)$ \medskip
      \item[iff\ ] there is a $j \in [1,M]$ such that for\ $s'\isdef s[i\to j]$,\\ $(\A, a_{s'(1)},\ldots,a_{s'(k)})\ \models\ \psi'(x_1, \ldots, x_k)$\medskip
      \item[iff\ ] there is a $j \in [1,M]$ such that\ $(\A, a_1,\ldots, a_M)\ \models\ (\psi')_{M, s[i\to j]}(y_1,\ldots,y_M)$\medskip
      \item[iff\ ] $(\A, a_1,\ldots,a_M)\ \models\ (\psi)_{M,s}(y_1, \ldots, y_M)$.
      \end{itemize}\medskip
    \item Accordingly, if $\psi = \forall x_i \; \psi'$ then
      \begin{gather*}
        (\psi)_{M,s} \ \isdef \quad \Und_{j=1}^M (\psi')_{M,s[i\to j]}.
      \end{gather*}
    \item Finally, assume that $\psi = \existm{0}{m} x_i \; \psi'$. 
      While it is straightforward to write,  in an analogous manner to the two latter cases, a formula with the desired semantics but with size exponential in $M$, we employ an inductive construction whose size is linear in $M$.

      More precisely, we let $(\psi)_{M, s} \isdef \gamma_0^{1, M}$, where $\gamma^{1,M}_0$ is the formula provided by the following claim:
  \end{itemize}
  \begin{clm}
    \label{claim:transform-existm}
    For all $j,j' \in [1,M]$ with $j\leq j'$ and each $p \in [0,m{-}1]$, there is a quantifier-free formula 
    $\gamma_p^{j,j'}(y_1,\ldots,y_M)$ such that for each $\sigma$-structure $\A$ with exactly $M$ elements and each enumeration $(a_1,\ldots,a_M)$ of $A$, the following equivalence holds:
    \begin{gather*}
      (\A, a_1,\ldots,a_M) \quad \models \quad \gamma^{j,j'}_p(y_1,\ldots,y_M)
    \end{gather*}
    if, and only if, the number of indices $\ell \in [j,j']$ such that
    \begin{gather*}
      (\A, a_1, \ldots, a_M) \quad \models \quad
      (\psi')_{M,s[i\to\ell]}(y_1,\ldots,y_M)
    \end{gather*}
    is congruent to $p$ modulo $m$.

    Furthermore, the formula $\gamma^{j,j'}_p$ is of size
    \begin{gather*}
      \size{(\psi')_{M,s}} \cdot (j'{-}j{+}1)^{\bigO(\log m)},
    \end{gather*}
    where the constant suppressed by the $O$-notation does not depend on the signature $\sigma$.
  \end{clm}
  To show that $\gamma_0^{1, M}$ satisfies the requirements of Lemma~\ref{lem:existential-formula}, let $\A$ be a $\sigma$-structure with exactly $M$ elements and let $(a_1,\ldots,a_M)$ be an enumeration of $A$. Then, due to Claim~\ref{claim:transform-existm}, the following equivalence holds:\medskip
  \begin{itemize}[label=[iff\ ]
  \item[] $(\A, a_{s(1)}, \ldots, a_{s(k)}) \ \models\ \existm{0}{m} x_i\;\psi'(x_1,\ldots,x_k)$ \medskip
  \item[iff\ ] the number of indices $\ell\in [1,M]$, such that for $s' \isdef s[i\to\ell]$, \\ $(\A, a_{s'(1)}, \ldots, a_{s'(k)})\ \models\ \psi'(x_1,\ldots,x_k)$, is congruent to $0$ modulo $m$\medskip
  \item[iff\ ] the number of indices $\ell\in [1,M]$, such that\ $(\A, a_1,\ldots, a_M) \ \models\ (\psi')_{M,s[i\to\ell]}(y_1,\ldots,y_M)$,\\ is congruent to $0$ modulo $m$ \\ (recall that $\psi'$ and $(\psi')_{M,s[i\to\ell]}$ satisfy the assumptions of Lemma~\ref{lem:existential-formula})\medskip
  \item[iff\ ] $(\A, a_1, \ldots, a_M)\ \models\ \gamma^{1,M}_0(y_1,\ldots,y_M)$.
  \end{itemize}
  \medskip
  By Claim~3, the size of $\gamma_0^{1,M}$ is in $ \size{(\psi')_{M,s}} \cdot M^{\bigO(\log m)} $. 
  It remains to prove Claim~\ref{claim:transform-existm}.
  \proof[Proof of Claim~\ref{claim:transform-existm}.]
    We start by defining, for each $j \in [1,M]$ and each $p\in [0,m{-}1]$, the formula $\gamma^{j,j}_p$:  
    We let $\gamma^{j,j}_0 \ \isdef \ \neg(\psi')_{M,s[i\to j]}$ and $\gamma^{j,j}_1 \ \isdef \ (\psi')_{M,s[i\to j]}$. For each $p > 1$, we let $\gamma^{j,j}_p$ be an unsatisfiable formula, e.g., $ \neg y_1{=}y_1$. By definition, these formulas satisfy Claim~\ref{claim:transform-existm}.
    
    For all $j,j'\in [1,M]$ with $j < j'$ and each $p \in [0,m{-}1]$, we let
    \begin{gather*}
      \gamma^{j,j'}_p \ \isdef \quad
      \Oder_{(p_1, p_2) \in P}
      \Bigl(\ 
      \gamma_{p_1}^{j,h}
      \ \und\ 
      \gamma_{p_2}^{h+1,j'}
      \ \Bigr),
    \end{gather*}
    where $h \isdef \bigl\lfloor \tfrac{j+j'}{2} \bigr\rfloor$ and 
    $P$ is the set of all tuples $(p_1,p_2)\in [0,m{-}1] \times [0,m{-}1]$ such that $p_1+p_2$ is congruent to $p$ modulo $m$.
    Note that $|P|=m$.
    
    For each $\sigma$-structure $\A$ with exactly $M$ elements and each enumeration $(a_1,\ldots,a_M)$ of $A$, the following equivalence holds:\medskip
    \begin{itemize}[label=[iff\ ]
    \item[] $(\A, a_1, \ldots, a_M) \ \models\  \gamma_p^{j,j'}(y_1,\ldots,y_M)$\medskip
    \item[iff\ ] there are $p_1,p_2 \in [0,m{-}1]$ with $p_1+p_2 \equiv p \mod m$ such that\\ $(\A, a_1,\ldots,a_M)\ \models\ \gamma^{j,h}_{p_1}(y_1,\ldots,y_M)$\ and\ $(\A, a_1,\ldots,a_M)\ \models\ \gamma^{h+1,j'}_{p_2}(y_1,\ldots,y_M)$\medskip
    \item[iff\ ] there are $p_1,p_2\in [0,m{-}1]$ such that $p_1+p_2 \equiv p \mod m$ and the number of indices $\ell$ in the intervals $[j,h]$ and $[h+1,j']$, such that\ $(\A, a_1,\ldots,a_M) \ \models\ (\psi')_{M,s[i\to\ell]}(y_1,\ldots,y_M)$, is congruent to~$p_1$ modulo $m$ and congruent to~$p_2$ modulo $m$, respectively,\medskip
    \item[iff\ ] the number of indices $\ell \in [j,j']$, such that\ $(\A,a_1,\ldots,a_M)\ \models\ (\psi')_{M,s[i\to\ell]}(y_1,\ldots,y_M)$, is congruent to $p$ modulo $m$.
    \end{itemize}
    \medskip
    It remains to show that for all $j,j'\in [1,M]$ and $p \in [0,m{-}1]$, 
    \begin{gather*}
      \size{\gamma_p^{j,j'}} \quad \in \quad \size{(\psi')_{M,s}} \cdot (j'{-}j{+}1)^{\bigO(\log m)}.
    \end{gather*}
    Note that the number of occurrences of $(\psi')_{M,s}$ in $\gamma_p^{j,j'}$ in the ``leaves'' of the inductive definition of $\gamma^{j,j'}_p$ 
 is in $m^{\bigO(\log (j'{-}j{+}1))}$. Moreover, taking into account the ``inner nodes'' and the ``leaves'' of the inductive definition of $\gamma^{j,j'}_p$ separately, we can bound the size of $\gamma^{j,j'}_p$ from above by
    \begin{gather}
      \label{expr:upper-bound-gamma}
      \size{\gamma^{j,j'}_p} \quad \in \quad
      T(j'{-}j{+}1) \ \ + \ \ \size{(\psi')_{M,s}} \cdot m^{\bigO(\log (j'{-}j{+}1))},
    \end{gather}
    where the function $T \colon \NNpos\to\NNpos$ is defined inductively by $T(1) \isdef 1$ and, for all $n > 1$, 
    \begin{gather}
      \label{expr:def-Tn}
      T(n) \ \isdef \quad a\cdot T(\lceil n / 2 \rceil)
    \end{gather}
    for a number $ a \geq 1$ of size $\bigO(m)$. For an upper bound on the growth of $T(n)$ we apply a variation of the well-known Master Theorem, which reads as follows:
    \begin{thm}[adapted from Theorem~4.1 of \cite{clrs}]
      Let $a \geq 1$ and let $b > 1$. Let $T \colon \NNpos \to \NN$ be defined by $T(1) \isdef 1$ and, for all $n > 1$, $ T(n) \isdef a \cdot T(\lceil n / b\rceil)$. 
      Then, $T(n) \in \Theta(n^{\log_b a})$.\qedhere
    \end{thm}    
    By choosing the number $a$ as in (\ref{expr:def-Tn}) and with $b \isdef 2$, we can conclude that
    \begin{gather*}
      T(n) \quad \in \quad \Theta(n^{\log a}) \quad \subseteq \quad n^{\bigO(\log m)}.
    \end{gather*}
    Replacing $T(j'{-}j{+}1)$ by $(j'{-}j{+}1)^{\bigO(\log m)}$ in (\ref{expr:upper-bound-gamma}) we obtain that
    \begin{gather*}
      \begin{split}
        \size{\gamma^{j,j'}_p} \quad \in 
        & \quad (j'{-}j{+}1)^{\bigO(\log m)} \ + \ \size{(\psi')_{M,s}} \cdot m^{\bigO(\log (j'{-}j{+}1))} \\
        \subseteq 
        & \quad \size{(\psi')_{M,s}} \cdot \bigl( (j'{-}j{+}1)^{\bigO(\log m)} \ + \ m^{\bigO(\log (j'{-}j{+}1))} \bigr)\\
        \subseteq
        & \quad \size{(\psi')_{M,s}} \cdot (j'{-}j{+}1)^{\bigO(\log m)} 
      \end{split}
    \end{gather*}
    This completes the proof of Claim~\ref{claim:transform-existm}.
  \qed
  For a bound on the size of $(\psi)_{M,s}$, note that the only size increasing steps in the inductive translation are the ones for the quantifiers, which increase the size of the formula by a factor of $M$, for first-order quantifiers, and by a factor in $M^{\bigO(\log m)}$, for modulo counting quantifiers. It follows that $\size{(\psi)_{M,s}} \in \size{\psi} \cdot M^{\bigO(q\cdot \log m)}$, where $q$ is the quantifier rank of $\psi$. If $\psi$ does not contain any modulo counting quantifier, $\size{(\psi)_{M,s}} \in \bigO(\size{\psi} \cdot N^q)$. It is easy to see that the inductive translation can be also carried out in time $\bigO(\size{(\psi)_{M,s}})$.
  This completes the proof of Lemma~\ref{lem:existential-formula}.
\qed

\smallskip

Theorem~\ref{thm:ep-upper-bound} is now obtained by a straightforward
combination of Lemma~\ref{lem:existential-sentence} and
Theorem~\ref{thm:ep-bound-minimal-model}.  
\proof[Proof of Theorem~\ref{thm:ep-upper-bound}.]
  Let $m\geq 2$ and let $\phi$ be an $\FOMOD{m}(\sigma)$-sentence of
  quantifier rank $q\geq 0$ that is preserved under extensions on
  $\mathfrak{C}_{\nu}$. Let
  \begin{align*}
    \begin{split}
      N \ \isdef \quad  N_\nu(m,q,\size{\sigma}) \quad \in \quad  m\cdot \Nsph{\Nsph{3^q}} 
      & \quad = \quad  m \cdot 2^{\nu({\Nsph{3^q}})^{\bigO(\size{\sigma})}}
    \end{split}
  \end{align*}
  be the upper bound on the
  size of $\mathfrak{C}_{\nu}$-minimal models of $\phi$, obtained from
  Theorem~\ref{thm:ep-bound-minimal-model}.

  By Lemma~\ref{lem:existential-sentence} there is an algorithm that
  constructs an existential $\FO(\sigma)$-sentence that is
  $\mathfrak{C}_{\nu}$-equivalent to $\phi$ in time
  \begin{gather*}
    \begin{split}
    \size{\phi} \cdot N^{\bigO(q\cdot\log m)} & \quad = \quad 
    \size{\phi} \cdot \Bigl( m \cdot 2^{\nu(\Nsph{3^q})^{\bigO(\size{\sigma})}} \Bigr)^{\bigO(q\cdot \log m)} \\
    & \quad\subseteq\quad
    \size{\phi} \cdot 2^{\bigO( (\log m)^2 \cdot q \cdot \nu(\Nsph{3^q})^{\bigO(\size{\sigma})})}\\
    & \quad \subseteq \quad
    \size{\phi} \cdot \Bigl( 2^{\nu(\Nsph{3^q})^{\bigO(\size{\sigma})}} \Bigr)^{(\log m)^2}\\
    & \quad = \quad
    \size{\phi} \cdot \Nsph{\Nsph{3^q}}^{(\log m)^2}.
    \end{split}
  \end{gather*}
  Here we make again use of the assumption that $\nu$ is strictly increasing.
  This concludes the proof of Theorem~\ref{thm:ep-upper-bound}.
\qed

\medskip
\subsection{Preservation under homomorphisms: Proof of
  Theorem \ref{thm:hp-upper-bound}}
\label{subsection:hp}
The combinatorial essence of the proof of
Theorem~\ref{thm:hp-upper-bound} is contained in the following
theorem.

\begin{thm}\label{thm:hp-bound-minimal-model}
  Let $\mathfrak{C}_{\nu}$ be a class of $\nu$-bounded
  $\sigma$-structures that is closed under disjoint unions and induced
  substructures.  Let $\phi$ be an
  $\FOMOD{m}(\sigma)$-sentence of quantifier rank $q\geq 0$ that is
  preserved under homomorphisms on~$\mathfrak{C}_{\nu}$.  There is a
  number
  \begin{gather*}
    N_\nu(q, \size{\sigma})\quad \in\quad   \Nsph{2{\cdot}3^q} \qquad\qquad \Bigl(= \quad
    2^{\nu(2{\cdot}3^q)^{\bigO(\size{\sigma})}}\ \Bigr)
  \end{gather*}
  such that every $\mathfrak{C}_{\nu}$-minimal model of $\phi$ has
  size at most $N_\nu(q, \size{\sigma})$.
\end{thm}
\proof
  The proof is similar to the proof of Lemma~7.1
  in~\cite{AjtaiGurevich1989}. However, it does not rely on Gaifman's
  theorem but uses Nurmonen's generalisation of Hanf's theorem, stated
  in Theorem~\ref{thm:Hanf-FOMOD}. Towards applying
  Theorem~\ref{thm:Hanf-FOMOD}, we let \mbox{$r\isdef 3^q$}, let
  $t\isdef q\cdot (\nu(r){+}1)+1$, and let~$s \in \Nsph{2r}$ be the number of non-isomorphic $2r$-spheres (with one centre) 
  realised in $\sigma$-structures in~$\mathfrak{C}_{\nu}$.

  Let $\A$ be a $\mathfrak{C}_{\nu}$-minimal model of $\phi$.  Towards
  a contradiction, assume that $|A| > s\cdot\nu(4r)$.  By
  Lemma~\ref{lem:scattered-sets},~$\A$~contains a
  $2r$\hbox{-}scattered set of size $s{+}1$.  Thus, since there are at
  most $s$ non-isomorphic $2r$\hbox{-}spheres realised in $\A$, there
  must be two elements $a_1,a_2\in A$ with disjoint and isomorphic
  $2r$-neighbourhoods.

  Let $\A' \isdef \A - \set{a_1}$.  Clearly, the $r$-spheres
  of elements in $A\setminus N_r^\A(a_1)$ are the same in $\A$ and in~$\A'$. 
  But the $r$-sphere of an element in $N_r^\A(a_1)$ might
  change when moving from $\A$ to $\A'$. However, by our choice of
  $a_1$ and $a_2$ we know that every $r$\hbox{-}sphere that is
  realised in $\A$ is also realised in $\A'$ (for elements outside the
  $r$-neighbourhood of $a_1$ this is obvious; and for elements
  $a'_1\in N_r^\A(a_1)$, the $r$-sphere of $a'_1$ in $\A$ is realised
  in $\A'$ by the corresponding element $a'_2$ in the~$r$\hbox{-}neighbourhood 
  of $a_2$).

  Now let $\B'$ be the disjoint union of $t{\cdot}m$ copies of $\A'$,
  and let $\B$ be the disjoint union of $\B'$ and of $t{\cdot}m$
  copies of $\A$.

  By construction, every $r$-sphere that is realised in $\B$ is also
  realised in $\B'$, and vice versa. Furthermore, the number of
  realisations of any $r$-sphere in $\B$ or $\B'$ is a multiple of
  $t{\cdot}m$.  In particular, $\B$ and $\B'$ satisfy the assumption
  of Theorem~\ref{thm:Hanf-FOMOD}, and therefore we have
  $\B\equiv^q_m\B'$.  Thus, $\B\models\phi$ iff $\B'\models\phi$.

  Furthermore, since $\mathfrak{C}_\nu$ is closed under taking induced
  substructures and disjoint unions, we know that $\B'$ and $\B$
  belong to $\mathfrak{C}_{\nu}$.  Obviously, there is a homomorphism
  that maps $\A$ to one of the copies of $\A$ in $\B$. Since $\phi$ is
  preserved under homomorphisms on~$\mathfrak{C}_{\nu}$ and
  $\A\models\phi$, we thus have $\B \models \phi$, and hence also
  $\B'\models\phi$.

  Recall that $\B'$ is a disjoint union of copies of $\A'$. By mapping
  each element of each copy of~$\A'$ to the corresponding element in
  $\A'$, we obtain a homomorphism from $\B'$ to $\A'$. Hence, since~$\phi$
   is preserved under homomorphisms and \mbox{$\B'\models\phi$},
  we obtain that also $\A' \models \phi$.  This, however, contradicts
  our assumption that $\A$ is a $\mathfrak{C}_{\nu}$-\emph{minimal}
  model of~$\phi$.  Therefore, $|A|\leq s\cdot\nu(4r)$. 
  We obtain that
  \begin{gather*}
    N_\nu(q, \size{\sigma}) \ \isdef \quad s\cdot \nu(4r)
    \quad \in \quad \Nsph{2{\cdot}3^q}.
  \end{gather*}
  This concludes the proof of
  Theorem~\ref{thm:hp-bound-minimal-model}.
\qed
\medskip

\noindent In the following lemma, we construct existential-positive
$\FO(\sigma)$\hbox{-}sentences for $\FOMOD{m}(\sigma)$-sentences that
are preserved under homomorphisms.  The proof is an algorithmic
version of the proof of Theorem~3.1 in
\cite{AtseriasDK06-homomorphism-preservation}.
\begin{lem}\label{lem:existential-positive-sentence}
  Let $\mathfrak{C}$ be a class of $\sigma$-structures that is
  decidable in time $t(n)$ for an arbitrary function $t\colon\NN\to\NN$.  There
  is an algorithm that, on input of a number $N \geq 0$ and an
  $\FOMOD{m}(\sigma)$-sentence $\phi$ of quantifier rank $q\geq 0$ constructs in time
  \begin{gather*}
    2^{N^{\bigO(\size{\sigma})}}
    \cdot
    \bigl( N^{\bigO(\size{\sigma} \cdot \size{\phi})} + t(\bigO(N^{\size{\sigma}})) \bigr)
  \end{gather*}
  an existential-positive $\FO(\sigma)$-sentence $\psi$ of size
  $2^{N^{\bigO(\size{\sigma})}}$, such that the following holds: If
  $\phi$ is preserved under homomorphisms on $\mathfrak{C}$ and every
  $\mathfrak{C}$-minimal model of $\phi$ has at most $N$ elements,
  then $\psi$ is $\mathfrak{C}$-equivalent to $\phi$.
\end{lem}
\proof
  For each finite $\sigma$-structure $\A$ let $\gamma_{\A}$ be the
  \emph{canonical conjunctive query} associated with $\A$. I.e.,
  $\gamma_{\A} = \exists x_1\cdots \exists x_{|A|}\, \theta_{\A}$
  where $x_1,\ldots,x_{|A|}$ are variables representing the elements
  $a_1,\ldots,\allowbreak a_{|A|}$ of $\A$'s universe $A$, and
  $\theta_{\A}$ is the conjunction of all atoms of the form
  $R(x_{i_1},\ldots,x_{i_r})$ where $R\in\sigma$, $r=\ar(R)$,
  $i_1,\ldots,i_r\in \set{1, \ldots, |A|}$, and
  $(a_{i_1},\ldots,a_{i_r})\in R^{\A}$. The well-known
  \emph{Chandra-Merlin Theorem} \cite{ChandraMerlin} states that for
  any $\sigma$-structure $\B$, there is a homomorphism from $\A$ to
  $\B$ if, and only if, $\B\models\gamma_{\A}$.  Clearly, given $\A$,
  the sentence $\gamma_{\A}$ can be constructed in time
  $\bigO(\size{\A})=\bigO(|A|^{\size{\sigma}})$.
  \smallskip

  On input of $\phi$ and $N$, the lemma's algorithm\smallskip
  \begin{enumerate}
  \item computes the set $M$ that consists all models of $\phi$\\ with
    universe $\set{1,\ldots, n}$, for $n \leq N$, that belong to
    $\mathfrak{C}$,\smallskip
  \item if $M=\emptyset$, it outputs the formula \ $\psi :=
    \textit{false}$,\smallskip
  \item if $M\neq\emptyset$, it outputs the formula \ $\psi :=
    \bigvee_{\A\in M} \gamma_{\A}$.\smallskip
  \end{enumerate}
  Clearly, $|M|$ is bounded by the number of all $\sigma$-structures with universe $\set{1,\ldots, n}$ and $n \leq N$. Hence, the size of $\psi$ is in
  \begin{gather*}
    \bigO(|M| \cdot N^{\size{\sigma}}) \quad \subseteq \quad
    2^{N^{\bigO(\size{\sigma})}}
  \end{gather*}
  Obviously, $\psi$ is an existential-positive
  $\FO(\sigma)$-sentence.

  Before giving details on the algorithm's step~1 and its running
  time, let us first show that $\psi$ is $\mathfrak{C}$\hbox{-}equivalent to
  $\phi$, provided that $\phi$ is preserved under homomorphisms on
  $\mathfrak{C}$ and that every $\mathfrak{C}$\hbox{-}minimal model of $\phi$
  has at most $N$ elements.  To this end, let $\B$ be an arbitrary
  $\sigma$-structure in~$\mathfrak{C}$.

  If $\B\models\psi$, then there is an $\A\in M$ such that
  $\B\models\gamma_\A$. Due to the Chandra-Merlin Theorem, there is a
  homomorphism from $\A$ to $\B$. As $\A\in\mathfrak{C}$ and
  $\A\models\phi$, and since $\phi$ is preserved under homomorphisms
  on $\mathfrak{C}$, we obtain that $\B\models\phi$.

  On the other hand, if $\B\models \phi$, then let $\A$ be a minimal
  induced substructure of $\B$ such that $\A\in\mathfrak{C}$ and
  $\A\models\phi$.  I.e., $\A$ is a $\mathfrak{C}$-minimal model of
  $\phi$. By assumption, $N$ is an upper bound on the size of the
  universe of $\A$.  Thus, by our choice of $M$, the set $M$ contains
  a structure $\A'$ that is isomorphic to $\A$.  Since $\A$ is a
  substructure of $\B$, the particular choice of the formula
  $\gamma_{\A'}$ implies that $\B\models \gamma_{\A'}$. Since $\A'\in
  M$, we obtain that also $\B\models\psi$.  In summary, this shows
  that $\psi$ is $\mathfrak{C}$-equivalent to $\phi$.
  
  \medskip

  Let us now turn to the algorithm's step~1 and the analysis of its
  time complexity.  To compute $M$, the algorithm enumerates all
  $\sigma$-structures $\A$ with $A= \set{1,\ldots,n}$ and $n\leq N$,
  and checks for each such $\A$ whether $\A\models \phi$ and $\A\in
  \mathfrak{C}$.

  By assumption, the question whether $\A\in\mathfrak{C}$ can be
  answered within time $t(\size{\A}) \in t(\bigO(N^{\size{\sigma}}))$.  Using
  the naive model checking algorithm for $\FOMOD{m}$, the question
  whether $\A\models\phi$ can be answered within time
  $\size{\A}^{\bigO(\size{\phi})} \subseteq
  N^{\bigO(\size{\sigma}\cdot\size{\phi})}$.  Since $2^{N^{\bigO(\size{\sigma})}}$ is an upper bound on the number of $\sigma$-structures with universe
  $\set{1,\ldots, n}$ and $n\leq N$, the entire computation of $M$
  takes time at most
  \begin{gather*}
    2^{N^{\bigO(\size{\sigma})}} \cdot
    \big(
    N^{\bigO(\size{\sigma}\cdot\size{\phi})}
    + t(\bigO(N^{\size{\sigma}}))
    \big). 
  \end{gather*}
  This completes the proof of Lemma~\ref{lem:existential-positive-sentence}.
  \qed
\medskip

\noindent Theorem~\ref{thm:hp-upper-bound} is now obtained by a straightforward
combination of Lemma~\ref{lem:existential-positive-sentence} and
Theorem~\ref{thm:hp-bound-minimal-model} (in the analogous way as
Theorem~\ref{thm:ep-upper-bound} was obtained by combining
Lemma~\ref{lem:existential-sentence} with
Theorem~\ref{thm:ep-bound-minimal-model}).
\medskip

\subsection{Closure properties.}
\label{subsection:closure-properties}

Our Theorems \ref{thm:ep-upper-bound} and \ref{thm:hp-upper-bound}
require that the considered classes be closed under disjoint unions
and induced substructures. In this section we provide simple examples which show that these closure properties are indeed necessary. Both examples
use graphs that are directed paths where some endpoints are colored
green. These are represented as structures over the signature $\sigma
\isdef \set{ E, G }$ as usual. That is, the binary relation symbol $E$
is interpreted by the edge relation and the unary relation symbol $G$
is interpreted by the set of green vertices.  In the following, a
vertex is a \emph{left endpoint} or a \emph{right endpoint} if it is
has, respectively, no ingoing edge or no outgoing edge. An
\emph{endpoint} is either a left or a right endpoint.  For $n\geq 1$,
a directed path on $n$ vertices where exactly the endpoints are
colored green will be denoted by ${\mathcal P}_n$ and a directed path
on $2n+1$ vertices where just the central vertex is colored green
will be denoted by ${\mathcal P}^C_n$.
\begin{thm}\label{thm:counterexample-disjoint-union}
  There is a class $\Class_1$ of $\sigma$-structures of degree at most $
  2$ that is closed under substructures but not under disjoint unions,
  and there is an $\FO(\sigma)$-sentence $\phi$ that is preserved
  under extensions and homomorphisms on $\Class_1$, but that has no
  $\Class_1$-equivalent existential $\FO(\sigma)$-sentence.
\end{thm}
\proof
  Let $\Class_1$ be the class that contains a $\sigma$-structure if there is an $n \geq 1$ such that the structure is isomorphic to a substructure of $\mathcal{P}_n$. By construction, $\Class_1$ is closed under substructures. It is not closed under disjoint unions, since e.g. the disjoint union of two copies of $\mathcal{P}_n$ is not a substructure of any $\mathcal{P}_m$. 

  There is an obvious $\FO(\sigma)$-sentence $\phi$ that is satisfied by a $\sigma$-structure $\A$ iff $|A| \geq 3$ and all endpoints of $\A$ are green. The models of $\phi$ that belong to $\Class_1$ are exactly the structures that are isomorphic to $\mathcal{P}_n$ for some $n \geq 3$, because each proper substructure $\A$ of $\mathcal{P}_n$ contains an endpoint that is not green. The sentence $\phi$ is preserved under homomorphisms (and hence also under extensions) on $\Class_1$ for the trivial reason that the only structure in $\Class_1$ to which there is an homomorphism from $\mathcal{P}_n$ is $\mathcal{P}_n$ itself.

  It remains to show that on $\Class_1$ the formula $\phi$ is not equivalent to an existential sentence. Assume towards a contradiction that $\phi$ is $\Class_1$-equivalent to an existential sentence $\psi \isdef \exists x_1 \cdots \exists x_k\; \xi$,  where $k \geq 1$ and where $\xi$ is quantifier-free.
  In particular, $\mathcal{P}_{k+3} \models \psi$ so that there are $a_1,\ldots,a_k \in P_{k+3}$ for which $(\mathcal{P}_{k+3}, a_1,\ldots,a_k) \models \xi$. 
  Let $\mathcal{P}$ be the substructure of $\mathcal{P}_{k+3}$ induced by $\set{a_1,\ldots, a_k}$. 
  Clearly, $\mathcal{P}\cong\mathcal{P}_{k+3}[\set{a_1,\ldots,a_k}]$. Thus $(\mathcal{P}, a_1,\ldots,a_k)\models\xi$ and so $\mathcal{P} \models \psi$. On the other hand, $\mathcal{P}$ contains at least one endpoint that is not colored green. Therefore, $\mathcal{P}\not\models\phi$. This contradicts our assumption that $\phi$ and $\psi$ are equivalent. 
\qed
\begin{thm}\label{thm:counterexample-substructures}
  There is a class $\Class_2$ of $\sigma$-structures of degree at most $2$ that is closed under disjoint unions but not under induced
  substructures, and there is an $\FO(\sigma)$-sentence that is
  preserved under extensions and homomorphisms on $\Class_2$, but
  that has no $\Class_2$-equivalent existential
  $\FO(\sigma)$-sentence.
\end{thm}
\proof
  Let $\Class_2$ be the class of all $\sigma$-structures that are
  disjoint unions of structures that are isomorphic to ${\mathcal P}_n$ or ${\mathcal P}^C_n$, for possibly different lengths $n\geq
  1$. By construction, $\Class_2$ is closed under disjoint unions. It
  is not closed under induced substructures, since e.g. ${\mathcal
    P}_3$ has an isolated vertex that is not colored green as an
  induced substructure, but such graphs do not belong to $\Class_2$.

  There is an obvious $\FO(\sigma)$-sentence $\phi$ that is satisfied
  by a $\sigma$-structure $\A$ iff it contains a green endpoint.  A
  structure belonging to $\Class_2$ satisfies $\phi$ iff it contains a
  copy of ${\mathcal P}_n$ for some $n \geq 1$, since the ${\mathcal
    P}^C_n$ do not contain green endpoints. The sentence $\phi$ is
  preserved under homomorphisms (and hence also under extensions) on
  $\Class_2$ because ${\mathcal P}_n$ cannot be mapped homomorphically
  to a ${\mathcal P}^C_m$ for any $m$ whatsoever, due to the two green
  endpoints.

  It remains to show that $\phi$ is not $\Class_2$-equivalent to an
  existential sentence. Assume to the contrary that 
  $\phi$ is $\Class_2$-equivalent to a sentence
  $\psi \isdef \exists  x_1 \cdots \exists x_k\ \xi$, where $k\geq 1$ and where $\xi$ is quantifier-free.
  Then ${\mathcal P}_{k+1}\models \psi$, i.e. there are $a_1, \ldots, a_k\in P_{k+1}$ for which
  $({\mathcal P}_{k+1},a_1, \ldots, a_k)\models \xi$. Let
  $M\isdef\set{a_1, \ldots,a_k}$. We partition $M$ into sets $L,R$, where
  $L$ is the (possibly empty) set of all vertices from $M$ that belong 
  to the connected component, in $\mathcal{P}_{k+1}[M]$, of
  the left endpoint of ${\mathcal P}_{k+1}$, 
  and $R \isdef M \setminus L$. Clearly, $L$ and $R$ are
  disconnected in ${\mathcal P}_{k+1}[M]$. In particular, the set $L$ cannot contain the right endpoint of $\mathcal{P}_{k+1}$. 

  Suppose that $L$ is empty and thus, $R = M$. There is a set $M' \subseteq P^C_k$ of vertices (``left of'' the central green vertex) in $\mathcal{P}^C_k$ such that $\mathcal{P}^C_k[M'] \cong \mathcal{P}_{k+1}[M]$. Hence, $\mathcal{P}^C_k \models \psi$. But clearly $\mathcal{P}^C_k[M'] \not\models\phi$. This is a contradiction.

  In the case that $L$ is not empty, the path $\mathcal{P}^C_k$ contains induced substructures that are isomorphic to $\mathcal{P}_{k+1}[L]$ and $\mathcal{P}_{k+1}[R]$, respectively (``right of'' and ``left of'' the central green vertex). Let $\mathcal{A}$ be the disjoint union of two copies $\A_1$ and $\A_2$ of $\mathcal{P}^C_k$. We map the elements of $L$ and $R$ to corresponding elements of $\A_1$ and $\A_2$, respectively. Now let $M'$ be the image of $M$ under this mapping. It is easy to verify that $\A[M']\cong\mathcal{P}_{k+1}[M]$. Hence $\A \models \psi$. But clearly $\A\not\models\phi$. This is a contradiction.
\qed

\section{Feferman-Vaught decompositions}\label{fv}
 
Throughout this section, let $\nu\colon\NN\to\NN$ be a fixed time-constructible
strictly increasing function and let $\mathfrak{C}_{\nu}$ be an
arbitrary class of $\nu$\hbox{-}bounded $\sigma$-structures for a finite relational signature $\sigma$. 

Let $(P_i)_{i\geq 1}$ be a sequence of unary relation symbols that are not already contained in  
$\sigma$. For every $s\geq 1$, by $\sigma_s$ we denote the signature $\sigma \cup\; \set{P_1,\ldots,P_s}$. 

Recall that a disjoint union $\B = \A_1 \dcup \cdots \dcup \A_s$ of
$\sigma$\hbox{-}structures $\A_1,\ldots,\A_s$  
involves injective functions $f_i \colon A_i \to B$ for each
$i\in[1,s]$ such that $f_1(A_1),\ldots,f_s(A_s)$ is a partition of
$B$.  
The \emph{disjoint sum $\A_1\oplus\cdots\oplus\A_s$ of
  $\A_1,\ldots,\A_s$} is a $\sigma_s$\hbox{-}structure $\A$ that
expands the disjoint union of $\A_1,\ldots,\A_s$ by the unary
relations $P^{\A}_i \isdef f_i(A_i)$, for all $i\in[1,s]$. Clearly,
$P_1^{\A},\ldots,P_s^{\A}$ is a partition of $A$, and for all $a\in P_i^{\A}$ and $b\in
P_j^{\A}$ with $i,j\in [1,s]$ and $i\not=j$,  there is no edge between
$a$ and $b$ in the Gaifman graph of $\A$.  

In Subsection~\ref{sec:fv-upper-bound} we provide an algorithm that computes, for every $s\geq 1$
and each $\FO(\sigma_s)$-formula $\phi$, a 
\emph{Feferman-Vaught decomposition} of $\phi$, that is, a decomposition into a
Boolean combination of $\FO(\sigma)$\hbox{-}formulas which is equivalent to
$\phi$ on each disjoint sum of $s$ structures in
$\mathfrak{C}_{\nu}$ (for a precise definition, see Subsection~\ref{subsection:disjoint-decompositions}). For functions~$\nu$ of exponential growth, this
algorithm has $3$\hbox{-}fold exponential time complexity in terms of the
input formula; for polynomial $\nu$, the time complexity is
$2$\hbox{-}fold exponential.  In Subsection~\ref{subsection:fv-upper-bound-tensorprod} we show how to extend the algorithm to products of structures obtained by applying \emph{transductions} to disjoint sums, e.g., to \emph{direct products} of structures as well as to \emph{cartesian products} and \emph{strong products} of graphs.

In Section~\ref{subsection:lower-bounds-for-feferman-vaught-decompositions} we show that the algorithm's time complexity is basically
optimal: For structures of degree~$3$, a $3$-fold exponential blow-up
of the decomposition in terms of the size of the input formula is
unavoidable, and for structures of degree $2$ there is still a
$2$\hbox{-}fold exponential blow-up. 

\medskip
\subsection{Disjoint decompositions}
\label{subsection:disjoint-decompositions}

Before presenting this section's main results, we give a precise
definition of the decompositions constructed by our algorithm. These
decompositions are a special case of so-called \emph{reduction
  sequences} \cite{Makowsky2004159}. They give conditions for the
validity of an $\FO(\sigma_s)$-formula in a disjoint sum of structures
in terms of a Boolean combination of $\FO(\sigma)$\hbox{-}formulas 
that speak about the component structures of  the disjoint sum. 

Let $s\geq 1$ and let $\ov{x}$ be a tuple of $n\geq 0$ variables.
  For each $i\in [1,s]$, let $\Delta_i$ be a finite set of
  $\FO(\sigma)$-formulas~$\delta$  with $\free(\delta)\subseteq
  \ov{x}$ and let $\beta$ be a propositional formula with variables
  from the set \mbox{$\AVARS_D \isdef \set{ \AVAR_{i,\delta} \colon
      i\in [1,s], \delta \in \Delta_i }$.} 
  The tuple \mbox{$D=(\Delta_1,\ldots,\Delta_s, \beta)$} is an
  \emph{$s$-reduction sequence over~$\ov{x}$}  
  (for short: \emph{reduction sequence}).
  The size $\size{D}$ of $D$ is defined as
 $\size{\beta} + \sum_{i=1}^s \sum_{\delta\in\Delta_i}\size{\delta}
 $, where $\size{\beta}$ is the size of the propositional formula
 $\beta$ when viewed as a word over the alphabet
 $\set{\nicht,\und,\oder,\impl,\gdw,(,)} \cup \AVARS_D$. 

  Let $\A_1,\ldots,\A_s$ be $\sigma$-structures and let $\ov{a}$ be a
  tuple $(a_1,\ldots,a_n)$ from $(A_1 \cup \cdots \cup A_s)^n$. We say that 
  $(\A_1,\ldots,\A_s,\ov{a})$ is a \emph{model of the reduction
    sequence $D$}, in symbols: $(\A_1,\ldots,\A_s,\ov{a}) \models D$,
  iff $\mu \models \beta$,  
  where $\mu \colon \AVARS_D \to \set{0, 1}$ is the truth assignment
  such that for each $i\in [1,s]$ and $\delta\in\Delta_i$, 
  \begin{gather*}
    \mu(\AVAR_{i,\delta}) \ \isdef \quad
    \begin{cases}
      \ 1, & \text{if $\free(\delta) \subseteq \ov{x}_i\ $ and  $\
        (\A_i,\ov{a}_{i})\models \delta(\ov{x}_{i})$},\\ 
      \ 0, & \text{otherwise.}
    \end{cases}
  \end{gather*}
  Here, $\ov{a}_i$ is the subsequence of $\ov{a}$ induced by all $a_j
  \in \ov{a} \cap A_i$ and $\ov{x}_i$ is the subsequence of $\ov{x}$
  induced by all $x_j$ such that $a_j \in \ov{a} \cap A_i$. 

  Let $\mathfrak{C}$ be a class of $\sigma$-structures and let
  $\phi(\ov{x})$ be an $\FO(\sigma_s)$-formula. An $s$-reduction
  sequence $D$ over $\ov{x}$ is an \emph{$s$-disjoint decomposition
    for $\phi(\ov{x})$ on $\mathfrak{C}$}  
(for short: \emph{disjoint decomposition for $\phi(\ov{x})$}) if for
every $s$-disjoint sum $\A = \A_1\oplus\cdots\oplus\A_s$ of  
 structures $\A_1,\ldots,\A_s \in \mathfrak{C}$  and all tuples $\ov{a}\in A^n$,
  \begin{gather*}
    (\A, \ov{a})\quad \models\quad  \phi(\ov{x}) 
    \qquad \text{iff} \qquad  
    (\A_1,\ldots,\A_s,\pi(\ov{a}))\quad \models\quad  D.
  \end{gather*} 
  Here, $\pi$ is the \emph{mapping} of the disjoint sum $\A$, i.e.,
  the mapping of the underlying disjoint union of  
  $\A$'s component structures.
  
  Intuitively, an $s$-disjoint decomposition for an
  $\FO(\sigma_s)$\hbox{-}formula $\phi(\ov{x})$ is a Boolean
  combination of $\FO(\sigma)$\hbox{-}formulas from sets $\Delta_i$,
  $i\in[1,s]$.  
This Boolean combination is equivalent to $\phi$ on every
$s$\hbox{-}disjoint sum $\A_1\oplus\cdots\oplus\A_s$ and, for each
$i\in[1,s]$, every $\FO(\sigma)$-formula from $\Delta_i$ is only
interpreted over the component $\A_i$ and with its free variables
assigned to elements from $\A_i$. 

\medskip
\subsection{An upper bound}
\label{sec:fv-upper-bound}
This section's main result (Theorem~\ref{thm:fv-upper-bound}) provides a 3-fold exponential algorithm
that computes a disjoint decomposition  
for an input $\FO(\sigma_s)$-formula $\phi(\ov{x})$ on
$\mathfrak{C}_\nu$. The algorithm proceeds as follows: First, $\phi(\ov{x})$ is turned in $3$-fold exponential time into a Boolean combination $\phi^H(\ov{x})$ of so-called \emph{Hanf-formulas}  (which will be defined below) such that the formula $\phi^H(\ov{x})$ is equivalent to $\phi(\ov{x})$ on disjoint sums of structures from $\Class_{\nu}$. To achieve this, we apply an algorithm by Bollig and Kuske \cite{BolligKuske-Hanf-2012}. In a second step, for each of the Hanf-formulas occurring in $\phi^H(\ov{x})$, a disjoint decomposition is computed in linear time (see Lemma~\ref{lem:fv-upper-bound-hnf}). Finally, these disjoint decompositions are combined into a disjoint decomposition $D=(\Delta_1,\ldots,\Delta_s,\beta)$ for $\phi(\ov{x})$ on $\Class_{\nu}$. In particular, also the formulas in $\Delta_i$ are \emph{Hanf-formulas}.

A \emph{Hanf-formula} with $n\geq 0$ free variables $\ov{x}$ is a
formula of the form $ \exists^{\geq k} y\; \sph_\tau(\overline{x}, y)$ 
where $\tau$ is the isomorphism type of a finite $r$-sphere (for
an~$r\geq 0$) with $n{+}1$ centres.   

Here, for a number $k\geq 1$ and a formula $\varphi(\ov{x},y)$ we write
\begin{gather*}
  \exists^{\geq k}y\ \varphi(\ov{x},y)  
\end{gather*}
as a shorthand for the formula 
\begin{gather*}
  \exists y_1\cdots\exists y_k \;\Big( \!\!\!\!
  \Und_{1\leq i<j\leq k}\!\!\!\!\!\nicht y_i{=}y_j \ \und \ 
  \forall y \;
  \big( \!\!\!
  \Oder_{1\leq i\leq k}\!\!\!y{=}y_i  \ \impl  \varphi(\ov{x},y)
  \big) 
  \Big).
\end{gather*}
Note that, given $k$, $y$ and $\varphi$, this formula can be
constructed in time $\lin{k^2+\size{\varphi}}$.
\smallskip

In \cite{BolligKuske-Hanf-2012}, Bollig and Kuske provided a $3$-fold exponential 
  algorithm that transforms a given $\FO(\sigma')$\hbox{-}formula $\psi$, for a finite relational signature $\sigma'$, 
 into a formula in Hanf normal form that is equivalent to $\psi$ on all
 $\sigma'$-structures of degree at most~$d$ (for~$d\geq 1$). Here, we apply their result in the
 slightly more general setting of $\nu$-bounded structures. In this setting, their proof
 yields the following (a proof can be found in the full version of
 \cite{HKS13-LICS}).

\begin{thm}[\cite{BolligKuske-Hanf-2012,HKS13-LICS}]\label{thm:hnf-upper-bound}
  Let $\sigma'$ be a finite relational signature, let $\nu\colon\NN\to\NN$ be a time-constructible and strictly increasing function, and let $\Class'_{\nu}$ be the class of all $\nu$-bounded $\sigma'$-structures.
  There is an
  algorithm which transforms an input $\FO(\sigma')$-formula~$\phi$ of
  quantifier rank $q\geq 0$ in time
  \begin{gather}\label{eq:Laufzeit-Hanf-Algo}
    2^{(\size{\phi}\cdot\nu(4^q))^{\bigO(\size{\sigma'})}}
  \end{gather}
  into a $\Class'_{\nu}$-equivalent formula $\phi^H$ in Hanf normal form.
  
  Moreover, each Hanf-formula occurring in $\phi^H$ is of the form  $ \exists^{\geq k}y\; \sph_\tau(\ov{x},y)$, where $k$ is at most $\size{\phi}\cdot (q{+}1)\cdot
  \nu(4^q)$, $\tau$ is a $\nu$-bounded $r$-sphere of
  radius $r\leq 4^q$, and $|\ov{x}|\leq |\free(\phi)|$.\qed
\end{thm}\medskip

\newcommand{\Hanf}[3]{\ensuremath{H_\nu(#1, #2, #3)}}
\noindent In the rest of this section we abbreviate the upper bound $2^{(\size{\phi} \cdot \nu(4^q))^{\bigO(\size{\sigma'})}}$ on the time complexity for the construction of Hanf normal forms from Theorem~\ref{thm:hnf-upper-bound} by the expression $\Hanf{\size{\phi}}{4^q}{\size{\sigma'}}$.

\smallskip

With these preparations, this section's main result can now be stated as follows:
\begin{thm}\label{thm:fv-upper-bound}
  Let $s\geq 1$. 
  There is an algorithm which, given an input $\FO(\sigma_s)$-formula
  $\phi(\ov{x})$ of quantifier rank $q\geq 0$, constructs in time 
  \begin{gather}\label{expr:fv-upper-bound}
    \Hanf{\size{\phi}}{4^q}{\size{\sigma}+s}
    \qquad\qquad \Bigl(=\quad 
    2^{(\size{\phi} \cdot \nu(4^q))^{\bigO(\size{\sigma}+s)}}\ \Bigr)
  \end{gather}
  a disjoint decomposition $(\Delta_1,\ldots,\Delta_s,\beta)$ for
  $\phi(\ov{x})$ on $\mathfrak{C}_{\nu}$. 
  Furthermore, the sets $\Delta_1,\ldots,\Delta_s$ contain only
  Hanf-formulas over the signature $\sigma$.
\end{thm}
Thus, if the function $\nu$ is exponential and $\phi$ is an $\FO(\sigma_s)$-formula, where $\sigma_s$ consists of exactly the relation symbols that occur in $\phi$, then $\phi$ can be decomposed in $3$-fold exponential time --- e.g., if $\nu=\nu_d$ for $d\geq 3$, then $\phi$ can be decomposed in time 
\begin{gather*}
  2^{d^{2^{\bigO(\size{\phi})}}}.
\end{gather*}
If $\nu$ is polynomial, then the transformation requires only $2$-fold exponential time, i.e. time
\begin{gather*}
  2^{2^{\bigO(\size{\phi}^2)}}.
\end{gather*}

The remainder of this subsection is devoted to the proof of Theorem~\ref{thm:fv-upper-bound}.
The first step of the algorithm employs the algorithm of Theorem~\ref{thm:hnf-upper-bound} to transform an $\FO(\sigma)$\hbox{-}formula $\phi$ into a $\Class_{\nu}$-equivalent formula $\phi^H$ in Hanf normal form.

For the second step of our algorithm, 
the following lemma provides an algorithm which constructs, given a
Hanf-formula, a disjoint decomposition for it; not only on
$\nu$\hbox{-}bounded $\sigma$\hbox{-}structures but on \emph{all} $\sigma$-structures.
\begin{lem}\label{lem:fv-upper-bound-hnf}
  Let $s\geq 1$. There is an algorithm which, given an input Hanf-formula $\psi(\ov{x})$ of the form $\exists^{\geq k} x_{n+1}\; \sph_{\tau}(\ov{x}, x_{n+1})$ with $n\geq 0$ free variables, 
 where $k\geq 1$ and, for an $r\geq 0$, $\tau$ is an  $r$-sphere with $n{+}1$ centres over the signature $\sigma_s$, constructs in time
  \begin{gather*}
    \bigO(s + \size{\psi})
  \end{gather*}
  a disjoint decomposition $(\Delta_1,\ldots,\Delta_s,\beta)$ for $\psi(\ov{x})$ on the class of all $\sigma$-structures.

  Furthermore, the sets $\Delta_1,\ldots, \Delta_s$ contain only
  Hanf-formulas over $\sigma$,
  and each $\Delta_i$ consists of at most one formula.
\end{lem}

\proof
  Let $n,r \geq 0$, $k\geq 1$, let $\ov{x} = (x_1,\ldots,x_n)$, and let $\tau=(\T,\ov{c})$ be an $r$\hbox{-}sphere with $n{+}1$ centres \hbox{$\ov{c} = (c_1,\ldots, c_{n+1})$}, i.e., $\T$ is a $\sigma_s$-structure whose elements have distance at most $r$ to $\ov{c}$.

  Recall that $\T$ is an induced substructure of an $s$-disjoint sum iff\smallskip

  \begin{enumerate}
  \item[(1)] every element in the universe of $\T$ is contained in exactly one of the sets $P_1^{\T},\ldots,P_s^{\T}$,  and\smallskip
  \item[(2)] if there is an edge between two nodes $a,b$ in the Gaifman graph of $\T$, then there is an $i\in [1,s]$ such that $a,b\in P_i^{\T}$.\smallskip
  \end{enumerate}
  We distinguish whether or not $\T$ is an induced substructure of an $s$\hbox{-}disjoint sum.

  \smallskip
  \textit{Case 1:\ } $\T$ is not an induced substructure of an $s$-disjoint sum. 
  Then, there is no $s$-disjoint sum~$\A$ with a tuple $\ov{a}\in A^{n+1}$ such that $\N_r^{\A}(\ov{a}) \cong \tau$.
  Therefore, the algorithm can output an arbitrary unsatisfiable $s$-disjoint decomposition.

  \smallskip
  \textit{Case 2:\ } $\T$ is an induced substructure of an $s$-disjoint sum. 
  \noindent In the following, we first describe the construction of an $s$-disjoint decomposition for $\psi(\ov{x})$. Afterwards we show the correctness of this construction and give an analysis of its running time.

  Since $\T$ is an induced substructure of an $s$-disjoint sum, for each $i\in[1,n{+}1]$, there is a $j\in[1,s]$ such that
  $c_i$ as well as  all elements in $N_r^{\T}(c_i)$ belong to
  $P_j^{\T}$. 
  For each $i\in [1,s]$, let $\ov{c}_i$ be the subsequence of $\ov{c}$ induced by all  $c_j \in \ov{c} \cap P_i^{\T}$, and let $\ov{x}_i$ be the subsequence of $\ov{x}$ induced by all $x_j$ such that $c_j\in \ov{c}\cap P_i^{\T}$.
  W.l.o.g., $ c_{n+1} \in P^{\T}_s$.

  The algorithm decomposes $\psi(\ov{x})$ into a disjoint decomposition $D\isdef(\Delta_1,
  \ldots,\Delta_s,\beta)$, which is  defined as follows:\smallskip
\begin{itemize}
 \item
  For each $i\in [1,s]$ where $\ov{c}\cap
  P_i^{\T}=\emptyset$, let \ $\Delta_i\isdef\emptyset$.\smallskip
 \item
  For each $i\in[1,s]$ where $\ov{c}\cap P_i^{\T}
  \not=\emptyset$, let \ $\Delta_i \isdef \set{\delta_i(\ov{x}_i)}$, \\
  where the $\FO(\sigma)$\hbox{-}formula $\delta_i$ will be defined
  later on.\smallskip
 \item
 $\beta$ is the propositional formula
  \begin{gather*}
    \Und \set{\ \AVAR_{i,\delta_i} \colon i \in [1,s]\ \ \text{such that}\ \ \ov{c}\,\cap\, P_i^{\T} \not=\emptyset\ }
  \end{gather*}
\end{itemize}
 It remains to define the formulas $\delta_{i}(\ov{x}_{i})$.
 Let \mbox{$\T_i \isdef \N_r^{\T}(\ov{c}_{i})$} and let $\tau_i$ be
 the $r$-sphere $(\T_i,\ov{c}_{i})$. Let \mbox{$\T'_i \isdef
  \bigl(\T_i\bigr)_{|\sigma}$} be the $\sigma$-reduct of $\T_i$, and
 let $\tau'_i$ be the $r$-sphere $\bigl( \T'_i, \ov{c}_{i} \bigr)$.
 Recall that we assume that $c_{n+1}\in P_s^{\T}$.
 For $i{=}s$ we let
 \begin{gather*}
    \delta_s(\ov{x}_{s})\ \isdef \quad \exists^{\geq k} x_{n+1}\  \sph_{\tau'_s}(\ov{x}_{s}, x_{n+1}).
 \end{gather*}
  For each $i \in [1,s{-}1]$, i.e., for each $i$ where $c_{n{+}1}
  \not\in P^{\T}_i$, we define the formula $\delta_i(\ov{x}_{i})$ by 
  \begin{gather*}
   \delta_i(\ov{x}_{i})\ \isdef \quad  \exists^{\geq 1} x_{n+1}\ 
    \sph_{ ( \T'_i, \ov{c}_{i} c) }(\ov{x}_{i}, x_{n+1}),
  \end{gather*}
  where $c$ is an arbitrary element from the tuple $\ov{c}_{i}$.
  Note that $\delta_i(\ov{x}_{i})$ is a Hanf-formula of signature
  $\sigma$ and equivalent to the formula
  $\sph_{\tau'_i}(\ov{x}_{i})$.

  \newclaims
  \begin{clm}\label{claim:fv-upper-bound-1}
    $(\Delta_1,\ldots,\Delta_s,\beta)$ is a disjoint decomposition for $\psi(\ov{x})$ on the class of all $\sigma$-structures.
  \end{clm}
 
  \uproof{Claim~\ref{claim:fv-upper-bound-1}}
  Let $\A=\A_1\oplus\cdots\oplus\A_s$ be the disjoint sum of $\sigma$-structures $\A_1,\ldots,\A_s$, let $\pi$ be the mapping of $\A$, and let $\ov{a}=(a_1,\ldots,a_n)\in A^n$. 
  For every $i\in[1,s]$, let $\ov{a}_i$ be the subsequence of $\ov{a}$ induced by all $a_j \in \ov{a} \cap P_i^{\A}$. By definition of a disjoint decomposition,
we have to show that 
  \begin{gather*}
    (\A,\ov{a})\quad \models\quad \exists^{\geq k} x_{n+1}\
    \sph_{(\T,\ov{c})}(\ov{x},x_{n+1}) \qquad \ (= \psi(\ov{x})) 
  \end{gather*}
  iff $(\A,\ov{a})$ is a model of $(\Delta_1,\ldots,\Delta_s,\beta)$. 
  That is,
  iff for all \mbox{$i \in [1,s]$} with $ \ov{c} \cap P_i^{\T}
  \not=\emptyset$, the conditions (a) and (b) hold, where \smallskip
  \begin{enumerate}
  \item[(a)]
    Each free variable of $\delta_i$ is interpreted by an element from
    $P_i^{\T}$, i.e., by definition of $\delta_i$, 
    for each $j\in [1,n]$, $N_r^{\A}(a_j)\subseteq P_i^{\A}$ iff $c_j \in P_i^{\T}$. \smallskip
  \item[(b)] $\A_i \models \delta_i[\pi(\ov{a}_i)]$.\smallskip
  \end{enumerate}
  
  Suppose that $(\A, \ov{a})$ satisfies
  $\psi(\ov{x})$ and let $i \in [1,s]$ such that $\ov{c}\cap P_i^{\T}
  \not=\emptyset$. Thus, for every $j\in[1,n]$, $\N_r^{\A}(a_j) \cong
  \N_r^{\T}(c_j)$. Since $\T$ is an induced substructure of a disjoint
  sum, this implies Condition~(a). 
  Furthermore, there are at least $ k$ elements $a\in A$ such that
  $\N_r^{\A}(a) \cong \N_r^{\T}(c_{n+1})$. Since $c_{n+1}\in P_s^{\T}$
  it follows that $N_r^{\A}(a)$ is a subset of $P_s^{\A}$. 

  Recall that $\T'_i$ is the $\sigma$-reduct of $\T_i$ and that
  $\N_r^{\A_{|\sigma}}(\ov{a}_i) $ is isomorphic to $
  \N_r^{\A_i}(\pi(\ov{a}_i))$.  
  Hence, also Condition~(b) holds as $\A_i$ satisfies $\delta_i[\pi(\ov{a}_i)]$ by
  definition of the corresponding formula $\delta_i(\ov{x}_i)$. 
  \smallskip

  For the other direction, suppose that the conditions (a) and (b) hold for all
  $i\in[1,s]$ with $\ov{c}\cap P_i^{\T}\not=\emptyset$. 
  Then, for each $i \in [1,s{-}1]$, Condition~(b) and the definition of
  $\delta_i$ imply that $\N_r^{\A_i}(\pi(\ov{a}_i))\cong \tau'_i$.  
  Furthermore, by Condition~(b) and the definition of $\delta_s$, there are at
  least $k$ elements $a\in A_s$ such that $\N_r^{\A_s}(\pi(\ov{a}_s
  a)) \cong \tau'_s$. 
  
  Recall that $\A$ is the disjoint sum of
  $\A_1,\ldots,\A_s$. Therefore, the $r$-neighbourhoods of all tuples~$\ov{a}_i$ and $\ov{a}_j$, for distinct \mbox{$i,j\in[1,s]$}, are
  disjoint and there are no edges in the Gaifman graph of~$\A$ between
  them.  
  Hence, for each tuple $\ov{a}$ and each element $a$ as chosen above,
  $\A \models \sph_{\tau}[\ov{a},a]$. Consequently, $(\A, \ov{a})$
  satisfies $\psi(\ov{x})$. 
  This completes the proof of Claim~\ref{claim:fv-upper-bound-1}. \qed

  \begin{clm}\label{claim:fv-upper-bound-2}
    The disjoint decomposition $(\Delta_1,\ldots,\Delta_s,\beta)$ can
    be computed in time $\bigO(s + \size{\psi})$. 
  \end{clm}

  \uproof{Claim~\ref{claim:fv-upper-bound-2}}
  Let \mbox{$\psi(\ov{x})\isdef \exists^{\geq
      k} x_{n+1}\; \sph_{\tau}(\ov{x},x_{n+1})$} be a Hanf-formula of signature
  $\sigma_s$, where \mbox{$k\geq 1$}, $n, r
  \geq 0$, and $\tau=(\T,\ov{c})$ is an $r$-sphere
   with~$n{+}1$ centres. On input of $\psi(\ov{x})$
   our algorithm proceeds as follows: \smallskip
  \begin{enumerate}
  \item[(1)] Decide if $\T$ is an induced substructure of an
    $s$-disjoint sum. If yes, continue to Step (2). Otherwise, return
    an unsatisfiable $s$-disjoint decomposition. \smallskip 
  \item[(2)] Construct the propositional formula $\beta$.\smallskip
  \item[(3)] For each $i\in[1,s]$, construct $\delta_i(\ov{x}_i)$.\smallskip
  \end{enumerate}\enlargethispage{\baselineskip}

  For Step (1), note that the structure $\T$ can easily be
  reconstructed from the formula $\sph_{\tau}$ in
  time~$\bigO(\size{\sph_{\tau}})$. The algorithm performs two passes
  over the structure $\T$.  In the first pass, it remembers
  occurrences of the elements of the universe of $\T$ in the sets
  $P_1^{\T},\ldots,P_s^{\T}$. This way it verifies that every element
  of $\T$ is in exactly one of these sets.  If this is the case, the
  algorithm continues with a pass over the other relations of $\T$ and
  checks that no two elements from different sets
  $P_1^{\T},\ldots,P_s^{\T}$ occur in a tuple of these relations. Both
  passes take time
  $\bigO(\size{\T}) \subseteq \bigO(\size{\sph_{\tau}})$.

  Step (2) takes time $\bigO(s)$, since the mapping of the
  constants~$\ov{c}$ to the sets $P_1^{\T}, \ldots, P_s^{\T}$ is
  already known as a byproduct of Step~(1). 

  In Step (3), the mapping of the elements  of $\T$ to the sets
  $P_1^{\T}$, \ldots, $P_s^{\T}$, gathered in Step (1), helps with
  constructing the formulas $\sph_{\tau'_i}$, for each $i\in[1,s]$,
  within a single pass over the formula~$\sph_{\tau}$. Therefore, the
  construction of all the formulas $\delta_i(\ov{x}_i)$ takes time
  $\bigO(\size{\psi})$. 

Altogether, the algorithm takes time $\bigO(s+\size{\psi})$. 
  This completes the proof of Claim~\ref{claim:fv-upper-bound-2} and
  Lemma~\ref{lem:fv-upper-bound-hnf}. 
\qed
\medskip

By using the algorithm for the construction of Hanf normal forms by Bollig and Kuske (Theorem~\ref{thm:hnf-upper-bound}) and
Lemma~\ref{lem:fv-upper-bound-hnf}, we can now prove
Theorem~\ref{thm:fv-upper-bound}. 
\smallskip
\proof[Proof of Theorem~\ref{thm:fv-upper-bound}.]
  Given an input $\FO(\sigma_s)$\hbox{-}formula $\phi(\ov{x})$ of
  quantifier rank $q\geq 0$ and with $n\geq 0$ free variables, our algorithm proceeds as follows:\smallskip
  \begin{enumerate}
  \item[(1)] 
    We employ Theorem~\ref{thm:hnf-upper-bound} to transform
    $\phi(\ov{x})$ into a formula
    $\phi^H(\ov{x})$ in Hanf normal form that is equivalent to $\phi(\ov{x})$ on the class of all $\nu$-bounded $\sigma_s$-structures. 
    This takes time $\Hanf{\size{\phi}}{4^q}{\size{\sigma} + s}$. Furthermore, the
    \mbox{Hanf-formulas} occurring in this formula are of the form
    $\exists^{\geq k} x_{n+1}\; \sph_{\tau}(\ov{x}, x_{n+1})$, where
    $k$ is at most $\size{\phi}\cdot (q{+}1)\cdot \nu(4^q)$, and $\tau$ is a
    $\nu$-bounded $r$-sphere of radius $r\leq 4^q$ with at most
    $n{+}1$ centres. Since $q{+}2 \leq \size{\phi}$, we have that
    \begin{gather*}
      k{+}1\quad \leq\quad \size{\phi}\cdot(q{+}1)\cdot\nu(4^q)+1 \quad \leq\quad \size{\phi}^2 \cdot \nu(4^q).
    \end{gather*}
  \item[(2)]
    For a suitable $L\geq 1$, let
     $\psi_1(\ov{x}), \ldots, \psi_L(\ov{x})$ be the
     Hanf-formulas occurring in $\phi^H(\ov{x})$.  
     For each $\ell\in[1,L]$, we use the construction described in the proof of Lemma~\ref{lem:fv-upper-bound-hnf} to
     compute a disjoint decomposition
     $(\Delta_{1,\ell},\ldots,\Delta_{s,\ell}, \beta_{\ell})$ for the Hanf-formula
     $\psi_{\ell}(\ov{x})$. Note
     that, for each $\ell\in[1,L]$ and each $i\in [1,s]$,
     $\Delta_{i,\ell}$ is either empty or consists of just a single
     Hanf-formula~$\delta_{i,\ell}$. 

    For each Hanf-formula $\exists^{\geq k}
    x_{n+1}\,\sph_{\tau}(\ov{x},x_{n+1})$ occurring in $\phi^H(\ov{x})$, this
    takes time 
    \begin{gather*} 
      \bigO(s + k^2 + \size{\sph_{\tau}})
      \quad
      \subseteq
      \quad
      (\size{\phi} \cdot \nu(4^q))^{\bigO(\size{\sigma}{+}s)}.
    \end{gather*}
    Since $L < \size{\phi^H}$ and $\size{\phi^H} \in \Hanf{\size{\phi}}{4^q}{\size{\sigma}{+}s}$, Step (2) takes  time  
    \begin{gather*}
      L \cdot (\size{\phi}\cdot\nu(4^q))^{\bigO(\size{\sigma} + s)}
      \quad \subseteq\quad  \Hanf{\size{\phi}}{4^q}{\size{\sigma} + s}.
    \end{gather*}
  \item[(3)]
    We output the disjoint decomposition
    $D\isdef(\Delta_1,\ldots,\Delta_s,\beta)$, where 
    $ \Delta_i \isdef \Delta_{i,1} \cup \cdots \cup \Delta_{i,L}$ for each $i\in[1,s]$, and
    the propositional formula $\beta$ is obtained from
    $\phi^H(\ov{x})$ by replacing each Hanf-formula
    $\psi_{\ell}(\ov{x})$, for every $\ell\in [1,L]$, by the
    propositional formula $\beta_{\ell}$.
    Again, this takes time $\Hanf{\size{\phi}}{4^q}{\size{\sigma} + s}$.\smallskip
  \end{enumerate}
 A ltogether, the running time of the algorithm is bounded by $\Hanf{\size{\phi}}{4^q}{\size{\sigma} + s}$.
  It is straightforward to verify that $D$ is indeed a disjoint
  decomposition for $\phi(\ov{x})$ on $\mathfrak{C}_\nu$, that is, for all structures $\A_1,\ldots,\A_s\in \mathfrak{C}_{\nu}$, their disjoint sum $\A=\A_1\oplus\cdots\oplus\A_s$ with mapping $\pi$, and all $\ov{a}\in A^n$, we have
  $(\A,\ov{a})\models \phi(\ov{x})$ iff
  $(\A_1,\ldots,\A_s,\pi(\ov{a}))\models D$.
  This completes the proof of Theorem~\ref{thm:fv-upper-bound}.
\qed

\medskip
\subsection{Generalised decompositions}
\label{subsection:fv-upper-bound-tensorprod}

In this section, we turn our attention from decompositions on disjoint sums of structures to decompositions speaking about more general products of structures. 
We show how to transfer the algorithm of Theorem~\ref{thm:fv-upper-bound} to an algorithm that produces decompositions on composite structures obtained by applying \emph{transductions}\footnote{also knows as \emph{first-order interpretations}, cf.\ \cite{EbbinghausFlum}} to disjoint sums.

Subsection~\ref{sec:transductions} provides the necessary background on transductions. Subsection~\ref{sec:decompositions-and-transductions} presents this section's main result, Corollary~\ref{cor:fv-upper-bound-generalised}, which lifts Theorem~\ref{thm:fv-upper-bound} from disjoint sums to more general decompositions defined via transductions. Subsection~\ref{sec:direct-products} presents an application of the result to the particular case of direct products (also called tensor products or cartesian products).

\subsubsection{Transductions}
\label{sec:transductions}
We consider two fixed finite relational signatures $\sigma$ and $\tau$.

Let $t\geq 1$ and let $\theta(x_1,\ldots, x_t)$ be an
$\FO(\sigma)$-formula with~$t$~free variables. Furthermore, assume that, 
 for each $R\in\tau$ of arity $r \isdef \ar(R)$, $\theta_R(\ov{y}_1\ldots\ov{y}_r)$
is an $\FO(\sigma)$-formula with $r\cdot t$ free variables from tuples
$\ov{y}_i \isdef (y_{i,1},\ldots, y_{i,t})$, for each $i\in [1,r]$. 
Then, the tuple $(\theta, (\theta_R)_{R\in\tau})$ is called a \emph{$t$-transduction from $\tau$ to $\sigma$} (or \emph{transduction}, if the parameters are given by the context).

Assume now that  $\Theta = (\theta, (\theta_R)_{R\in\tau})$ is a $t$-transduction from~$\tau$ to~$\sigma$ as described above.
For every $\sigma$-structure $\A = (A, (R^{\A})_{R\in\sigma})$, the \emph{application $\Theta(\A)$ of the transduction $\Theta$ to $\A$} is a $\tau$\hbox{-}structure \mbox{$\B = (B, (R^{\B})_{R\in\tau})$} whose universe $B$ consists of exactly the tuples $(a_1,\ldots,a_t)\in A^t$ such that $ \A \models \theta[a_1,\ldots,a_t]$, and where each relation symbol $R \in \tau$ with arity $r$ is interpreted by the set of all tuples $(\ov{a}_1,\ldots,\ov{a}_r)$ in $B^r$, such that $\A \models \theta_R[\ov{a}_1\ldots\ov{a}_r]$.

On the other hand, a $t$-transduction $\Theta$ from $\tau$ to $\sigma$ can be
applied to $\FO(\tau)$-formulas.
For every $\FO(\tau)$-formula $\phi(\ov{x})$ with $n\geq 0$ free
variables $\ov{x} = (x_1,\ldots,x_n)$, the \emph{application
  $\Theta(\phi)$ of the transduction $\Theta$ to $\phi(\ov{x})$} is an
$\FO(\sigma)$-formula $\psi(\ov{x}_1\ldots\ov{x}_n)$ with $n{\cdot}t$
new free variables from variable tuples $\ov{x}_i \isdef
(x_{i,1},\ldots,x_{i,t})$,  $i\in[1,n]$, that is defined inductively
as follows:  \smallskip

\begin{itemize}
\item 
If $\phi(\ov{x}) = R(x_1,\ldots, x_r)$, for a relation symbol $R\in\tau$ with arity $r$, then
\begin{gather*}
  \Theta(\phi)(\ov{x}_1\ldots\ov{x}_r)\ \isdef\quad  
  \Und_{i=1}^r \theta(\ov{x}_i)
  \ \ \und\ 
  \theta_R(\ov{x}_1\ldots\ov{x}_r).
\end{gather*}

\item If $\phi$ is of the form $x_1{=}x_2$ then 
\begin{gather*}
  \Theta(\phi)(\ov{x}_1\ov{x}_2)\ \isdef\quad 
  \theta(\ov{x}_1) \ \und \ 
   \Und_{j=1}^t x_{1,j}{=}x_{2,j}.
\end{gather*}

\item For the Boolean connectives the translation distributes,
i.e. \mbox{$\Theta(\neg \phi) \isdef \neg \Theta(\phi)$} and, for each
\mbox{$\star \in \set{ \und,\oder,\impl,\gdw}$}, we have \mbox{$\Theta(\phi \star
  \psi) \isdef \Theta(\phi) \star \Theta(\psi)$}.  \smallskip

\item If $\phi(\ov{x}) = \exists y\; \psi(\ov{x},y)$ then let \mbox{$\ov{y} = (y_{1},\ldots,y_{t})$} be a tuple of new variables and let 
\begin{gather*}
  \Theta(\phi)(\ov{x}_1\ldots\ov{x}_n)
  \ \isdef\quad
  \exists \ov{y}\ \bigl( \theta(\ov{y}) \ \und\ \Theta(\psi)(\ov{x}_1\ldots\ov{x}_n\ov{y})\bigr).
\end{gather*}
\item Finally, for $\phi(\ov{x}) = \forall y \; \psi(\ov{x},y)$, let $\Theta(\phi) \isdef \Theta(\neg \exists y\; \neg \psi)$.
\end{itemize}
Obviously, the application of a fixed transduction to an input
$\FO$\hbox{-}formula can be carried out in time linear in the size of
the input formula. The following
lemma makes this precise and gives a bound on the quantifier rank of
the resulting formula.  

\begin{lem}\label{lem:transduction-complexity}
  Let $t\geq 1$ and let $\Theta = (\theta, (\theta_R)_{R\in\tau})$ be a $t$\hbox{-}transduction from $\tau$ to $\sigma$.
  Let $k$ and $p$ be the maximum size and the maximum quantifier rank of the formulas $\theta$ and $\theta_R$, for each $R\in\tau$.
  For every $\FO(\tau)$-formula $\phi$, the $\FO(\sigma)$-formula
  $\Theta(\phi)$ can be constructed in time
  $\bigO((t{+}k)\cdot\size{\phi})$. Furthermore, if $q$ is the
  quantifier rank of $\phi$ then $\Theta(\phi)$ has quantifier rank at most~$t{\cdot} q + p$.
\end{lem}

\proof
  Let $\phi$ be an $\FO(\tau)$-formula of quantifier rank \mbox{$q\geq 0$}. By induction on the shape of $\phi$, we show that there is a suitable number $c \geq 1$, that has to be greater than some values obtained during the course of the induction, such that $\size{\Theta(\phi)} \leq c\cdot (t{+}k)\cdot \size{\phi}$ and $\qr(\Theta(\phi)) \leq t\cdot q + p$. \smallskip
  
  \begin{itemize}
  \item If $\phi = R(x_1,\ldots, x_r)$, for a relation symbol $R\in\tau$ with arity $r$, then $\size{\Theta(\phi)} \leq c\cdot k \cdot (r{+}1)$. Of course, $r{+}1 \leq \size{\phi}$. Therefore,  $\size{\Theta(\phi)} \leq c\cdot (t{+}k)\cdot  \size{\phi}$.
  Furthermore, the quantifier rank of $\Theta(\phi)$ is at most $p$. \smallskip

  \item For a Boolean combination $\phi$, the induction step is obvious.\smallskip

  \item
    Assume that $\phi = Q y\,\psi(\ov{x},y)$ for a quantifier $Q \in \set{\exists,\forall}$. In this case, we have that $ \size{\Theta(\phi)} 
    \leq c\cdot (t{+}k) + \size{\Theta(\psi)}$. Since $\size{\Theta(\psi)} \leq c\cdot (t{+}k)\cdot (\size{\phi}{-}1)$, it follows that $\size{\Theta(\phi)}$ is at most $c\cdot (t{+}k)\cdot \size{\phi}$. 
    Furthermore, by construction of $\Theta(\phi)$, we have that $\qr(\Theta(\phi))$ is at most $t + \max\set{p, \qr(\Theta(\psi))}$. Because $\qr(\Theta(\psi)) \leq t\cdot (q{-}1) + p$  it follows that
    $\qr(\Theta(\phi)) \leq t\cdot q + p$. \smallskip
  \end{itemize}
  Clearly, the formula $\Theta(\phi)$ can be constructed in time  $\bigO((t{+}k)\cdot\size{\phi})$, that is linear in the size of~$\phi$.
\qed

\medskip

\noindent The following proposition relates the
application of transductions to structures and formulas to each
other. The proof is an immediate consequence of the definition of
$\Theta(\A)$ and $\Theta(\varphi)$ (see, e.g., \cite{EbbinghausFlum}).
\begin{prop}\label{prop:transduction-property}
  Let $t\geq 1$ and let $\Theta = (\theta, (\theta_R)_{R\in\sigma})$ be a $t$-transduction from $\tau$ to $\sigma$. 
  For every $\FO(\tau)$-formula $\phi(\ov{x})$ with $n\geq 0$ free variables $\ov{x}=(x_1,\ldots,x_n)$, each $\sigma$-structure $\A$ and each tuple $(\ov{a}_1,\ldots,\ov{a}_n) \in (\theta(\A))^n$, the following is true:
  \begin{align*}
    &\ \bigl(\Theta(\A), \ov{a}_1, \ldots, \ov{a}_n\bigr) &\ \models &\quad \phi(x_1,\ldots,x_n) \\
    \text{iff} \quad
    &\ \bigl(\A, \ov{a}_1\ldots\ov{a}_n\bigr) &\ \models &\quad \Theta(\phi)(\ov{x}_1\ldots\ov{x}_n),
  \end{align*}
  where, for each $i\in [1,n]$, $\ov{x}_i \isdef (x_{i,1},\ldots,x_{i,t})$.\qed
\end{prop}\medskip

\noindent Note here the difference between the expressions $(\ov{a}_1, \ldots, \ov{a}_n)$ and $\ov{a}_1 \ldots \ov{a}_n$, where $\ov{a}_i$ denotes an arbitrary tuple (for every $i\in[1,n]$): While $(\ov{a}_1,\ldots,\ov{a}_n)$ denotes the tuple of length $n$ whose elements are exactly the tuples $\ov{a}_i$ for $i\in[1,n]$, the expression $\ov{a}_1\ldots\ov{a}_n$ represents the concatenation of these tuples, i.e., a tuple of length $|\ov{a}_1| + \cdots + |\ov{a}_n|$.

\subsubsection{Decompositions obtained by transductions}
\label{sec:decompositions-and-transductions}

This section's main result is a corollary to Theorem~\ref{thm:fv-upper-bound} and reads as follows:
\begin{cor}\label{cor:fv-upper-bound-generalised}
  Let $\sigma$ and $\tau$ be finite relational signatures and let $\mathfrak{D}_{\nu}$ be the class of all $\nu$\hbox{-}bounded $\tau$\hbox{-}structures.
  Let $s,t\geq 1$ and let $\Theta = (\theta, (\theta_R)_{R\in\tau})$ be a $t$-transduction from $\tau$ to $\sigma_s$.
  There is an algorithm which, given an input $\FO(\tau)$-formula $\phi(\ov{x})$ of quantifier rank $q\geq 0$ and with $n\geq 0$ free variables, constructs in time
  \begin{gather*}
    \Hanf{\size{\phi} \cdot (t{+}k)}{4^{t \cdot q + p}}{\size{\sigma} + s}
    \qquad\qquad \Bigl( =\quad 
    2^{(\size{\phi}\cdot (t{+}k) \cdot \nu(4^{t\cdot q + p}))^{\bigO(\size{\sigma}{+}s)}} \ \Bigr)
  \end{gather*}
  a reduction sequence $D=(\Delta_1,\ldots,\Delta_s,\beta)$ over $\ov{x}$, such that for all structures  $\A_1,\ldots,\A_s\in\mathfrak{D}_{\nu}$ and every tuple $(\ov{a}_1,\ldots,\ov{a}_n)$ in  $(\theta(\A_1\oplus\cdots\oplus  \A_s))^n$, the following is true:
  \begin{align*}
    &\ \bigl(\Theta(\A_1\oplus\cdots\oplus\A_s), (\ov{a}_1, \ldots, \ov{a}_n)\bigr) &\ \models &\quad \phi(\ov{x}) \\
    \text{iff} \quad
    &\ \bigl(\A_1, \ldots, \A_s, \pi(\ov{a}_1\ldots\ov{a}_n)\bigr) &\ \models &\quad D,
  \end{align*}
  where $\pi$ is the mapping of the disjoint sum $\A_1\oplus\cdots\oplus\A_s$.

  Here, $k$ and $p$ are the maximum size and maximum quantifier rank of the formulas $\theta$ and $\theta_R$, for every $R\in\sigma$.
  Furthermore, the sets $\Delta_1,\ldots,\Delta_s$ consist of Hanf-formulas over $\sigma$.
\end{cor}

\proof
  Let $\phi(\ov{x})$ be an $\FO(\tau)$-formula of quantifier rank~$q\geq 0$ and with $n \geq 0$ free variables \mbox{$\ov{x} = (x_1,\ldots,x_n)$.}
  The algorithm proceeds in two steps: \smallskip
  \begin{enumerate}
  \item\label{item:FV-UpperBound-Generalised-1} 
    In a first step, the algorithm applies the transduction $\Theta$ to the input formula $\phi(\ov{x})$ to obtain an $\FO(\sigma_s)$\hbox{-}formula $\Theta(\phi)(\ov{x}_1\ldots\ov{x}_n)$ where, for each $i\in[1,n]$, $\ov{x}_i$ is a tuple  $(x_{i,1},\ldots,x_{i,t})$ of free variables. From Proposition~\ref{prop:transduction-property} it follows that for all $\sigma_s$-structures~$\A$ and all $(\ov{a}_1,\ldots,\ov{a}_n)$ in  $(\theta(\A))^n$, we have that
    \begin{align*}
      &\ \bigl(\Theta(\A), (\ov{a}_1, \ldots, \ov{a}_n)\bigr) &\ \models &\quad \phi(x_1,\ldots,x_n) \\
      \text{iff} \quad
      &\ \bigl(\A, \ov{a}_1\ldots\ov{a}_n\bigr) &\ \models &\quad \Theta(\phi)(\ov{x}_1\ldots\ov{x}_n).
    \end{align*}
    By Lemma~\ref{lem:transduction-complexity}, this step requires time \mbox{$\bigO((t{+}k)\cdot\size{\phi})$}, where $k$ is the maximum size of the $\FO(\sigma_s)$\hbox{-}formulas~$\theta$ and $\theta_R$, for every $R\in\tau$. 
 Also,  \mbox{$\size{\Theta(\phi)} \in \bigO((t{+}k)\cdot \size{\phi})$} and
 \mbox{$\qr(\Theta(\phi)) \leq t{\cdot}q + p$}, where $p$ is the maximum
 quantifier rank of the $\FO(\sigma_s)$-formulas~$\theta$ and $\theta_R$, for every $R\in\tau$.\smallskip
  \item\label{item:FV-UpperBound-Generalised-2}
    In the second step, the algorithm constructs a disjoint decomposition
    $D=(\Delta_1,\ldots,\Delta_s,\beta)$ for $\Theta(\phi)(\ov{x}_1\ldots\ov{x}_n)$ on
    $\mathfrak{D}_{\nu}$. I.e., for all
    $\A_1,\ldots,\A_s\in\mathfrak{D}_{\nu}$ and the disjoint sum $\A$
    of $\A_1,\ldots,\A_s$ with its mapping $\pi$ and for each tuple $\ov{a} \in
    A^{n\cdot t}$, we have
    \begin{gather*}
      (\A,\ov{a})\quad \models\quad \Theta(\phi)(\ov{x}_1 \ldots \ov{x}_n) 
      \qquad \text{iff} \qquad 
      (\A_1,\ldots,\A_s,\pi(\ov{a}))\quad \models\quad  D.
    \end{gather*} 
    By Theorem~\ref{thm:fv-upper-bound}, the construction of $D$ from $\Theta(\phi)$ takes time
    \begin{gather*}
      \Hanf{\size{\Theta(\phi)}}{4^{\qr(\Theta(\phi))}}{\size{\sigma} + s}
      \quad \subseteq \quad
      \Hanf{\size{\phi} \cdot (t{+}k)}{4^{t \cdot q + p}}{\size{\sigma} + s}
    \end{gather*}
  \end{enumerate}
  It follows from Step (1) and Step (2) that, for every disjoint sum $\A$ of structures $\A_1,\ldots,\A_s\in\mathfrak{D}_{\nu}$ and all $(\ov{a}_1,\ldots,\ov{a}_n)$ in $ (\theta(\A))^n$, it holds that
  \begin{align*}
    &\ (\Theta(\A),(\ov{a}_1,\ldots,\ov{a}_n)) &\ \models &\quad \phi(\ov{x}) \\
    \text{iff} \quad &\ 
    (\A_1,\ldots,\A_s,\pi(\ov{a}_1\ldots \ov{a}_n)) &\ \models &\quad D
  \end{align*} 
  and that construction of $D$ from $\phi(\ov{x})$ altogether takes time
  \begin{gather*}
    \Hanf{\size{\phi} \cdot (t{+}k)}{4^{t \cdot q{+}p}}{\size{\sigma} + s}.
  \end{gather*}
  This concludes the proof of Corollary~\ref{cor:fv-upper-bound-generalised}.
\qed

\subsubsection{Decompositions on direct products}
\label{sec:direct-products}

We exemplify the application of Corollary~\ref{cor:fv-upper-bound-generalised} with the following result on \emph{direct products} of structures. 
Let $s\geq 1$. For $\sigma$\hbox{-}structures $\A_1, \ldots, \A_s$, the
direct product $\A_1 \otimes  \cdots \otimes \A_s$ is the
$\sigma$-structure $\A = (A, (R^{\A})_{R\in \sigma})$, where the
universe $A$ is the set $A_1\times\cdots\times A_s$, 
and for each relation symbol \mbox{$R\in\sigma$} of arity $r\geq 1$, the relation $R^{\A}$~is the set of all tuples $( (a_{1,1},\ldots,
a_{1,s}), \ldots,(a_{n,1},\ldots, a_{r,s}))$ in $A^r$ such that, for
each $i\in[1,s]$, the tuple $(a_{1,i},\ldots,a_{r,i})$ belongs to the relation $R^{\A_i}$.  
\begin{cor}\label{cor:fv-upper-bound-tensorprod}
  Let $s\geq 1$.  There is an algorithm which, given an input
  $\FO(\sigma)$-formula $\phi(\ov{x})$ of quantifier rank $q\geq 0$
  and with $n\geq 0$ free variables, constructs in time 
  \begin{gather*}
    \Hanf{s \cdot \size{\phi}}{4^{s \cdot q}}{\size{\sigma} + s}
    \qquad\qquad \Bigl(= \quad
    2^{(s \cdot \size{\phi} \cdot \nu(4^{s\cdot q}))^{\bigO(\size{\sigma}{+}s)}} \Bigr)
  \end{gather*}
  a reduction sequence $D = (\Delta_1,\ldots,\Delta_s,\beta)$ over
  $\ov{x}$, such that for all  $\A_1,\ldots,\A_s \in
  \mathfrak{C}_{\nu}$  and all $(\ov{a}_1,\ldots,\ov{a}_n)$ in \mbox{$
    (A_1 \times \cdots \times A_s)^n$}, 
  \[
    \begin{array}{llcl}
     & \bigl(\A_1 \otimes \cdots \otimes \A_s, (\ov{a}_1, \ldots,
    \ov{a}_n)\bigr) & \models &\quad \phi(\ov{x}) 
    \smallskip\\ 
    \text{iff} \qquad
    & \bigl(\A_1, \ldots, \A_s, \ov{a}_1\ldots\ov{a}_n\bigr) & \models &\quad D.
    \end{array}
  \]
  Furthermore, the sets $\Delta_1,\ldots,\Delta_s$  consist of
  Hanf-formulas over $\sigma$.
\end{cor}
\proof
Consider the following $s$-transduction $\Theta \isdef (\theta, (\theta_R)_{R\in\sigma})$ from~$\sigma$ to~$\sigma_s$ with
\begin{gather*}
  \theta(x_1,\ldots, x_s)\ \isdef\quad \Und_{i=1}^s P_i(x_i)
\end{gather*}
and, for each $R\in\sigma$ of arity $r\isdef \ar(R)$,
\begin{gather*}
  \theta_R(\ov{x}_1\ldots\ov{x}_r)\ \isdef\quad
  \Und_{i=1}^s R(x_{1,i},\ldots,x_{r,i}),
\end{gather*}
where $\ov{x}_j = (x_{j,1},\ldots,x_{j,s})$ for each $j\in [1,r]$.
The formulas of $\Theta$ correspond to the definition of the (tensor) product of $\sigma$-structures. I.e., for all $\sigma$-structures $\A_1,\ldots,\A_s$ and the disjoint sum $\A$ of $\A_1,\ldots,\A_s$ with mapping $\pi$, 
\begin{gather*}
  \Theta(\A) \quad \cong \quad \A_1 \otimes \cdots \otimes \A_s
\end{gather*}
via the isomorphism $f \colon \theta(\A) \to (A_1 \times \cdots \times A_s)$ defined by $f(\ov{a}) \isdef \pi(\ov{a})$, for all $\ov{a}\in\theta(\A)$.
 
Therefore, by Proposition~\ref{prop:transduction-property},  for every $\FO(\sigma)$-formula $\phi(\ov{x})$ with a tuple $\ov{x}$ of  $n\geq 0$ free variables $(x_1,\ldots,x_n)$, the following holds:
Let $\A_1,\ldots,\A_s$ be $\sigma$-structures and let $\A$ be the disjoint sum of $\A_1,\ldots,\A_s$ with mapping $\pi$. Let $(\ov{a}_1, \ldots, \ov{a}_n)\in (\theta(\A))^n$. Then,\smallskip
\begin{align*}
  &\ \bigl(\A_1 \otimes \cdots \otimes \A_s, (\pi(\ov{a}_1), \ldots, \pi(\ov{a}_n))\bigr)&\!\!\!\!\!\models &\ \phi(\ov{x}) \\ 
  \text{iff} \,  
  &\ \bigl(\Theta(\A), (\ov{a}_1,\ldots,\ov{a}_n)\bigr)&\!\!\!\!\!\models &\ \phi(\ov{x}) \\ 
  \text{iff} \, 
  &\ \bigl(\A, \ov{a}_1\ldots\ov{a}_n\bigr)&\!\!\!\!\!\models &\ \Theta(\phi)(\ov{x}_1\ldots\ov{x}_n),
\end{align*}
where $\ov{x}_i \isdef (x_{i,1},\ldots,x_{i,s})$ for each $i\in[1,n]$.

The proof concludes with an application of Corollary~\ref{cor:fv-upper-bound-generalised} to this transduction.
Note that the formulas $\theta$ and $\theta_{R}$, for each $R\in\sigma$, are quantifier-free. Hence, by Corollary~\ref{cor:fv-upper-bound-generalised}, it takes time
\begin{gather*}
  \Hanf{s \cdot \size{\phi}}{4^{s \cdot q}}{\size{\sigma} + s}
\end{gather*}
to compute a reduction sequence $ D = (\Delta_1,\ldots,\Delta_s, \beta)$ over $\ov{x}$, such that for each disjoint sum $\A$ of structures \mbox{$\A_1,\ldots,\A_s\in\mathfrak{C}_{\nu}$} and all $(\ov{a}_1,\ldots,\ov{a}_n) \in ( \theta(\A))^n$,
  \begin{align*}
    &\ \bigl(\Theta(\A), (\ov{a}_1, \ldots, \ov{a}_n)\bigr) &\ \models &\quad \phi(\ov{x}) \\ 
    \text{iff} \quad
    &\ \bigl(\A_1, \ldots, \A_s, \pi(\ov{a}_1\ldots\ov{a}_n)\bigr) &\ \models &\quad D,
  \end{align*}
  where $\pi$ is the mapping of the disjoint sum $\A$.
  Observe that a tuple $\ov{a}$ is in $\theta(\A_1\oplus\cdots\oplus\A_s)$ iff $\pi(\ov{a})$ is in $A_1 \times \cdots \times A_s$. Hence, for all $\A_1,\ldots,\A_s\in \mathfrak{C}_{\nu}$ and all $(\ov{a}_1,\ldots,\ov{a}_n)\in (A_1 \times \cdots \times A_s)^n$, 
  \begin{align*}
    &\ \bigl(\A_1\otimes\cdots\otimes\A_s, (\ov{a}_1, \ldots, \ov{a}_n)\bigr) &\ \models &\quad \phi(\ov{x}) \\ 
    \text{iff} \quad
    &\ \bigl(\A_1, \ldots, \A_s, \ov{a}_1\ldots\ov{a}_n\bigr) &\ \models &\quad D.
  \end{align*} 
  This concludes the proof of Corollary~\ref{cor:fv-upper-bound-tensorprod}.
\qed

Note that Corollary~\ref{cor:fv-upper-bound-tensorprod} also holds when the direct product of structures is replaced by other graph theoretical products, e.g., the \emph{cartesian product} and the \emph{strong product} of graphs (cf.\ \cite{HN2004}).

\section{Lower Bounds.}\label{section:lower-bounds}
In this section, we prove lower bounds corresponding to the upper bounds of the main results
of sections \ref{section:PreservationTheorems} and \ref{fv} concerning preservation theorems and Feferman-Vaught decompositions.
A key ingredient to all our lower bounds is an encoding of large numbers by bounded degree trees that allows to compare numbers by using small $\FO$-formulas
In Subsection~\ref{section:binary-tree-encodings}, we introduce this encoding. 
The remaining subsections~\ref{subsection:lower-bounds-for-preservation-theorems} and~\ref{subsection:lower-bounds-for-feferman-vaught-decompositions} contain our lower bounds for preservation theorems and Feferman-Vaught decompositions, respectively.

\subsection{Binary tree encodings}
\label{section:binary-tree-encodings}
We recall an encoding of numbers by \emph{binary trees}, i.e., trees where every node has at most two children, which was already used in \cite{HKS13-LICS}. This binary tree encoding is an adaptation of an encoding of numbers by trees of unbounded degree from Chapter~10 in~\cite{FlGr06}.

Consider the signature $\set{E}$ that consists of just a single binary relation symbol $E$. A \emph{forest} is a disjoint union of finite directed rooted trees.
The \emph{height} of a forest $\F$ is the length of a longest directed path in $\F$. For each node $a$ of a forest $\F$, $\F_a$ is the subtree of $\F$ induced by all nodes reachable by a directed path from~$a$. For a tree $\T$ and a number $d\geq 0$, we write $\T[\leq d]$ to denote the subtree of $\T$ induced by all nodes of $\T$ that are reachable from the root of $\T$ by a directed path of length at most $d$. A tree $\B$ is a \emph{complete binary tree} if all leaves of $\B$ have the same height and every non-leaf node has exactly two children. 

For numbers $i, n\in\NN$, we write $\bit(i,n)$ to denote the $i$-th bit in the binary representation of $n$. I.\,e., $\bit(i, n) = 1$ iff $\lfloor \tfrac{n}{2^i} \rfloor$ is odd.

We define the (non-elementary) function 
$\Tower\colon \NN\to\NN$ by $\Tower(0) \isdef 1$ and
\begin{gather*}
  \Tower(h) \ \isdef \quad 2^{\Tower(h{-}1)}
  \quad \text{for all $h \geq 1$.}
\end{gather*}
I.\,e.,
 $\Tower(h)$ is a tower of $2s$ of height $h$.

\begin{figure}
  \begin{center}
    \begin{tikzpicture}
      [
      every node/.style={circle,draw,semithick, inner sep=0.07cm, minimum size=0.3cm},
      edge from parent/.style={draw,-latex,semithick},
      level distance=1cm,
      level 1/.style={sibling distance=5.04cm},   
      level 2/.style={sibling distance=2.52cm},   
      level 3/.style={sibling distance=1.26cm},   
      level 4/.style={sibling distance=2.016cm},  
      level 5/.style={sibling distance=1.008cm},  
      level 6/.style={sibling distance=0.882cm},  
      level 7/.style={sibling distance=0.882cm},  
      level 8/.style={sibling distance=0.756cm}   
      ]
      \node {$42$}
      child {
        node {}
        child {
          node {}
          child {
            node {}
          }
          child {
            node {}
            child {
              node {$5$}
              child {
                node {}
                child {
                  node {$0$}
                }
                child {
                  node {$2$}
                  child {
                    node {$1$}
                    child {
                      node {$0$}
                    }
                  }
                }
              }
              child {
                node {}
              }
            }
          }
        }
        child {
          node {}
          child {
            node {}
          }
          child {
            node {}
          }
        }
      }
      child {
        node {}
        child {
          node {}
          child {
            node {}
            child {
              node {$1$}
              child {
                node {}
              }
              child {
                node {}
                child {
                  node {$0$}
                }
              }
            }
            child {
              node {$3$}
              child {
                node {}
                child {
                  node {$1$}
                  child {
                    node {$0$}
                  }
                }
              }
              child {
                node {}
                child {
                  node {$0$}
                }
              }
            }
          }
          child {
            node {}
          }
        }
        child {
          node {}
          child {
            node {}
          }
          child {
            node {
            }
          }
        }
      };   
    \end{tikzpicture}
  \end{center}
  \caption{A binary tree from the set $\mathfrak{B}_1(42)$, i.e., a binary tree
    encoding of the number $42$ with parameter $1$. The numbers depicted
    within some of the nodes are \emph{not} part of the tree encoding;
    they are just indicated here to illustrate which number is encoded
    by the subtree starting at the respective node.}
  \label{fig:bintree42}
\end{figure}
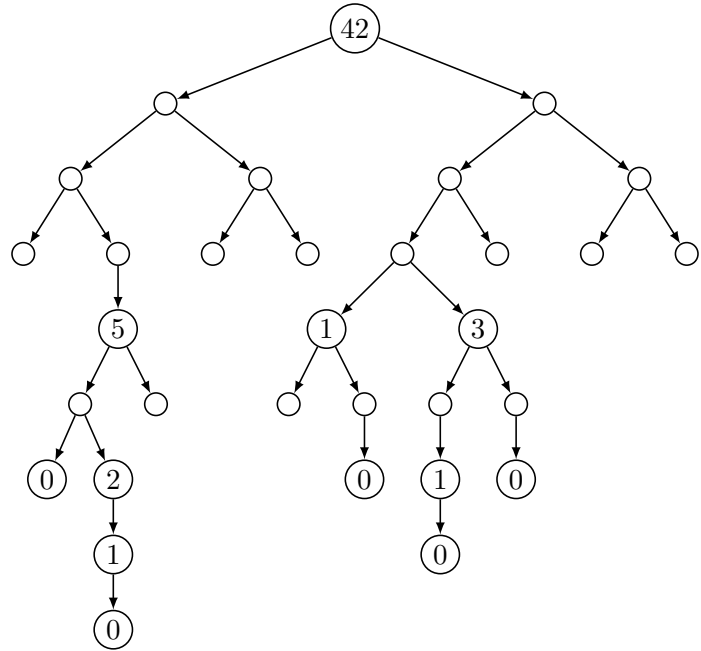

\smallskip
For each $h \geq -1$ and $i \in [0,\Tower(h{+}3){-}1]$, we define inductively a set $\mathfrak{B}_h(i)$ of binary trees that (each) encode the (binary expansion of the) number $i$.\medskip
\begin{description}
\item[$h=-1$] For $i \in [0,\Tower(2){-}1] = \set{0,1,2, 3}$, the set $\mathfrak{B}_{-1}(i)$ contains exactly the binary trees that are isomorphic to the binary tree $\B_{-1}(i)$ depicted in Figure~\ref{fig:bintreesStart}.
\begin{figure}[!ht]
  \centering
  \subfigure{
    \begin{tikzpicture}
      [
      every node/.style={circle,draw,semithick, inner sep=0.07cm, minimum size=0.3cm},
      edge from parent/.style={draw,-latex,semithick},
      level distance=0.84cm,
      level 1/.style={sibling distance=3.36cm},
      level 2/.style={sibling distance=1.68cm},
      level 3/.style={sibling distance=0.84cm},
      level 4/.style={sibling distance=1.344cm},
      level 5/.style={sibling distance=0.672cm},
      level 6/.style={sibling distance=0.588cm},
      level 7/.style={sibling distance=0.588cm},
      level 8/.style={sibling distance=0.504cm}
      ]
      \path[use as bounding box] (-1.7,1.3) rectangle (1.7,-2.3);
      \node[draw=none] at (0, 1) { $  \B_{-1}(0) $};
      \node { $ 0 $};
    \end{tikzpicture}
  }
  \subfigure{
    \begin{tikzpicture}
      [
      every node/.style={circle,draw,semithick, inner sep=0.07cm, minimum size=0.3cm},
      edge from parent/.style={draw,-latex,semithick},
      level distance=1cm,
      level 1/.style={sibling distance=3.36cm},
      level 2/.style={sibling distance=1.68cm},
      level 3/.style={sibling distance=0.84cm},
      ]
      \path[use as bounding box] (-1.7,1.3) rectangle (1.7,-2.3);
      \node[draw=none] at (0, 1) { $  \B_{-1}(1) $};
      \node { $ 1 $ }
      child {
        node { $ 0 $ }
      };
    \end{tikzpicture}
  }
  \subfigure{
    \begin{tikzpicture}
      [
      every node/.style={circle,draw,semithick, inner sep=0.07cm, minimum size=0.3cm},
      edge from parent/.style={draw,-latex,semithick},
      level distance=1cm,
      level 1/.style={sibling distance=3.36cm},
      level 2/.style={sibling distance=1.68cm},
      level 3/.style={sibling distance=0.84cm},
      ]
      \path[use as bounding box] (-1.7,1.3) rectangle (1.7,-2.3);
      \node[draw=none] at (0, 1) { $  \B_{-1}(2) $};
      \node { $ 2 $ }
      child {
        node { $ 1 $ }
        child {
          node { $ 0 $ }
        }
      };
    \end{tikzpicture}
  }
  \subfigure{
    \begin{tikzpicture}
      [
      every node/.style={circle,draw,semithick, inner sep=0.07cm, minimum size=0.3cm},
      edge from parent/.style={draw,-latex,semithick},
      level distance=1cm,
      level 1/.style={sibling distance=1cm},
      level 2/.style={sibling distance=1cm},
      ]
      \path[use as bounding box] (-1.7,1.3) rectangle (1.7,-2.3);
      \node[draw=none] at (0, 1) { $ \B_{-1}(3) $};
      \node { $ 3 $ }
      child {
        node { $ 1 $ }
        child {
          node { $ 0 $ }
        }
      }
      child {
        node { $ 0 $ }
      }
      ;
    \end{tikzpicture}
  }
  \caption{
    The binary trees $\B_{-1}(0)$, $\B_{-1}(1)$, $\B_{-1}(2)$, and $\B_{-1}(3)$ are tree encodings for the numbers $0$, $1$, $2$, and $3$, respectively, as defined in~\cite{FlGr06}. Note that the numbers depicted within the nodes are \emph{not} part of the tree encoding; they are just indicated here to illustrate which number is encoded by the subtree starting at the respective node.}
  \label{fig:bintreesStart}
\end{figure}
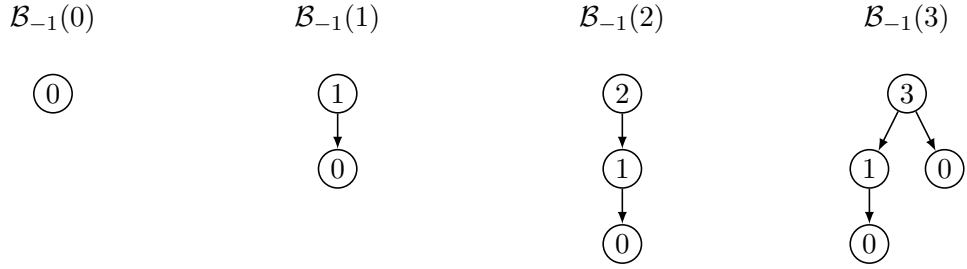
  
\medskip
\item[$h\geq0$] For $i \in [0,\Tower(h{+}3){-}1]$, the set $\mathfrak{B}_h(i)$ consists of all binary trees $\B$ that satisfy each of the following properties:\medskip
  \begin{itemize}
  \item $\B[\leq \Tower(h{+}1){-}1]$ is a complete binary tree of height $\Tower(h{+}1){-}1$.\medskip
  \item For every  $j \in [0, \Tower(h{+}2){-}1]$ with $\bit(j, i)=1$, there is a node $b$ of height $\Tower(h{+}1)$ in $\B$ such that $\B_b \in \mathfrak{B}_{h-1}(j)$. \medskip
  \item For every node $b$ of height $\Tower(h{+}1)$ in $\B$, there is a $j \in [0,\Tower(h{+}2){-}1]$ such that $\B_b \in \mathfrak{B}_{h-1}(j)$ and $\bit(j,i)=1$. \medskip
  \end{itemize}
\end{description}
Each tree in $\mathfrak{B}_h(i)$ is called a \emph{binary tree encoding of $i$ with parameter $h$}. An example of a binary tree encoding of $i = 42$ with parameter $h=1$ is depicted in Figure~\ref{fig:bintree42}.

\smallskip
An induction on the parameter $h$ shows that every number $i \in [0,\Tower(h{+}3){-}1]$ has at least one binary tree encoding with parameter $h$. 
The following easy observation gives an upper bound on the height of binary tree encodings.
\begin{lemma}
  \label{lem:height-Bhi}
  For each $h > 1$ and every $i \in [0, \Tower(h{+}3)-1]$, each binary tree in $\mathfrak{B}_h(i)$ has height less than $2 \cdot \Tower(h{+}1)$. 
\end{lemma}
\proof
  Let $h > 1$ and $i \in [0, \Tower(h{+}3)-1]$. By definition of $\mathfrak{B}_h(i)$ (see also Figure~\ref{fig:bintree42}) it is easy to convince oneself that each binary tree in $\mathfrak{B}_h(i)$ has height less than 
  \begin{gather*}
    2\ +\ \sum_{k=0}^h \Tower(k{+}1) 
    \quad < \quad 2 \cdot \Tower(h{+}1).
  \end{gather*}
  The latter inequality can be shown by a straightforward induction.
\qed\medskip

\noindent An adaptation of Lemma~3.2 (see also Chapter 10 in \cite{FlGr06}), Lemma~3.3, and Lemma~3.4 of \cite{DGKS07}  allows to express arithmetic relations between binary tree encodings of numbers by ``small'' $\FO(E)$-formulas:
\newcommand{\less}{\textit{less}}
\newcommand{\suc}{\textit{succ}}
\newcommand{\enc}{\textit{enc}}
\newcommand{\eq}{\textit{eq}}
\newcommand{\LOBtrees}{\ensuremath{\mathfrak{B}^{<}}}
\newcommand{\LOBforests}{\ensuremath{\mathfrak{BF}^{<}}}
\newcommand{\GBE}{\ensuremath{\mathfrak{G}}}

\begin{lemma}\label{lem:arithmetic-formulas}
  For $h \geq -1$, there are $\FO(E)$-formulas $\enc_h(x)$, $\fmin_h(x)$, $\eq_h(x,y)$, $\less_h(x,y)$, $\suc_h(x,y)$ and $\fmax_h(x)$ of size $\bigO(\Tower(h))$ (for $h\geq 0$) such that for each  binary forest $\F$ and all nodes $a,b\in F$, 
  \begin{align*}
    \F\ \models &\ \enc_h[a] & & \text{iff} \quad \F_a \in
    \mathfrak{B}_h(i)\ \\ 
    & & & \text{for an $i\in [0,\Tower(h{+}3){-}1]$}, \\
    \intertext{and if there are $i,j \in [0, \Tower(h{+}3){-}1]$ such
      that $\F_a\in \mathfrak{B}_h(i)$  
     and $\F_b\in \mathfrak{B}_h(j)$, 
     then} 
    \F\ \models &\ \fmin_h[a] & & \text{iff} \quad i=0,  \\
    \F\ \models &\ \eq_h[a,b] & & \text{iff} \quad i=j,  \\
    \F\ \models&\ \less_h[a,b] & & \text{iff} \quad i < j, \\
    \F\ \models&\ \suc_h[a,b] & & \text{iff} \quad i{+}1 = j, \quad\text{and} \\
    \F\ \models&\ \fmax_h[a] & & \text{iff} \quad i = \Tower(h{+}3)-1.
  \end{align*}
\end{lemma}
\proof
  For each $d\geq 0$, there is an $\FO(E)$-formula $\delta_{\leq d}(x,y)$ of size $\bigO(\log d)$ expressing in a binary forest that there is a directed path of length at most $d$ from node $x$ to node $y$. For $d=0$ and~$d=1$, the formulas $\delta_{\leq 0}(x,y)$ and $\delta_{\leq 1}(x,y)$ can be chosen as $x{=}y$ and $x{=}y \oder E(x,y)$, respectively. For~$d\geq 1$, define
\begin{gather*}
  \begin{split}
  \delta_{\leq 2d}(x, y) \ \isdef\quad &
  \exists z 
  \forall x'
  \forall y'\; \Bigl( \;
  \bigl(\;(x'{=}x \und y'{=}z)\ \oder\ (x'{=}z \und y'{=}y)\;\bigr) \ \impl\ \delta_{\leq d}(x', y')\; \Bigr) \\
  \delta_{\leq 2d+1}(x,y)  \ \isdef\quad &
  \exists z\; \big(\delta_{\leq 1}(x,z)\ \und \ \delta_{\leq 2d}(z,y)\big)\,.
  \end{split}
\end{gather*}
It is easy to check that $\delta_{\leq d}$ has size in $\bigO(\log d)$. 

Furthermore, 
let $\delta_{=d}(x,y) \isdef \delta_{\leq d}(x,y) \und \neg\delta_{\leq d{-}1}(x,y)$ be the $\FO(E)$-formula expressing that there is a directed path of length exactly $d$ from node $x$ to node $y$. Of course, the size of $\delta_{=d}(x,y)$ is linear in $\delta_{\leq d}$ and therefore also in $\bigO(\log d)$. 

  Consider  for each $d\geq 0$  the following $\FO(E)$-formula:
  \begin{gather*}
    \begin{split}
      \gamma_d(x) \ \isdef \quad & \exists y\ \delta_{=d}(x, y) \ \und \\
      & 
      \begin{split}
         \forall y\ \bigl( 
        & \delta_{\leq d-1}(x, y) \ \impl \\
        &\; \exists z_0\; \exists z_1\; (E(y,z_0) \und E(y,z_1) \und \neg z_0{=}z_1) \bigr).
      \end{split}
    \end{split}
  \end{gather*}
  For each $d\geq 0$, the formula $\gamma_d(x)$ states that there is a node $y$ that is reachable from $x$ by a path of length exactly $d$ and that every node $y$ that is reachable from $x$ by a path of length less than $d$ has exactly two children. Hence, for every binary forest $\F$ and each $a\in F$, we have that~$\F \models \gamma_d[a]$ iff $\F_a[\leq d]$ is a complete binary tree of depth $d$.
  Since $\delta_{=d}$ and $\delta_{\leq d{-}1}$ have size in $\bigO(\log d)$, also $\gamma_d$ has size in $\bigO(\log d)$.

  Furthermore, for each $i \in \set{0, 1,2, 3}$, it is straightforward to define a formula $\enc_{-1,i}(x)$ that is satisfied if $x$ is the root node of a binary tree that is isomorphic to the binary tree $\B_{-1}(i)$ depicted in Figure~\ref{fig:bintreesStart}. 

   With this preparation, the $\FO(E)$-formula $\enc_h(x)$ can be defined corresponding to the definition of binary tree encodings with parameter $h$.
   Let 
   \begin{gather*}
     \enc_{-1}(x) \ \isdef \quad  \enc_{-1,0}(x) \ \oder \  \enc_{-1,1}(x) \ \oder \  \enc_{-1,2}(x) \ \oder \  \enc_{-1,3}(x).
   \end{gather*}
   Clearly, the formula $\enc_{-1}(x)$ is satisfied if $x$ is the root node of a binary tree that is isomorphic to the binary tree~$\B_{-1}(i)$, for one of the numbers $i\in\set{0,1,2,3}$. 

   For $h \geq 0$, we let
  \begin{gather*}
    \enc_h(x)\ \isdef\quad  \gamma_{\Tower(h+1)-1}(x) \ \und \ \forall y \bigl( \delta_{=\Tower(h+1)}(x,y)\ \impl\ \enc_{h-1}(y) \bigr). 
  \end{gather*}
  For each $h \geq 0$, 
  the size of the formulas $\gamma_{\Tower(h+1)-1}$ and $\delta_{=\Tower(h+1)}$ is linear in $\Tower(h)$. An easy induction shows that, for each $h\geq 0$,
  \begin{gather*}
    \sum_{i=0}^h \Tower(i)\quad <\quad 2\cdot\Tower(h).
  \end{gather*}
  Hence, for $h\geq 0$,  the formula $\enc_h$ has size in $\bigO(\Tower(h))$.

  We let $\fmin_{-1}(x) \isdef \enc_{-1,0}(x)$. For $h \geq 0$, we let $\fmin_h(x) \isdef \neg \exists y\;\delta_{=\Tower(h+1)}(x, y)$. Clearly, $\min_h$ is of size $\bigO(\Tower(h))$. 

The construction of the formulas $\eq_h$, $\less_h$, $\suc_h$ and $\fmax_h$  are easy adaptations of Lemma~3.2 in \cite{DGKS07} (Lemma~10.21 in \cite{FlGr06}) and Lemma~3.4 in~\cite{DGKS07}. Their construction is best understood by keeping in mind the binary expansions of the numbers, encoded by binary trees.

Consider the formula
\begin{gather*}
  \feq_{-1}(x, y) \ \isdef \quad \Oder_{i=0}^3\ \Bigl(\,\enc_{-1,i}(x) \ \und \ \enc_{-1,i}(y)\,\Bigr). 
\end{gather*}\enlargethispage{2\baselineskip}
Note that the formula $\feq_{-1}(x, y)$ is satisfied if there is an $i \in \set{0,1,2,3}$ such that $x$ and $y$ are root nodes of two binary trees isomorphic to the binary tree $\B_{-1}(i)$ depicted in Figure~\ref{fig:bintreesStart}. 

For each~$h\geq 0$, we let
  \begin{gather*}
    \begin{split}
       \feq_h(x,y) \ \isdef \quad & \bigl( \exists x'\; \delta_{=\Tower(h+1)}(x,x') \ \gdw \ \exists y'\;\delta_{=\Tower(h+1)}(y,y')\bigr) \ \und \\
      & 
      \begin{split}
        \forall x'\; \Bigl( 
        & \delta_{=\Tower(h+1)}(x,x') \ \impl \\
        &
        \begin{split}
          \exists y'\; \Bigl( 
          & \delta_{=\Tower(h+1)}(y,y') \ \und \\
          &
          \begin{split}
            \forall y''\; \bigl( 
            & \delta_{=\Tower(h+1)}(y,y'') \ \impl \\
            &
            \begin{split}
              \exists x''\; \bigl( 
              & \delta_{=\Tower(h+1)}(x,x'') \ \und \\
              &
              \begin{split}
                \forall u \forall v \bigl( 
                & ((u{=}x'\ \und\ v{=}y')\ \oder \\
                & \; (u{=}x''\ \und\ v{=}y''))\\
                & \impl \ \feq_{h-1}(u,v)\bigr)\bigr)\bigr)\Bigr)\Bigr)
              \end{split}
            \end{split}
          \end{split}
        \end{split}
      \end{split}
    \end{split}
  \end{gather*}
  As for $\enc_h$, it can easily be seen that for each $h\geq 0$, the size of $\feq_h$ is in $\bigO(\Tower(h))$.
  \smallskip
  
  The following formula $\less_{-1}(x, y)$ is satisfied if there are numbers $i,j \in \set{0,\ldots,3}$ with~$i < j$ such that $x$ and $y$ are root nodes of binary trees that are isomorphic to the binary trees $\B_{-1}(i)$ and~$\B_{-1}(j)$, respectively:
  \begin{gather*}
    \less_{-1}(x, y) \ \isdef \quad
    \Oder_{0 \leq i < j \leq 3} \Bigl(\; \enc_{-1,i}(x) \ \und \ \enc_{-1,j}(y)\;\Bigr).
  \end{gather*}
  For every $h \geq 0$, let
  \begin{gather*}
    \begin{split}
      \less_h(x,y)\ \isdef \quad \exists y' \Bigl( 
      &  \delta_{=\Tower(h+1)}(y,y') \ \und \\
      & \forall x' \; \bigl(  \delta_{=\Tower(h+1)}(x,x')\ \impl\  \neg\eq_{h-1}(x',y')\bigr) \ \und \\
      & 
      \begin{split}
        \forall x''\; \bigl( 
        & (  \delta_{=\Tower(h+1)}(x,x'')\ \und\ \less_{h-1}(y',x'') ) \\
        & \impl \ \exists y''\; ( \delta_{=\Tower(h+1)}(y,y'')\ \und\  \feq_{h-1}(y'',x''))\bigr)\Bigr)
      \end{split}
    \end{split}
  \end{gather*}
  There is a number $c > 0$ such that, for all $h\geq 0$,
  \begin{gather*}
    \size{\less_h}\quad \leq\quad c\ +\ \sum_{i=0}^h\; (c\ +\ 4 \cdot \size{\delta_{=\Tower(i+1)}}\ +\ 2\cdot \size{\feq_{i-1}} ).
  \end{gather*}
  Since, for $i\geq 0$, the formulas $ \delta_{=\Tower(i+1)}$ and $\feq_{i-1}$ have size in $\bigO(\Tower(h))$, the formula $\less_h$ has size in $\bigO(\Tower(h))$.
  \smallskip

  Similarly to the formula $\less_{-1}$, a formula $\suc_{-1}$ can be chosen as
  \begin{gather*}
    \suc_{-1}(x, y) \ \isdef \quad
    \Oder_{0 \leq i \leq 2} \Bigl(\; \enc_{-1,i}(x) \ \und \ \enc_{-1,i+1}(y)\;\Bigr).
  \end{gather*}
  For each $h \geq 0$, we let
  \begin{gather*}
    \begin{split}
      \suc_h(x,y)\ \isdef \quad \exists y' \Bigl(
      &  \delta_{=\Tower(h+1)}(y,y') \\[1mm]
      \und\quad 
      & \forall y'' \bigl( (  \delta_{=\Tower(h+1)}(y,y'')\ \und\ \neg \feq_{h-1}(y'',y')) \ \impl\ \less_{h-1}(y',y'')\bigr) \\[1mm]
      \und\quad 
      & \forall x' \bigl(  \delta_{=\Tower(h+1)}(x,x')\ \impl\ \neg \eq_{h-1}(x',y')\bigr) \\[1mm]
      \und\quad 
      &  
      \forall y'' \bigl( (  \delta_{=\Tower(h+1)}(y,y'')\ \und\  \less_{h-1}(y',y'')) \\[0.5mm]
      & \quad\quad \impl\ \exists x'' ( \delta_{=\Tower(h+1)}(x,x'')\ \und\ \eq_{h-1}(x'',y''))\bigr) \\[1mm]
      \und\quad
      &
      \forall x'' \bigl( (  \delta_{=\Tower(h+1)}(x,x'')\ \und\  \less_{h-1}(y',x'')) \\[0.5mm]
      & \quad\quad \impl\ \exists y''(  \delta_{=\Tower(h+1)}(y,y'')\ \und\  \feq_{h-1}(y'',x''))\bigr) \\[1mm]
      \und\quad
      & \Bigl(\neg\fmin_{h-1}(y') \\[0.5mm]
      &\ \impl\ \bigl( \exists x'\;(  \delta_{=\Tower(h+1)}(x,x')\ \und\  \fmin_{h-1}(x')) \\[0.5mm]
      & \ \qquad\und\ 
      \forall x'\; \bigl( ( \delta_{=\Tower(h+1)}(x,x')\ \und\ \less_{h-1}(x',y')) \\[0.5mm]
      & \qquad\qquad\quad\ \ \impl\ 
       \exists z\; (\suc_{h-1}(x',z) \\
       & \qquad\qquad\qquad\qquad\ \ \; \und\ (z{=}y'\; \oder\;    \delta_{=\Tower(h+1)}(x,z)))\bigr)\bigr)\Bigr)\Bigr)
    \end{split}
  \end{gather*}
  Recall that the formulas $ \delta_{=\Tower(h+1)}$, $\feq_h$, and $\less_h$ have size in $\bigO(\Tower(h))$. Hence, a simple induction shows that for $h\geq 0$, also $\suc_h$ has size in $\bigO(\Tower(h))$.
  \smallskip

  Finally, let $\fmax_{-1}(x) \isdef \enc_{-1,3}(x)$; and for each $h \geq 0$, let
  \begin{gather*}
    \begin{split}
      \fmax_{h}(x) \ \isdef \quad & \exists y\; \bigl( \delta_{=\Tower(h+1)}(x,y) \ \und \ \fmin_{h-1}(y)\bigr) \ \und \\[1mm]
      & \forall y\; \bigl( \delta_{=\Tower(h+1)}(x,y) \\[0.5mm]
      &\quad \ \impl \ ( \fmax_{h-1}(y) 
        \ \oder\ \exists z\; ( \delta_{=\Tower(h+1)}(x,z)\ \und\ \suc_{h-1}(y,z)))\bigr).
    \end{split}
  \end{gather*}
  Again, a simple induction shows that, for each $h\geq 0$, the size of $\fmax_h$ is in $ \bigO(\Tower(h))$.
  This concludes the proof of Lemma~\ref{lem:arithmetic-formulas}.
\qed

\subsection{Lower bounds for preservation theorems}
\label{subsection:lower-bounds-for-preservation-theorems}
The upper bounds of Section~\ref{section:PreservationTheorems} are complemented by the following two lower bounds
for certain classes of finite acyclic structures of degree $3$ and
first-order sentences that are
preserved under extensions (homomorphisms): We show that even under
these restrictions, a $3$-fold exponential blow-up in terms of the
size of the input sentence is unavoidable when constructing the
equivalent existential (existential-positive) first-order sentence.
\begin{thm}\label{thm:ep-lower-bound}
  Let $\sigma \isdef \set{S_0,S_1,V_0,V_1}$, where $S_0$, $S_1$ are
  binary and $V_0$, $V_1$ are  unary relation symbols, and let
  $\mathfrak{C}$ be the class of all finite ordered binary forests $\F$ of
  signature~$\sigma$, 
  where $V_0^{\F}$ and~$V_1^{\F}$ may be arbitrary subsets of the universe. 
  There is a real number $\epsilon > 0$ and a sequence $(\phi_h)_{h > 1}$ of $\FO(\sigma)$-sentences
  of increasing size such that 
  for each $h > 1$ the following holds: \smallskip
  \begin{enumerate}
  \item $\phi_h$ is preserved under extensions on $\mathfrak{C}$, and\smallskip
  \item 
    every existential $\FO(\sigma)$-sentence that is equivalent to $\phi_h$ on $\mathfrak{C}$ has size at least 
    \begin{gather*}
      2^{2^{2^{\epsilon\cdot\size{\phi_h}}}}.
    \end{gather*}
  \end{enumerate}
\end{thm}
\begin{thm}\label{thm:hp-lower-bound}
  Let $\sigma' \isdef \set{S_0,S_1}\cup\setc{V_M}{M\subseteq\set{0,1}}$,
  where $S_0$, $S_1$ are binary relation symbols and, for each
  $M\subseteq\set{0,1}$, $V_M$ is a unary relation symbol. Let
  $\mathfrak{C}'$ be the class of all finite ordered binary forests~$\F$ over $\sigma'$,
  where $(V_M^{\F})_{M\subseteq\set{0,1}}$ is a partition of the universe.
There is a real number~$\epsilon > 0$ and a sequence $(\phi'_h)_{h > 1}$ of
$\FO(\sigma')$-sentences of increasing size
such that for each~$h > 1$ the following holds:\smallskip
  \begin{enumerate}
  \item $\phi'_h$ is preserved under homomorphisms on
    $\mathfrak{C}'$, and\smallskip
  \item
    every existential-positive $\FO(\sigma')$-sentence that is equivalent to $\phi'_h$ on $\mathfrak{C}'$ has size at least
    \begin{gather*}
      2^{2^{2^{\epsilon\cdot\size{\phi'_h}}}}.
    \end{gather*}
  \end{enumerate}\medskip
\end{thm}

\noindent In Theorem~\ref{thm:ep-lower-bound} and Theorem~\ref{thm:hp-lower-bound}, an \emph{ordered binary forest} of signature $\sigma$ or $\sigma'$ is a structure whose Gaifman graph is a forest and where the binary relation symbols $S_0$ and $S_1$ are interpreted as the left and right successor relation and every node is allowed to have at most one left successor and at most one right successor. An \emph{ordered binary tree} is an ordered binary forest with only one connected component.

The proofs of Theorem~\ref{thm:ep-lower-bound} and
Theorem~\ref{thm:hp-lower-bound}, which can be found below,
use the encoding of numbers by binary trees.
The main challenge here is to find sequences of sentences that not only have
large minimal models but are also preserved under extensions and homomorphisms, respectively. Towards this end, the auxiliary unary relation symbols in $\sigma$ and $\sigma'$ are
introduced to interpret binary tree encodings in ordered binary
forests. 
Both proofs rely on the following observation:
\begin{lem}\label{lem:lower-bound-existential-sentence}
  Let $\sigma$ be a relational signature and let~$\mathfrak{C}$ be a class of $\sigma$-structures that is closed under induced substructures.
  For each $\FO(\sigma)$-sentence $\phi$ and each $N\geq 1$ the following holds: If $\phi$ has a~$\mathfrak{C}$-minimal model of size at least $N$, then every existential $\FO(\sigma)$-sentence that is $\mathfrak{C}$-equivalent to~$\phi$ has size greater than~\mbox{$N$}.
\end{lem}
\proof
  Let $\phi$ be an $\FO(\sigma)$-sentence and let $\A$ be a
  $\mathfrak{C}$\hbox{-}minimal model of $\phi$ with at least $N$ elements. 
For contradiction, assume that $\psi$ is an existential $\FO(\sigma)$-sentence of size at most $N$ that is~$\mathfrak{C}$\hbox{-}equivalent to $\phi$. In particular, $\psi$ has less than $N$ quantifiers. 

Since $\phi$ and $\psi$ are $\mathfrak{C}$-equivalent, $\A$ is also a
model of $\psi$. But since $\psi$ is an existential sentence
with less than $N$ quantifiers, there is an induced substructure $\B$ of
$\A$ of size less than $N$, such that~$\B$ is a model of $\psi$.
Since $\mathfrak{C}$ is closed under induced substructures, $\B$
belongs to $\mathfrak{C}$. And since $\phi$ and $\psi$ are
$\mathfrak{C}$-equivalent, $\B$ also is a model of $\phi$.
However, this contradicts the assumption that $\A$ is a $\mathfrak{C}$\hbox{-}\emph{minimal} model of $\phi$.
\qed
\medskip

Recall the $\FO(E)$-formulas provided by Lemma~\ref{lem:arithmetic-formulas},
which define arithmetic relations between binary tree encodings of numbers.
It is straightforward to use these formulas to construct, for each $h\geq
1$, an $\FO(E)$-sentence of size $\bigO(\Tower(h))$ that has a minimal
model of size at least $\Tower(h{+}3)$. 
However, it is far less obvious to construct
sentences that, at the same time, are preserved under extensions 
(on finite binary forests).
To achieve this, we interpret binary tree encodings in complete
\emph{ordered} binary forests of suitable height. This way, we make
sure that no extension of such binary forests can modify the binary
tree encodings.  

Consider the signature $\sigma \isdef \set{S_0,S_1,V_0,V_1}$ where
$S_0$ and $S_1$ are binary relation symbols and $V_0$ and $V_1$ are
unary relation symbols. 
Recall that in each ordered binary forest $\F$ over the signature $\sigma$, the unary relations $V_0^{\F}$ and $V_1^{\F}$ may be interpreted by
arbitrary sets of nodes of $\F$.
In the following, we denote the class of all finite ordered binary forests over $\sigma$ by $\mathfrak{C}$. Note that every structure in
$\mathfrak{C}$ has degree at most $3$. 

\emph{Complete} ordered binary trees are defined in the obvious way.
Also, we adapt the notions $\T[{\leq}d]$ and $\F_a$ from
Section~\ref{section:binary-tree-encodings} from unordered binary trees $\T$ and
forests $\F$ to \emph{ordered} binary trees $\B$ and forests $\F$ (with
$d\geq 0$ and $a\in F$) in the obvious way. Thus, $\B[{\leq}d]$ is the
subtree of~$\B$ induced by all nodes of $\B$ that are reachable from
the root of $\B$ by a directed path of length at most~$d$. And $\F_a$
is the subtree of $\F$ induced by all nodes reachable by a directed
path from $a$.

Recall the definition of transductions from Section~\ref{sec:transductions} and
consider the following transduction $\Theta \isdef (\theta,\theta_E)$ from $\set{E}$ to $\sigma$, defined by $\theta(x) \isdef x{=}x$ and
\begin{gather*}
  \theta_E(x,y)\ \isdef\ \Oder_{i\in \set{0,1}} \bigl(S_i(x,y) \ \und \ V_i(x)\bigr).
\end{gather*}
 The transduction $\Theta$ makes use of the unary relations $V_0$
 and~$V_1$ to interpret binary forests in ordered binary forests. More
 specifically, for an ordered binary forest $\F$, the structure $\Theta(\F)$ is the  (unordered) binary forest consisting of all nodes from $\F$ and, for all nodes  \mbox{$a,b\in F$}, there is an edge from $a$ to~$b$ iff $b$ is the left
 successor of~$a$ in $\F$ and $a \in V_0^{\F}$, or $b$ is the right
 successor of $a$ in $\F$ and~$a\in V_1^{\F}$. 

\smallskip

\proof[Proof of Theorem~\ref{thm:ep-lower-bound}.]
  Recall that each extension of a structure $\A$ contains $\A$ as an \emph{induced} substructure. 
  Consider a complete ordered binary tree $\A\in\mathfrak{C}$ of
  height $d\geq 1$ and let $a$ be its root node. 
  Assume that $\Theta(\A)_a$ is a binary tree of height at most
  $d{-}1$. Note that every leaf $b$ of $\Theta(\A)_a$ has a left and a
  right successor in $\A$ but is neither contained in $V_0^{\A}$ nor in
  $V_1^{\A}$. Therefore, by construction of the transduction $\Theta$,
  for each extension $\B$ of $\A$ in $\Class$ we have that $\Theta(\A)_a$ and $\Theta(\B)_a$ are isomorphic. 

  We will use this observation to protect binary tree encodings,
  interpreted in complete ordered binary trees, against modifications
  by extensions of the underlying structure. Recall that we know by 
  Lemma~\ref{lem:height-Bhi} that for each $h > 1$ and every 
  $i \in [0,\Tower(h{+}3){-}1]$, each binary tree from
  $\mathfrak{B}_h(i)$ has height less than $2\cdot\Tower(h{+}1)$. 

  Similarly to the $\FO(E)$-formula $\gamma_d(x)$, defined in the proof of
  Lemma~\ref{lem:arithmetic-formulas}, for each $d\geq 0$, there is an
  $\FO(\sigma)$-formula $\gamma^<_d(x)$ of size $\bigO(\log d)$ that
  is satisfied by a node $a$ of an ordered binary forest $\A$ iff
  $\A_a[{\leq} d]$ is a complete ordered binary tree of height $d$. 

  In the following, the transduction $\Theta$ is applied to the
  $\FO(E)$-formulas provided by Lemma~\ref{lem:arithmetic-formulas}. 
  Let  $(\phi_h)_{h > 1}$ be the sequence of
  $\FO(\sigma)$\hbox{-}sentences that is defined, for each $h > 1$,
  by  
  \begin{gather*}
    \begin{split}
      \phi_h\ \isdef \quad & \exists x \; \bigl(\enc^<_h(x)\ \und\ \Theta(\fmin_h)(x)\bigr) \ \  \und \\
      &
      \begin{split}
        \forall x \; \bigl( 
        & \enc^<_h(x)\ \ \impl\ \\
        & \bigl(\Theta(\fmax_h)(x)\ \oder\ \exists y \ (\enc^<_h(y)\ \und\ \Theta(\suc_h)(x,y))\bigr)\bigr), 
      \end{split}
    \end{split}
  \end{gather*}
  where
  \begin{gather*}
    \enc^<_h(x)\ \isdef\quad \Theta(\enc_h)(x)\ \und\  \gamma^<_{2\cdot\Tower(h{+}1)}(x).   
  \end{gather*}
  The formula $\phi_h$
 expresses that there is a node $x$ that encodes the number $0$ as a binary tree encoding with parameter $h$, and for each number $i < \Tower(h{+}3)-1$ that is encoded by a node $x$, there also exists a node $y$ that encodes the number $i+1$. 

  For the fixed transduction $\Theta$ it follows from
  Lemma~\ref{lem:transduction-complexity} that, for each $\FO(E)$-formula $\phi$, the size of $\Theta(\phi)$ is linear in the size of $\phi$. Therefore, by Lemma~\ref{lem:arithmetic-formulas} and since $\gamma^<_{2\cdot\Tower(h{+}1)}$ has size in~$\bigO(\Tower(h))$, the formula $\phi_h$ also has size in $\bigO(\Tower(h))$.

  For the remainder of the proof, we fix an arbitrary $h > 1$.
  The following claim follows from the choice of the sentence $\phi_h$ and from Lemma~\ref{lem:lower-bound-existential-sentence}.
 \newclaims
  \begin{clm}\label{claim:ep-lower-bound-1}
    Every existential $\FO(\sigma)$-sentence that is equivalent to $\phi_h$ on $\mathfrak{C}$ has size at least $\Tower(h{+}3)$.
  \end{clm}
   \uproof{Claim~\ref{claim:ep-lower-bound-1}} It is easy to see that there are structures in $\Class$ that satisfy $\phi_h$. Furthermore, by definition of the subformulas of $\phi_h$ (see Lemma~\ref{lem:arithmetic-formulas}), each model $\A\in\mathfrak{C}$ of $\phi_h$ has to contain at least $\Tower(h{+}3)$ pairwise distinct nodes $a_0,\ldots,a_{\Tower(h{+}3)-1}$ 
such that, for each \mbox{$i \in [0,\Tower(h{+}3){-}1]$}, the binary tree  $\Theta(\A)_{a_i}$ is a binary tree encoding with parameter $h$ of the number $i$, i.e., the binary tree $\Theta(\A)_{a_i}$ belongs to the set $\mathfrak{B}_h(i)$. 

Together with Lemma~\ref{lem:lower-bound-existential-sentence}, this observation completes the proof of Claim~\ref{claim:ep-lower-bound-1}. \qed
   \smallskip

  \begin{clm}\label{claim:ep-lower-bound-2}
    $\phi_h$ is preserved under extensions on $\mathfrak{C}$.
  \end{clm}
  \uproof{Claim~\ref{claim:ep-lower-bound-2}}
  Let $\A\in\mathfrak{C}$ be a model of $\phi_h$. 
  By definition of $\phi_h$, there are pairwise distinct nodes~$a_0,\ldots,a_{\Tower(h{+}3)-1}$ in $\A$ such that, for each $i\in [0,\Tower(h{+}3){-}1]$, the binary tree
  $\Theta(\A)_{a_i}$ belongs to the set $\mathfrak{B}_h(i)$.

    Let $\B\in\mathfrak{C}$ be an extension of $\A$. 
    By construction of $\phi_h$, for each $i \in
    [0,\Tower(h{+}3){-}1]$, the substructure $\A_{a_i}[\leq 2{\cdot}\Tower(h{+}1)]$ is
    a complete ordered binary tree. Therefore, $\Theta(\A)_{a_i}$ and~$\Theta(\B)_{a_i}$ are isomorphic and thus, also
    $\Theta(\B)_{a_i}$ belongs to the set $\mathfrak{B}_h(i)$. On the other hand, let $b$
    be a node from $\B$ such that $\B \models \enc^<_h[b]$. Then,
    there is an \mbox{$i \in [0,\Tower(h{+}3){-}1]$} such that
    $\Theta(\B)_b$ belongs to $\mathfrak{B}_h(i)$ and hence, either
    $\B\models\Theta(\fmax_h)[b]$ or $\B \models
    \Theta(\suc_h)[b,a_{i+1}]$.  
    Altogether, it follows that $\B\models\phi_h$. This completes the proof of Claim~\ref{claim:ep-lower-bound-2}.\qed 

    \smallskip
     Since $(\phi_h)_{h > 1}$ is a sequence of $\FO(\sigma)$-sentences of increasing size $\bigO(\Tower(h))$, there is a real number $\epsilon > 0$ such that $\epsilon \cdot \size{\phi_h} \leq \Tower(h)$, for every $h > 1$.
 By Claim~\ref{claim:ep-lower-bound-1}, every existential $\FO(\sigma)$-sentence that is equivalent to $\phi_h$ on $\mathfrak{C}$ has size at least
    \begin{gather*}
      \Tower(h{+}3)\quad =\quad 2^{2^{2^{\Tower(h)}}}
      \quad \geq\quad 2^{2^{2^{\epsilon \cdot \size{\phi_h}}}}.
    \end{gather*}    
    This concludes the proof of Theorem~\ref{thm:ep-lower-bound}.
\qed
\medskip 

For the proof of Theorem~\ref{thm:hp-lower-bound} we need to
construct $\FO$-sentences with large minimal models that are preserved under
homomorphism. 
Let \mbox{$\sigma' \isdef
  \set{S_0,S_1}\cup\setc{V_M}{M\subseteq\set{0,1}}$} be a signature, where $S_0$ and $S_1$
are binary relation symbols and, for each $M\subseteq\set{0,1}$, $V_M$
is a unary relation symbol. Consider the class $\mathfrak{C}'$ of all
finite \emph{ordered and colored binary forests} over $\sigma'$, i.e.,
finite $\sigma'$-structures $\A$ where the binary relations
$S_0^{\A}$ and $S_1^{\A}$ correspond to the left and right successor
relation of a forest,
every node is allowed to have at most one left and at most one right
successor, and the unary relations $V_M^{\A}$, for $M\subseteq
\set{0,1}$, are a partition of the universe of $\A$.
Note that, for each $\A \in \Class'$ and each $\B\in\Class'$, if there is a homomorphism $h$ from $\A$ to $\B$ then for every $a\in A$ and each $M \subseteq \set{0,1}$, it holds that $a \in V^{\A}_M$ iff $h(a)\in V^{\B}_M$.

We only sketch a proof of Theorem~\ref{thm:hp-lower-bound}, which is very similar to the one of Theorem~\ref{thm:ep-lower-bound}.

Let $\Theta'$ be the transduction $(\theta', \theta'_E)$ from $\set{E}$ to $\sigma'$ that is defined by $\theta'(x) \isdef x{=}x$ and
\begin{gather*}
  \theta'_E(x,y) \ \isdef \ \Oder_{i\in\set{0,1}} \Bigl(S_i(x,y)\quad \und\
  \Oder_{M\subseteq\set{0,1}, i\in M} V_M(x)\Bigr).
\end{gather*}

\proof[Proof of Theorem~\ref{thm:hp-lower-bound}.]
  For colored and ordered binary trees and forests from $\mathfrak{C}'$, we make use of the same notation already introduced for ordered binary forests.
  Consider a complete colored and ordered binary tree \mbox{$\A\in\mathfrak{C}'$} of height $d\geq 1$ and let $a$ be its root node. Assume that $\Theta(\A)_a$ is a binary tree of height at most $d{-}1$.
  Observe that every homomorphism $h$ from $\A$ to a colored and ordered binary forest $\B\in\mathfrak{C}'$ is injective. Furthermore, for each $\B\in\mathfrak{C}'$ for which there is a homomorphism $h$ from $\A$ to $\B$, the binary trees $\Theta(\A)_a$ and $\Theta(\B)_{h(a)}$ are isomorphic. 

  Define a sequence $(\phi'_h)_{h > 1}$ of $\FO(\sigma')$-formulas similar to the sequence $(\phi_h)_{h> 1}$ in the proof of Theorem~\ref{thm:ep-lower-bound}, with the only modification being the application of the transduction $\Theta'$ instead of $\Theta$. Of course, for each $h > 1$, also $\phi'_h$ has size $\bigO(\Tower(h))$.

  Let $h > 1$. 
  Since every existential-positive sentence is an existential sentence and the class $\mathfrak{C}'$ is closed under induced substructures we can follow the lines of the proof of Theorem~\ref{thm:ep-lower-bound} to show that every existential-positive $\FO(\sigma)$-sentence that is equivalent to $\phi'_h$ on $\mathfrak{C}'$ has size at least~$\Tower(h{+}3)$.

  Similarly, using the observation above, some small adaptations to the
  proof of Theorem~\ref{thm:ep-lower-bound} suffice to show that
  $\phi'_h$ is preserved under homomorphisms on $\mathfrak{C}'$.  
  This concludes the proof of Theorem~\ref{thm:hp-lower-bound}.
\qed

\subsection{Lower bounds for Feferman-Vaught decompositions}
\label{subsection:lower-bounds-for-feferman-vaught-decompositions}
The following two lower bounds show that for structures of degree $3$
or $2$ the algorithm of Theorem~\ref{thm:fv-upper-bound} for the construction of reduction sequences for Feferman-Vaught decompositions is basically
optimal. 
Recall that a \emph{binary forest} is a disjoint union of directed
trees of signature $\set{E}$, where every node has at most~$2$
children. 
\begin{thm}\label{thm:fv-lower-bound-BF}
  There is a real number $\epsilon > 0$ and a sequence of $\FO(E)$-sentences
  $(\phi_h)_{h\geq 1}$ of increasing size such that, for every
  \mbox{$h\geq 1$}, every $2$-disjoint decomposition for $\phi_h$ on
  finite binary forests has size at least
  \begin{gather*}
    2^{2^{2^{\epsilon \cdot
        \size{\phi_h}}}}.
  \end{gather*}
\end{thm}\medskip

\noindent Let $\sigma' \isdef \set{S, L_0, L_1}$ be the signature consisting of
a binary relation symbol $S$ and two unary relation symbols $L_0$ and
$L_1$. A \emph{labeled chain} is a finite $\sigma'$-structure
$\mathcal{C}$ whose $\set{S}$-reduct is a chain of finite length,
i.e., a finite directed path, and where the sets $L_0^{\mathcal{C}}$
and $L_1^{\mathcal{C}}$ are disjoint subsets of the universe of $\C$. 
The class of all $\sigma'$\hbox{-}structures that are disjoint unions
of finitely many labeled chains is denoted by $\mathfrak{UC}$. Note
that all structures in $\mathfrak{UC}$ have degree at most two. 

\begin{thm}\label{thm:fv-lower-bound-UC}
  There is a real number $\epsilon > 0$ and a sequence of
  $\FO(\sigma')$-sentences $(\phi'_h)_{h\geq 1}$ of increasing size
  such that, for every \mbox{$h \geq 1$}, every $2$-disjoint
  decomposition for $\phi'_h$ on $\mathfrak{UC}$ has size at least 
  \begin{gather*}
    2^{2^{\epsilon \cdot \size{\phi'_h}}}.
  \end{gather*}
\end{thm}\medskip

\noindent Both theorems are corollaries to a generalisation of Proposition~6.7 in~\cite{GoeJL12}.
For proving our lower bounds, we use the following Lemma \ref{lemma:fv-lower-bound-generalised},
along with suitable encodings of numbers by binary trees (for
Theorem~\ref{thm:fv-lower-bound-BF}) and labeled paths (for
Theorem~\ref{thm:fv-lower-bound-UC}).
The lemma is proved by a simple counting argument and distills the
combinatorial essence of 
the proof of Proposition~6.7 in \cite{GoeJL12}. 

\begin{lem}\label{lemma:fv-lower-bound-generalised}  
Let $\sigma$ be a relational signature and let $\sigma_2 \isdef \sigma\cup \set{P_1,P_2}$, where $P_1$ and $P_2$ are unary relation symbols that are not contained in $\sigma$. Let $\mathfrak{C}$ be a class
of $\sigma$-structures, and let $\phi$ be an $\FO(\sigma_2)$\hbox{-}sentence. 
Let $H\geq 1$. If there are $\sigma$-structures $\A_0,\ldots,
 \A_{2^H-1} \in \mathfrak{C}$ such that for all $i,j\in [0,2^H{-}1]$
it holds that 
\begin{gather*}
    \A_i \oplus \A_j \quad \models\quad \phi
    \qquad\text{iff}\qquad
    i = j,
\end{gather*}
then every $2$-disjoint decomposition for $\phi$ on $\mathfrak{C}$ has
size at least $H$.
\end{lem}

\proof
  Assume that  $\A_0,\ldots,\A_{2^H-1}$ are structures from $\mathfrak{C}$ such that for all $i,j\in [0,2^H{-}1]$,
  \begin{gather}\label{eq:AiAjPhiEq}
    \A_i \oplus \A_j\quad \models\quad \phi
    \qquad\text{iff}\qquad i = j.
  \end{gather}
  For contradiction, assume that there is a $2$-disjoint decomposition
  $D\isdef (\Delta_1,\Delta_2,\beta)$ for $\phi$ on $\mathfrak{C}$ of size less than $H$. Thus, $\Delta_1$ and $\Delta_2$ are finite sets of
  $\FO(\sigma)$-sentences and $\beta$ is a propositional formula with
  variables from the set $\AVARS_D \isdef \set{ \AVAR_{k,\delta} \colon k \in
    [1,2], \delta \in \Delta_k}$, and for all $i,j\in [0,2^H{-}1]$ we
  have 
  \begin{gather}\label{eq:AiAjPhiBeta}
    \A_i\oplus\A_j\quad \models\quad \phi
    \qquad\text{iff}\qquad
    \mu_{i,j}\quad \models\quad \beta,
  \end{gather}
  where $\mu_{i,j} \colon \AVARS_D \to \set{0,1}$ assigns variables of $\beta$ such that for all $\delta_1\in\Delta_1$,
  \begin{gather*}
    \mu_{i,j}(\AVAR_{1,\delta_1}) = 1 
    \qquad\text{iff}\qquad 
    \A_i\quad \models\quad \delta_1,
  \end{gather*}
  and for all $\delta_2\in\Delta_2$,
  \begin{gather*}
    \mu_{i,j}(\AVAR_{2,\delta_2}) = 1 
    \qquad\text{iff}\qquad 
    \A_j\quad \models\quad \delta_2.
  \end{gather*}
  From \eqref{eq:AiAjPhiBeta} and \eqref{eq:AiAjPhiEq} as just mentioned above, we know that
  for all $i,j\in [0,2^{H}{-}1]$,
  \begin{gather}\label{eq:MuijEq}
    \mu_{i,j}\quad \models\quad \beta
    \qquad\text{iff}\qquad 
    i=j.
  \end{gather}
  
  Note that the number of variables of $\beta$ is less than $H$, and hence the
  number of distinct variable assignments is less than $2^H$.
  Thus, there
  exist $i, j\in [0,2^H{-}1]$ with $i\not=j$ such that $\mu_{i,i} =
  \mu_{j,j}$. We let $\mu\isdef \mu_{i,i}$. Clearly, due to
  \eqref{eq:MuijEq}, we have 
  \begin{equation}\label{eq:MuBeta}
    \mu\quad\models\quad\beta.
  \end{equation}
  Along the definition of $\mu_{i,j}$, and using the fact that $\mu=\mu_{i,i}=\mu_{j,j}$,  
  it is straightforward to see that $\mu_{i,j}=\mu$.
  Thus, from \eqref{eq:MuBeta} and \eqref{eq:AiAjPhiBeta} we obtain
  that $\A_i \oplus \A_j \models \phi$. This, however, contradicts~\eqref{eq:AiAjPhiEq}.
\qed
\medskip

Using the above lemma, along with the binary tree encodings introduced in Subsection~\ref{section:binary-tree-encodings}
we obtain a proof of Theorem~\ref{thm:fv-lower-bound-BF}.
Let us remark that the $\FO$-formulas we use within the proof
do not even make use of the unary relation symbols $P_1$ and $P_2$. 
\smallskip
\proof[Proof of Theorem~\ref{thm:fv-lower-bound-BF}.]
  We use the $\FO(E)$-formula $\feq_h(x,y)$ from Lemma~\ref{lem:arithmetic-formulas}.
  For every $h \geq 1$, let 
  \begin{gather*}
      \phi_h \ \isdef \quad \forall x\ \Bigl( \Root(x)\ \impl\ \exists y\ \bigl( \Root(y)\ \und\ \feq_h(x, y)\ ¸\und\ \neg x{=}y\bigr) \Bigr).
  \end{gather*}
  The formula $\Root(x)$ is satisfied by exactly the root nodes of the trees in a forest; it can easily defined by \mbox{$\Root(x)\isdef\neg\exists y\; E(y,x)$}.
  Because $\feq_h$ has  size in $\bigO(\Tower(h))$, also $\phi_h$ has size in~$\bigO(\Tower(h))$.
  
  Let $h \geq 1$ and let $H \isdef \Tower(h{+}3)$. For every $i \in
  [0,H{-}1]$, let $\mathcal{B}_{h,i}$ be a binary tree encoding of $i$
  with parameter $h$, i.e., let $\mathcal{B}_{h,i}\in\mathfrak{B}_h(i)$
  (see Section~\ref{section:binary-tree-encodings} above for the definition of~$\mathfrak{B}_h(i)$).
    Furthermore, for each $i \in [0,2^H{-}1]$, let $\A_{h,i}$ be the disjoint union of all binary tree encodings $\mathcal{B}_{h,i'}$ for $i' \in [0,H{-}1]$ such that $\text{Bit}(i',i)=1$.
  
  It is easy to verify that for numbers $i,j\in [0,2^H{-}1]$
  \begin{gather*}
    \A_{h,i} \oplus
    \A_{h,j}\quad \models\quad \phi_h
    \qquad\text{iff}\qquad i = j.
  \end{gather*}
  By Lemma~\ref{lemma:fv-lower-bound-generalised}, every $2$-disjoint decomposition for $\phi_h$ on the class of binary forests has size at least $H=\Tower(h{+}3)$.
  Analogously to the final step of the proof of Theorem~\ref{thm:ep-lower-bound}, and since $\phi_h$ has size in $\bigO(\Tower(h))$, we can conclude that
  \begin{gather*}
    H\quad \geq\quad 2^{2^{2^{\epsilon\cdot \size{\phi_h}}}}, 
  \end{gather*}
  for a suitable real number $\epsilon >0$.    
  This completes the proof of Theorem~\ref{thm:fv-lower-bound-BF}.
\qed
\medskip

For the proof of Theorem~\ref{thm:fv-lower-bound-UC}
we employ the obvious encodings of numbers by strings (cf., e.g.,
\cite{FrickGrohe-FO-MSO-revisited,HKS13-LICS}): 
Let $\Sigma = \set{0,1}$. For $h\geq 1$ and $i \in [0,2^{2^h}{-}1]$ let
$\bin_{2^h}(i)$ denote the binary expansion of $i$ of length $2^h$. 
Strings $w\in\Sigma^{+}$ are represented by structures $\B_w\in\mathfrak{UC}$ in the usual way: the universe of $\B_w$ is the set of positions of the string $w$, the relation $S^{\B_w}$ is the successor relation on the positions of $w$, and $L^{\B_w}_a$ consists, for each $a\in\Sigma$, of all positions of $w$ that carry the letter $a$.

We will use structures $\B_w$ for strings $w=\bin_{2^h}(i)$ with \mbox{$i \in [0,2^{2^h}{-}1]$}, and we will rely on the following result of~\cite{FrickGrohe-FO-MSO-revisited}.
\begin{lem}[{Lemma 20 in \cite{FrickGrohe-FO-MSO-revisited}}]
  \label{lem:eq-Deg-2}
  There is a sequence of $\FO(\sigma')$-formulas $(\feq'_h(x, y))_{h\geq 1}$ of size
  $\size{\feq'_h(x,y)} \in \bigO(h)$ such that for all structures $\mathcal{B}\in\mathfrak{UC}$ and all nodes $a,b$ of $\mathcal{B}$ the following holds: If $a$ and $b$ are the starting positions of labeled chains isomorphic to $\mathcal{B}_{\text{bin}_{2^h}}(i)$ and $\mathcal{B}_{\text{bin}_{2^h}}(j)$, respectively, for $i,j \in [0,2^{2^h}{-}1]$, then
  \begin{align*}
     \hspace{4.5cm} & (\mathcal{B}, a, b)\quad \models\quad \feq'_h(x, y) \qquad\text{iff}\qquad i=j. & & \hspace{3.27cm}\qEd
  \end{align*}
\end{lem}

\medskip
We are now ready to prove Theorem~\ref{thm:fv-lower-bound-UC}.
\smallskip
\proof[Proof of Theorem~\ref{thm:fv-lower-bound-UC}.]
  For every $h \geq 1$, let 
  \begin{gather*}
    \phi'_h \ \isdef \quad \forall x\ \Bigl( \Root'(x)\ \impl\ \exists y\ \bigl( \Root'(y)\ \und\  \feq'_h(x, y)\ \und\ \neg x{=}y\bigr) \Bigr),
  \end{gather*}
  where $\Root'(x) \isdef \neg \exists y\ S(y,x)$.
Because $\feq'_h$ has size in $\bigO(h)$, also $\phi'_h$ has size in $\bigO(h)$.
   
  Let $h \geq 1$ and let $H \isdef 2^{2^h}$. 
  For each $i \in [0, 2^H{-}1]$, let $\A_{h,i}$ be the disjoint union of all labeled chains  $\mathcal{B}_{\text{bin}_{2^h}}(i')$ for $i' \in [0,H{-}1]$ such that $\text{Bit}(i',i)=1$.
  It is easy to verify that for all numbers $i,j \in [0, 2^H{-}1]$ it holds that
  \begin{gather*}
    \A_{h,i} \oplus
    \A_{h,j}\quad \models\quad \phi'_h
    \qquad\text{iff}\qquad i = j.
  \end{gather*}
  Therefore, by Lemma~\ref{lemma:fv-lower-bound-generalised}, every $2$-disjoint
  decomposition for $\phi'_h$ on $\mathfrak{UC}$ has size at least
  \begin{gather*}
    H\quad =\quad 2^{2^{h}} \quad \geq \quad 2^{2^{\epsilon\cdot \size{\phi_h}}}, 
  \end{gather*}
  for a suitable real number $\epsilon >0$.    
  This completes the proof of Theorem~\ref{thm:fv-lower-bound-UC}.
\qed

\medskip

\section{Concluding remarks}
\label{section:conclusion}
In this section, we give a short summary of our main results and some directions for further research. For this, we fix a relational signature $\sigma$ and a class $\Class_d$ of $\sigma$-structures of degree $\leq d$, for a $d \geq 3$. 

Our first two main results, to which we will refer in the following with (PE) and (PH), are algorithmic versions of two preservations theorems that are restricted to the class $\Class_d$. Both require the class $\Class_d$ to be closed under induced substructures and disjoint unions. \smallskip
\begin{itemize}[label=(PH)]
\item[(PE)] For each sentence $\phi$ of $\FOMOD{m}(\sigma)$ that is preserved under extensions on $\Class_d$, a $\Class_d$-equivalent existential $\FO(\sigma)$-sentence can be constructed in $5$\hbox{-}fold exponential time. \medskip 
\item[(PH)] For each sentence $\phi$ of $\FOMOD{m}(\sigma)$ that is  preserved under homomorphisms on $\Class_d$, a $\Class_d$-equivalent existential-positive $\FO$-sentence can be constructed in $4$-fold exponential time, provided that $\Class_d$ is decidable in $1$-fold exponential time. \smallskip
\end{itemize}
For (PE) and (PH) we have shown that a $3$-fold exponential blow-up of the computed existential or existential-positive sentence is unavoidable. 

Our third main result is an algorithmic version of the Feferman-Vaught theorem for disjoint sums (and, using transductions, other products of structures) of $\sigma$-structures of bounded degree.\smallskip
\begin{itemize}[label=(PH)]
\item[(FV)] For each formula $\phi(\ov{x})$ of $\FO(\sigma_s)$, where $s \geq 1$, a disjoint decomposition with respect to disjoint sums and direct products of $\Class_d$-structures can be computed in $3$-fold exponential time.\smallskip
\end{itemize}
Furthermore, a matching lower bound shows that our algorithm for (FV) is basically optimal. 

For most of our results it is known that there is no hope to considerably extend the class of structures to which they apply. Notably, concerning (PE) and (FV), non-elementary lower bounds on trees of unbounded degree are known from \cite{DGKS07}.

An obvious task for future research is to close the gap between the upper and lower bounds concerning (PE) and (PH). Another direction for future work is to consider other generalised quantifiers (instead of modulo counting quantifiers) and study to what extent corresponding generalisations of our results (PE), (PH) and (FV) can be achieved.

\section*{Acknowledgement}
We would like to thank the anonymous referees for their valuable comments.
Furthermore, the second author would like to thank \mbox{Dietrich} Kuske for inspiring
discussions on Feferman-Vaught decompositions. 

\bibliographystyle{halpha}

\vspace{-20 pt}
\end{document}